\documentclass[traditabstract,longauth]{aa} 
\usepackage{multirow}
\usepackage{xcolor}
\usepackage[breaklinks, colorlinks, citecolor=blue, linkcolor=blue]{hyperref}
\usepackage{caption}
\usepackage{subcaption}
\usepackage{threeparttable}
\usepackage{rotating}
\usepackage{rotfloat}
\usepackage{float}
\usepackage{moresize}
\usepackage{wrapfig}
\usepackage{multicol}
\usepackage{wrapfig}
\usepackage[normalem]{ulem}
\usepackage{graphicx}
\usepackage{txfonts}
\usepackage[T1]{fontenc}
 \usepackage{graphicx}  
\usepackage{amsmath}    
 \usepackage{color}
\usepackage{amssymb}
\usepackage{epstopdf}
\usepackage{natbib}
\bibpunct{(}{)}{;}{a}{}{,} 
\usepackage{url}

\newcommand{\vsini}{$V \sin i_\star$}

\newcommand{\teff}{$T_{\rm eff}$}
\newcommand{\logg}{log\,$g_\star$}

\newcommand{\feh}{[Fe/H]}
\newcommand{\cah}{[Ca/H]}
\newcommand{\mgh}{[Mg/H]}
\newcommand{\nah}{[Na/H]}
\newcommand{\sih}{[Si/H]}

\newcommand{\kms}{km\,s$^{-1}$}
\newcommand{\ms}{m~s$^{-1}$}

\newcommand{\gc}{g~cm$^{-3}$}

\newcommand{\Lsun}{$L_{\odot}$}                          
\newcommand{\Msun}{$M_{\odot}$}
\newcommand{\Rsun}{$R_{\odot}$}
 
\newcommand{\mearth}{$M_{\oplus}$}
\newcommand{\rearth}{$R_{\oplus}$}
\newcommand{\rjup}{$R_\mathrm{J}$}
\newcommand{\mjup}{$M_\mathrm{J}$}

\newcommand{\Porb}{$P_{\rm orb}$} 
\newcommand{\mstar}{$M_\star$}
\newcommand{\lstar}{$L_\star$}
\newcommand{\rstar}{$R_\star$}
\newcommand{\rhostar}{$\rho_{\mathrm{*}}$}

\newcommand{\rplanet}{$R_{\mathrm{p}}$}

\newcommand{\tic}{TIC~372172128}
\newcommand{\ticb}{TIC~372172128.01}
\newcommand{\targeta}{TOI-2196}
\newcommand{\targetone}{TOI-2196.01}
\newcommand{\targetb}{TOI-2196~b}

\newcommand{\bjdtdb}{\ensuremath{\rm {BJD_{TDB}}}}
 
\newcommand{\smassariadne}[1][]{$1.032 \pm 0.038$}   
\newcommand{\smassariadnegrav}[1][]{$1.0 \pm 0.1$}   
\newcommand{\smassisochrones}[1][]{$1.012 \pm 0.034$}   
\newcommand{\smasstorres}[1][]{$1.027 \pm 0.070$}   
\newcommand{\smassparam}[1][]{$0.982 \pm 0.037$}   
\newcommand{\smassparamonefive}[1][]{$0.982 \pm 0.037$}   
\newcommand{\smasssouthworth}[1][]{$1.072 \pm 0.073$}   
\newcommand{\smassspechmatch}[1][]{$1.006\pm0.080$}   
\newcommand{\smassbasta}[1][]{$0.973 \pm 0.053$}   

\newcommand{\sradiusariadne}[1][]{$1.043 \pm 0.017$}   
\newcommand{\sradiusisochrones}[1][]{$1.038 \pm 0.008$}   
\newcommand{\sradiustorres}[1][]{$1.022\pm 0.073$}  
\newcommand{\sradiusparam}[1][]{$1.049\pm 0.034$}  
\newcommand{\sradiussouthworth}[1][]{$1.062 \pm 0.040$}   
\newcommand{\sradiusspechmatch}[1][]{$1.158\pm0.120$}   
\newcommand{\sradiusbasta}[1][]{$1.045\pm0.025$}   
\newcommand{\srgaia}[1][]{$1.103^{+0.027}_{-0.068}$}  

\newcommand{\srhoariadne}[1][]{$1.25 \pm 0.09$}   
\newcommand{\srhoisochrones}[1][]{$1.28 \pm 0.06$}   
\newcommand{\srhospechmatch}[1][]{$0.91 \pm 0.29$}   
\newcommand{\srhoparam}[1][]{$1.23 \pm 0.14$}   
\newcommand{\srhobasta}[1][]{$1.20 \pm 0.11$}   
\newcommand{\srhotorres}[1][]{$1.4 \pm 0.3$}   
 
\newcommand{\steffariadne}[1][]{$5634 \pm 31$} 
\newcommand{\sloggariadne}[1][]{$4.42 \pm 0.04$} 
\newcommand{\sfehariadne}[1][]{$0.13 \pm 0.05$} 
\newcommand{\Lumariadne}[1][]{$0.99 \pm 0.04$} 
\newcommand{\Avariadne}[1][]{$0.03 \pm 0.02$} 

\newcommand{\steffbasta}[1][]{$5715 \pm 54$} 
\newcommand{\sloggbasta}[1][]{$4.39 \pm 0.04$} 
\newcommand{\sfehbasta}[1][]{$0.15 \pm 0.09$} 

\newcommand{\steffsme}[1][]{$5552 \pm 85$} 
\newcommand{\sloggCasme}{$4.42 \pm 0.05$} 
\newcommand{\sloggMgsme}[1][]{$4.41 \pm 0.08$} 
\newcommand{\scahsme}[1][]{$0.15 \pm 0.06$} 
\newcommand{\sfehsme}[1][]{$0.14 \pm 0.05$} 
\newcommand{\snasme}[1][]{$0.20 \pm 0.08$} 
\newcommand{\smghsme}[1][]{$0.18 \pm 0.09$} 
\newcommand{\ssihsme}[1][]{$0.15 \pm 0.08$} 
\newcommand{\svsini}[1][]{$2.0 \pm 0.4$} 
\newcommand{\svmic}[1][]{$1.0$} 
\newcommand{\svmac}[1][]{$2.8$} 
\newcommand{\velsme}[1][$\mathrm{km\,s^{-1}}$]{$xxx \pm x$} 

\newcommand{\steffspechmatch}[1][]{$5623 \pm 110$} 
\newcommand{\sloggspechmatch}[1][]{$4.22 \pm 0.12$} 
\newcommand{\sfehspechmatch}[1][]{$0.18 \pm 0.09$} 
\newcommand{\sagespechmatch}[1][]{$8.1 \pm 1.5$} 

\newcommand{\sloggparam}[1][]{$4.36 \pm 0.04$} 

\newcommand{\steffgaia}[1][]{$5462^{+176}_{-66} $} 
\newcommand {\lstargaia}[1][]{$0.98\pm0.01$} 
\newcommand{\parallaxgaia}{$3.7882\pm0.0132$} 
\newcommand{\velgaia}[1][$\mathrm{km\,s^{-1}}$]{$35.51 \pm 0.56$} 
\newcommand{\pmra}{$16.326 \pm 0.011$} 
\newcommand{\pmdec}{$-20.168 \pm 0.013$} 

\newcommand{\ageisochrones}[1][]{$5.4 \pm 1.8$} 
\newcommand{\ageparam}{$7.1\pm 3.4$} 
\newcommand{\ageariadne}[1][]{$4.5 \pm 2.0$} 
 
\newcommand{\smass}[1][]{$1.032 \pm 0.038$} 
\newcommand{\sradius}[1][]{$1.043 \pm 0.017$}
\newcommand{\stemp}[1][]{$5634 \pm 31$}
\newcommand{\Tzerob}[1][]   {$2036.5126_{-0.0016}^{+0.0019}$} 
\newcommand{\Pb}[1][]   {$1.1947268_{ - 9.3e-06}^ {+7.9e-06}$} 
\newcommand{\eb}[1][ ]   {$0.0_{-0.0}^ {+ 0.0}$} 
\newcommand{\wb}[1][]   {$90.0_{-0.0}^ {+0.0}$} 
\newcommand{\bb}[1][ ]   {$0.712_{-0.036}^{+0.034}$} 
\newcommand{\arb}[1][ ]   {$4.61\pm 0.10$} 
\newcommand{\rrb}[1][ ]   {$0.0308 _{ - 0.0012}^ {+0.0013 }$} 
\newcommand{\kb}[1][]   {$15.16_{-0.69}^{+0.67}$} 
\newcommand{\mpb}[1][]   {$26.0 \pm 1.3$} 
\newcommand{\rpb}[1][]   {$3.51 \pm 0.15$} 
\newcommand{\Tperib}[1][]   {$2036.5126_{-0.0016}^{+0.0019}$} 
\newcommand{\rpbSC}[1][$R_{\oplus}$]   {$3.61 \pm 0.26$~#1} 
\newcommand{\rpbSAAO}[1][$R_{\oplus}$]   {$3.37 _{ - 0.57 } ^ { + 0.36 }$~#1} 
\newcommand{\rpbCTIO}[1][$R_{\oplus}$]   {$3.42 _{ - 0.36 } ^ { + 0.58 }$~#1} 
\newcommand{\rpbSSO}[1][$R_{\oplus}$]   {$3.34 \pm 0.37$~#1} 

\newcommand{\prvb}[1][]   {$-8.76e-06 \pm 1.1e-07$} 
\newcommand{\ib}[1][]   {$81.11_{-0.55}^{+0.57}$} 
\newcommand{\ab}[1][]   {$0.02234 \pm 0.00060$} 
\newcommand{\depthbSC}[1][]   {$950.7_{-74.4}^{+78.8}$} 
\newcommand{\RMbSC}[1][]   {$1.33_{-0.11}^{+0.12}$} 
\newcommand{\insolationb}[1][]   {$2000\pm 100$} 
\newcommand{\tsmb}[1][ ]   {$25.8_{-3.3}^{+3.6}$} 
\newcommand{\denstrb}[1][]   {$1.295_{-0.081}^{+0.086}$} 
\newcommand{\densspb}[1][]   {$1.283 _{-0.076}^{+0.081}$} 
\newcommand{\Teqb}[1][]   {$1860\pm 20$} 
\newcommand{\ttotb}[1][]   {$1.504 _{ - 0.059 } ^ { + 0.054 }$} 
\newcommand{\tfulb}[1][]   {$1.325 _{ - 0.072 } ^ { + 0.065 }$} 
\newcommand{\tegb}[1][]   {$0.0892 _{ - 0.0076 } ^ { + 0.0087 }$} 
\newcommand{\deltamagb}[1][]   {$1.56 _{ - 0.21} ^ { + 0.23}$} 
\newcommand{\denpb}[1][]   {$3.31_{ - 0.43} ^ { + 0.51}$} 
\newcommand{\grapb}[1][]   {$2100\pm 200$} 
\newcommand{\grapparsb}[1][]   {$2100 \pm 200$} 
\newcommand{\jspb}[1][]   {$30\pm 2$}
\newcommand{\qoneSC}[1][]   {$0.27\pm 0.10$}
\newcommand{\qtwoSC}[1][]   {$0.47\pm 0.10$} 
\newcommand{\qoneSAAO}[1][]   {$0.53 \pm 0.10$} 
\newcommand{\qtwoSAAO}[1][]   {$0.17\pm 0.10$}
\newcommand{\qoneCTIO}[1][]   {$0.51\pm 0.10$} 
\newcommand{\qtwoCTIO}[1][]   {$0.18\pm 0.09$} 
\newcommand{\qoneSSO}[1][]   {$0.54 \pm 0.10$} 
\newcommand{\qtwoSSO}[1][]   {$0.18 \pm 0.09$} 
\newcommand{\HARPSN}[1][]   {$0.1008 \pm 0.0079$} 
\newcommand{\jHARPSN}[1][]   {$1.2 _{ - 0.80} ^ { + 0.77}$} 
\newcommand{\jtrSC}[1][]   {$0.002433 \pm 1.7e-05$} 
\newcommand{\jtrSAAO}[1][]   {$0.001706 _{ - 5.0e-05 } ^ { + 5.4e-05 }$} 
\newcommand{\jtrCTIO}[1][]   {$0.001376 _{ - 4.8e-05} ^ { + 5.1e-05}$} 
\newcommand{\jtrSSO}[1][]   {$0.000939 _{ - 3.7e-05} ^ { + 3.9e-05}$} 
\newcommand{\ltrend}[1][]   {$-0.238 \pm 0.019$} 
 
\begin{document} 
 
 \titlerunning{TOI-2196~b: Rare planet in the hot Neptune desert transiting a G-type star}
   \title{TOI-2196~b: Rare planet in the hot Neptune desert transiting a G-type star}
\author{Carina~M.~Persson\inst{\ref{OSO}} 
\and
Iskra~Y.~Georgieva\inst{\ref{OSO}} 
\and
Davide~Gandolfi\inst{\ref{Torino}} 
\and
Lorena~Acu\~na\inst{\ref{Marseille}} 
\and
Artem~Aguichine\inst{\ref{Marseille}} 
\and
Alexandra~Muresan\inst{\ref{Chalmers}}  %
\and
 Eike~Guenther\inst{\ref{Tautenburg}} 
\and
John~Livingston\inst{\ref{AstrobiologycentreTokyo},\ref{NAOTokyo},\ref{SOKENDAI}} 
\and
Karen~A.~Collins\inst{\ref{HarvardSmithsonian}} 
\and
Fei~Dai\inst{\ref{Pasadena}}  
\and
Malcolm~Fridlund\inst{\ref{OSO}, \ref{Leiden}} 
\and
Elisa~Goffo\inst{\ref{Torino},\ref{Tautenburg}} 
\and
James~S.~Jenkins\inst{ \ref{PortalesChile},\ref{AstronomyChile}}
\and
Petr~Kab\'ath\inst{\ref{Ondrojev}} 
\and
Judith~Korth\inst{\ref{Chalmers}}
\and
Alan~M.~Levine\inst{\ref{MIT}} 
\and
Luisa~M.~Serrano\inst{\ref{Torino}} 
\and
Jos\' e~Vines\inst{\ref{AstronomyChile}}
\and
Oscar~Barrag\' an\inst{\ref{Oxford}} 
\and
Ilaria~Carleo\inst{\ref{IAC}}
\and
Knicole~D.~Colon\inst{\ref{Goddard}} 
\and
William~D.~Cochran\inst{\ref{McDonald}} 
\and
Jessie~L.~Christiansen\inst{\ref{Caltech}} 
\and
Hans~J.~Deeg\inst{\ref{IAC}, \ref{LaLaguna}} 
%
\and
Magali~Deleuil\inst{\ref{Marseille}} 
\and
Diana~Dragomir\inst{\ref{NewMexico}}  
\and
Massimiliamo~Esposito\inst{\ref{Tautenburg}} 
\and
Tianjun~Gan\inst{\ref{Beijing}}  
\and
Sascha~Grziwa\inst{\ref{Cologne}}  
\and
Artie~P.~Hatzes\inst{\ref{Tautenburg}} 
\and
Katharine Hesse\inst{\ref{MIT}} 
\and
Keith~Horne\inst{\ref{StAndrews}} 
%
%
\and
Jon~M.~Jenkins\inst{\ref{NASAames}}
\and
John~F.~Kielkopf\inst{\ref{Louisville}}
\and
P.~Klagyivik\inst{\ref{FreieUniversitatBerlinGeo}}  
\and
Kristine~W.~F.~Lam\inst{\ref{DLR}} 
\and 
David~W.~Latham\inst{\ref{HarvardSmithsonian}}   
\and 
Rafa~Luque\inst{\ref{UniversityChicago}} 
\and
Jaume Orell-Miquel\inst{\ref{IAC}, \ref{LaLaguna}} 
\and
Annelies~Mortier\inst{\ref{CavendishUK},\ref{KavliUK}}
\and
Olivier~Mousis\inst{\ref{Marseille}} 
\and
Noria~Narita\inst{\ref{AstrobiologycentreTokyo},\ref{IAC},\ref{KomabaTokyo}} 
\and
Hannah~L.~M.~Osborne\inst{\ref{MullardVincent}} 
\and
Enric~Palle\inst{\ref{IAC}}
\and
Riccardo~Papini\inst{\ref{Firenze}} 
%
\and
George~R.~Ricker\inst{\ref{MIT}} 
\and
Hendrik~Schmerling\inst{\ref{Cologne}} 
\and
Sara~Seager\inst{\ref{MIT}, \ref{PlanetSciencesMIT}, \ref{AeronauticsMIT}} 
\and
Keivan~G.~Stassun\inst{\ref{Vanderbilt}}  
\and
Vincent~Van~Eylen\inst{\ref{MullardVincent}} 
\and
Roland~Vanderspek\inst{\ref{MIT}} %
\and  Gavin~Wang\inst{\ref{TsinghuaBeijing}} 
%
\and Joshua~N.~Winn\inst{\ref{Princeton}} 
%
\and
Bill~Wohler\inst{\ref{NASAames},\ref{SETI}} 
\and Roberto~Zambelli\inst{\ref{Magra}} 
\and
Carl~Ziegler\inst{\ref{Austin}} 
 }     
 \offprints{carina.persson@chalmers.se}
    \institute{ Department of Space, Earth and Environment, Chalmers University of Technology, Onsala Space Observatory, SE-439 92 Onsala, Sweden. \label{OSO}  
       \email{\url{carina.persson@chalmers.se}}
   \and Dipartimento di Fisica, Universita degli Studi di Torino, via Pietro Giuria 1, I-10125, Torino, Italy. \label{Torino}  
   \and
   Aix Marseille Universit\'e, Institut Origines, CNRS, CNES, LAM, Marseille, France.  \label{Marseille} 
   \and
   Department of Space, Earth and Environment, Chalmers University of Technology, Chalmersplatsen 4, 412 96 Gothenburg, Sweden. \label{Chalmers}  
   \and  Th\"uringer Landessternwarte Tautenburg, Sternwarte 5, 07778 Tautenburg, Germany.  \label{Tautenburg}
        \and Astrobiology Center, 2-21-1 Osawa, Mitaka, Tokyo 181-8588, Japan.  \label{AstrobiologycentreTokyo}    
   \and National Astronomical Observatory of Japan, 2-21-1 Osawa, Mitaka, Tokyo 181-8588, Japan.  \label{NAOTokyo}
  \and Department of Astronomy, The Graduate University for Advanced Studies (SOKENDAI), 2-21-1 Osawa, Mitaka, Tokyo, Japan.  \label{SOKENDAI}
        \and Center for Astrophysics \textbar \ Harvard \& Smithsonian, 60 Garden Street, Cambridge, MA 02138, USA. \label{HarvardSmithsonian}
    \and Division of Geological and Planetary Sciences, 1200 E California Blvd, Pasadena, CA, 91125, USA. \label{Pasadena}   
     \and Leiden Observatory, University of Leiden, PO Box 9513, 2300 RA, Leiden, The Netherlands. \label{Leiden} 
       \and N\'ucleo de Astronom\'ia, Facultad de Ingenier\'ia y Ciencias, Universidad Diego Portales, Av. Ej\'ercito 441, Santiago, Chile. \label{PortalesChile}  
     \and Centro de Astrof\'isica y Tecnolog\'ias Afines (CATA), Casilla 36-D, Santiago, Chile.  \label{AstronomyChile} 
     \and Astronomical Institute of the Czech Academy of Sciences, Fri\v cova 298, 25165, Ond\v rejov, Czech Republic. \label{Ondrojev} 
            \and Department of Physics and Kavli Institute for Astrophysics and Space Research, Massachusetts Institute of Technology, Cambridge, MA 02139, USA.   \label{MIT}
     \and Sub-department of Astrophysics, Department of Physics, University of Oxford, Oxford, OX1 3RH, UK. \label{Oxford}
          \and Instituto de Astrof\' isica de Canarias, C. Via Lactea S/N, E-38205 La Laguna, Tenerife, Spain.\label{IAC}
     \and  NASA Goddard Space Flight Center, Exoplanets and Stellar Astrophysics Laboratory (Code 667), Greenbelt, MD 20771, USA.  \label{Goddard}
         \and  McDonald Observatory and Center for Planetary Systems Habitability, The University of Texas, Austin Texas USA. \label{McDonald}  
      \and Caltech/IPAC-NASA Exoplanet Science Institute, 770 S. Wilson Avenue, Pasadena, CA 91106, USA.    \label{Caltech} 
 \and Universidad de La Laguna, Dept. de Astrof\'isica, E-38206 La Laguna, Tenerife, Spain. \label{LaLaguna}  
     \and    Department of Physics and Astronomy, University of New Mexico, 210 Yale Blvd NE, Albuquerque, NM 87106, USA.  \label{NewMexico} 
     \and Department of Astronomy and Tsinghua Centre for Astrophysics, Tsinghua University, Beijing 100084, China. \label{Beijing} 
     \and  Rheinisches Institut für Umweltforschung an der Universität zu Köln, Aachener Strasse 209, D-50931 Köln, Germany. \label{Cologne}  %
     \and SUPA Physics and Astronomy, University of St. Andrews, Fife, KY16 9SS Scotland, UK. \label{StAndrews}   
\and NASA Ames Research Center, Moffett Field, CA 94035, USA. \label{NASAames}
  \and Department of Physics and Astronomy, University of Louisville, Louisville, KY 40292, USA. \label{Louisville}    
  \and Freie Universit\"at Berlin, Institute of Geological Sciences, Malteserstr. 74-100, 12249 Berlin, Germany. \label{FreieUniversitatBerlinGeo}
  \and Institute of Planetary Research, German Aerospace Center (DLR), Rutherfordstrasse 2, D-12489 Berlin, Germany. \label{DLR} %
  \and Department of Astronomy \& Astrophysics, University of Chicago, Chicago, IL 60637, USA.  \label{UniversityChicago}
    \and Astrophysics Group, Cavendish Laboratory, University of Cambridge, J.J. Thomson Avenue, Cambridge CB3 0HE, UK. \label{CavendishUK}   
   \and Kavli Institute for Cosmology, University of Cambridge, Madingley Road, Cambridge CB3 0HA, UK. \label{KavliUK}    
   \and Komaba Institute for Science, The University of Tokyo, 3-8-1 Komaba, Meguro, Tokyo 153-8902, Japan.  \label{KomabaTokyo}    
    \and Mullard Space Science Laboratory, University College London, Holmbury St Mary, Dorking, Surrey RH5 6NT, UK. \label{MullardVincent}
  \and Wild Boar Remote Observatory, San Casciano in val di Pesa, Firenze, 50026 Italy. \label{Firenze}    
  \clearpage
  \and Department of Earth, Atmospheric, and Planetary Sciences, Massachusetts Institute of Technology, Cambridge, MA 02139, USA \label{PlanetSciencesMIT}
  \and Department of Aeronautics and Astronautics, Massachusetts Institute of Technology, Cambridge, MA 02139, USA  \label{AeronauticsMIT}
 \and Department of Physics \& Astronomy, Vanderbilt University, Nashville, TN 37235, USA.  \label{Vanderbilt}  
  \and Tsinghua International School, Beijing 100084, China.  \label{TsinghuaBeijing}  
  \and Department of Astrophysical Sciences, Princeton University, Princeton, NJ 08544, USA \label{Princeton}
  \and SETI Institute, Mountain View, CA 94043, USA. \label{SETI} 
   \and Societ\'a Astronomica Lunae, Castelnuovo Magra, Italy. \label{Magra}    
  \and Department of Physics, Engineering and Astronomy, Stephen F. Austin State University, 1936 North St, Nacogdoches, TX 75962, USA.  \label{Austin} 
}   

   \date{Received  27 May 2022; accepted 11 July 2022}

 
  \abstract
  {The hot Neptune desert is a  region hosting a small number of short-period  Neptunes in the radius-instellation diagram.  
  Highly irradiated planets are usually either small ($R \lesssim  2$~\rearth) and rocky or they are gas giants with radii of $\gtrsim1$~\rjup. 
  Here, we report on   the intermediate-sized planet  \targetb~(TIC 372172128.01)   on a 1.2~day orbit around a 
G-type star ($V = 12.0$, [Fe/H] = 0.14 dex) discovered by the Transiting Exoplanet Survey Satellite in sector~27. 
We  collected 41  radial velocity measurements with the HARPS spectrograph 
to confirm the planetary nature of the transit signal and to determine the mass. The  
radius of \targetb~is \rpb~\rearth, which, combined with the  mass of  26.0\,$\pm\,1.3$~\mearth,~ results in a   bulk density of \denpb~\gc. 
Hence, the radius  implies that this planet is a sub-Neptune, although the density is twice than that of Neptune.
A significant trend in the HARPS radial velocity measurements points to the presence of a distant companion 
with a lower limit on the period and mass of  220~days and 0.65~\mjup, respectively, assuming zero eccentricity. 
The short period of planet~b 
implies a high equilibrium temperature of  1860\,$\pm\,20$~K, for zero albedo and   isotropic emission.  
This  places the planet in the hot Neptune desert,      joining a group of 
very few planets in this parameter space discovered in recent years. These planets suggest that   
the hot Neptune desert may be divided   in two parts for planets with equilibrium temperatures of   $\gtrsim 1800$~K: 
a hot sub-Neptune desert  devoid of 
   planets with radii of $\approx 1.8-3$~\rearth\  and a sub-Jovian desert for radii of $\approx 5-12$~\rearth. 
   More planets in this parameter space are needed to further investigate this finding.  
Planetary interior structure models of \targetb~are consistent with a  H/He atmosphere  mass fraction   between 
0.4~\% and 3~\%,  with a mean value of 0.7~\%   
on top of a rocky interior.  
We estimated the amount of mass   this planet might have lost at a young age and we find that while the 
mass loss could have been significant, the planet had not changed in terms of character: it was born as a small volatile-rich planet and it remains one at present.
  }

   \maketitle


\section{Introduction} \label{Section: introduction}
With the large number of \emph{Kepler} \citep{2010Sci...327..977B}  planets, evidence of a   
bimodal  population has emerged, made up of    small planets   with a gap in the size 
distribution  between $\sim1.5$ and 2~\rearth,~often referred to as the radius gap 
\citep{2017AJ....154..109F, VanEylen2018, 2021MNRAS.507.2154V, 2022arXiv220110020P}. This feature was predicted before the 
observational discovery by several groups as a result of envelope mass loss due to photoevaporation 
 \citep{2013ApJ...776....2L, 2013ApJ...775..105O, 2014ApJ...795...65J, 2016ApJ...831..180C}. Other studies have shown that 
 core-powered envelope mass loss  could also carve out a   gap in the radius distribution of small planets   \citep{2018MNRAS.476..759G, 2019MNRAS.487...24G}.

 The two populations on either side of the radius gap are:  super-Earths,  commonly defined as 
having radii of \mbox{$1.2 \lesssim R/R_{\oplus} \lesssim 1.7$} and believed 
to be mainly rocky, and volatile-rich 
sub-Neptunes with radii  of \mbox{$1.7 \lesssim R/R_{\oplus} \lesssim 4$}. 
According to available models, a part of the  super-Earth population could, in fact,  be   remnant   cores of sub-Neptunes stripped of their  atmospheres. 
 Consequently, mass loss plays an important role in the first few hundred million years of exoplanet evolution  \citep{2006ApJ...649.1004A,2018A&A...612A..25K}.

\begin{table}[!t]
\caption{Basic parameters for \targeta.}
\begin{center}
 \resizebox{0.9\columnwidth}{!}{%
\begin{tabular}{lll} 
\hline\hline
     \noalign{\smallskip}
Parameter    & Value   \\
\noalign{\smallskip}
\hline
\noalign{\smallskip}
\multicolumn{2}{l}{\emph{Main Identifiers}} \\
\noalign{\smallskip}
TIC  & 372172128 \\
2MASS &  J20492158-7029058 \\
WISE &  J204921.59-702906.1 \\
TYC & 9325-00163-1 \\
UCAC4 & 098-095039 \\
Gaia   &        6375983988633147392     \\
\noalign{\smallskip}
\hline
\noalign{\smallskip}
\multicolumn{2}{l}{\emph{Equatorial coordinates}} \\
\noalign{\smallskip} 
R.A. $(J2000.0)$ & $20\fh49\fm 21\fs57$     \\
Dec. $(J2000.0)$ & -70$\fdg 29\farcm05\farcs95$    \\
\noalign{\smallskip}
\hline
\noalign{\smallskip}
\multicolumn{2}{l}{\emph{Magnitudes}} \\
\textit{TESS} & $11.3643 \pm 0.0060$ &  \\
Johnson $ B$ &   $12.6740\pm0.0160$     \\
Johnson $V$  & $11.9530\pm0.0120$      \\
 $G \tablefootmark{a}$  &   $11.8175\pm0.0002$ \\
 $G_{RP}\tablefootmark{a}$      &   $11.3030\pm0.0007$ \\
$G_{BP} \tablefootmark{a}$      &   $12.1712\pm0.0014$ \\
 $g$ & $12.2770\pm0.0200$ \\
 $r$ & $11.7500\pm0.0150$ \\
  $i$ & $11.6300\pm0.0060$ \\
$J$  &   $10.743\pm 0.024$     \\
$H$    &   $10.452\pm0.026$      \\
$K$   &     $10.346\pm0.023$   \\
WISE $W1$   &     $10.326\pm0.022$   \\
WISE $W2$   &     $10.344\pm0.019$   \\
\noalign{\smallskip}
\hline
\noalign{\smallskip}  
Parallax$\tablefootmark{a}$  (mas) &\parallaxgaia\,   \\  
Systemic velocity$\tablefootmark{a}$ (\kms) & \velgaia   \\   
$\mu_{RA}\tablefootmark{a}$ (mas~yr$^{-1}$) & \pmra   \\   
$\mu_{Dec}\tablefootmark{a}$ (mas~yr$^{-1}$) & \pmdec   \\  
\noalign{\smallskip}
\hline
\noalign{\smallskip}
\teff$\tablefootmark{b}$ (K)  & \stemp \\
\mstar$\tablefootmark{b}$  (\Msun) & \smass \\
\rstar$\tablefootmark{b}$ (\Rsun)  & \sradius \\
\rhostar$\tablefootmark{b}$ (\gc)  & \srhoariadne \\
\lstar$\tablefootmark{b}$ (\Lsun)  & \Lumariadne \\ 
\logg$\tablefootmark{b}$    &  \sloggariadne \\
\feh$\tablefootmark{b}$    & \sfehsme \\
\cah$\tablefootmark{b}$    & \scahsme\\
\mgh$\tablefootmark{b}$   & \smghsme \\
\nah$\tablefootmark{b}$    & \snasme \\
\sih$\tablefootmark{b}$    & \ssihsme \\
\vsini$\tablefootmark{b}$ (\kms) &\svsini\\
Age$\tablefootmark{b}$ (Gyr) & \ageariadne \\
    \noalign{\smallskip} \noalign{\smallskip}
\hline 
\end{tabular}
}
\tablefoot{
\tablefoottext{a}{Gaia eDR3.} 
\tablefoottext{b}{This work (Sect.~\ref{Subsection: Stellar modelling}).} 
}
\end{center}
\label{Table: Star basic parameters}
\end{table}

 Despite the abundance of small planets, there is  a  dearth of   hot sub-Neptunes and Neptunes  
  in the radius-instellation (or equilibrium temperature) diagram, namely, the so-called hot Neptune desert or sub-Jovian desert
  \citep{2011ApJ...727L..44S, 2011A&A...528A...2B,2014ApJ...787...47S, 2016A&A...589A..75M}, or (otherwise) the hot super-Earth desert 
  as referred to by \citet{2016NatCo...711201L} for smaller planets with radii between 2.2 and 3.8~\rearth. 
  The observed period distribution already  drops  for \mbox{$P_{\rm orb} < 3$~days}     
  despite a strong selection bias due to easy detection which indicates that short period planets are  rare.  
  Most    planets facing high instellation  
are  either small and rocky   with masses below 10~\mearth~and radii $\lesssim2$~\rearth, or massive gas
giants with radii $\gtrsim1$~\rjup.  

 \begin{figure}[!ht]
 \centering
        \includegraphics[scale=0.2]{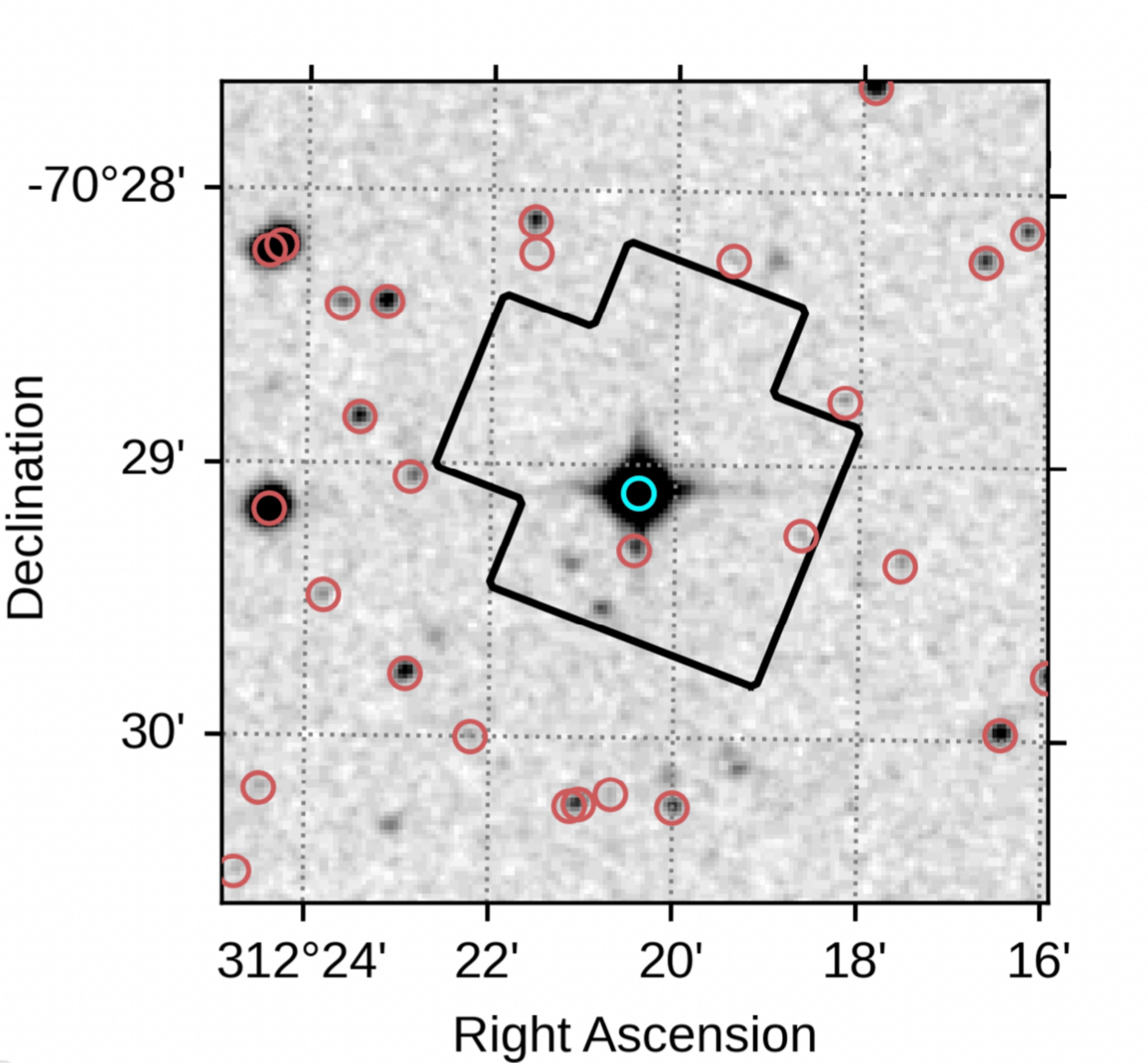} 
   \caption{3\arcmin\,$\times$\,3\arcmin\,DSS2 image (red filter) centered on \targeta~(cyan circle). The SPOC photometric aperture of sector 27 is outlined in black while Gaia DR2 sources within 2\arcmin~from the target are marked by the red circles.}
      \label{Figure: DSS with TESS apertures}
 \end{figure}
 
Up until a few years ago, the hot Neptune desert was almost completely empty in terms of observed planets. 
New discoveries have begun to uncover a  population of planets in the desert, although their small number does not allow for the exact circumstances of their existence to be defined. 
 High precision radius and mass measurements of these planets are crucial to  constrain  theoretical models of their formation and evolution.
 The lack of planets in this parameter space   suggests  
difficulties of retaining  an extended atmosphere in   strong irradiation environments  \citep{LopFor14}, 
possibly indicating different formation and evolution 
mechanisms, or high-eccentricity migration \citep{Mazeh13, 2018MNRAS.479.5012O}. 

Within the hot Neptune desert, there are currently only three known planets with equilibrium 
temperatures above 1800~K\footnote{Assuming a Bond albedo of zero and a heat   redistribution factor of unity (isotropic emission).}, corresponding to  a 1.3~day orbit for sun-like stars, 
with a precision of 10~\% and 30~\% or better in measured radii and masses: K2-100~b  \citep{2019MNRAS.490..698B}, TOI-849~b \citep{2020Natur.583...39A}, and LTT 9779~b \citet{2020NatAs...4.1148J}. 
Two additional planets with radii measured to a precision of 10~\% or better  but without measured masses are also known: 
 K2-278~b \citep{2018AJ....156..277L} and Kepler-644~b \citep{2018ApJ...866...99B}.

 This paper presents the discovery and characterisation of the intermediate-sized planet \targetb~(\ticb)   
 in the hot Neptune desert 
 discovered by The Transiting Exoplanet Survey Satellite \citep[\textit{TESS}; ][]{2015JATIS...1a4003R} in 2020. 
 Following the discovery, our international KESPRINT\footnote{KESPRINT is an international 
consortium devoted to the characterisation and research of exoplanets discovered with space-based missions, \url{http://kesprint.science}.}  
collaboration performed 
follow-up radial velocity observations of this planet candidate to confirm the planetary nature and determine its mass. 
The star's equatorial coordinates  together with other  
basic parameters are listed in Table~\ref{Table: Star basic parameters}. 

We present the observations in Sect.~\ref{Section: Observations}  and the data analysis in  Sect.~\ref{Section: Data analysis}. 
In  Sect.~\ref{Section: Discussion},  we discuss  the hot Neptune  desert,   the planet  interior,
and atmospheric mass loss. We end the paper with our     conclusions in   Sect.~\ref{Section: Conclusions}. 
  
%

 \begin{figure*}
  \centering
  \includegraphics[width=0.95\textwidth]{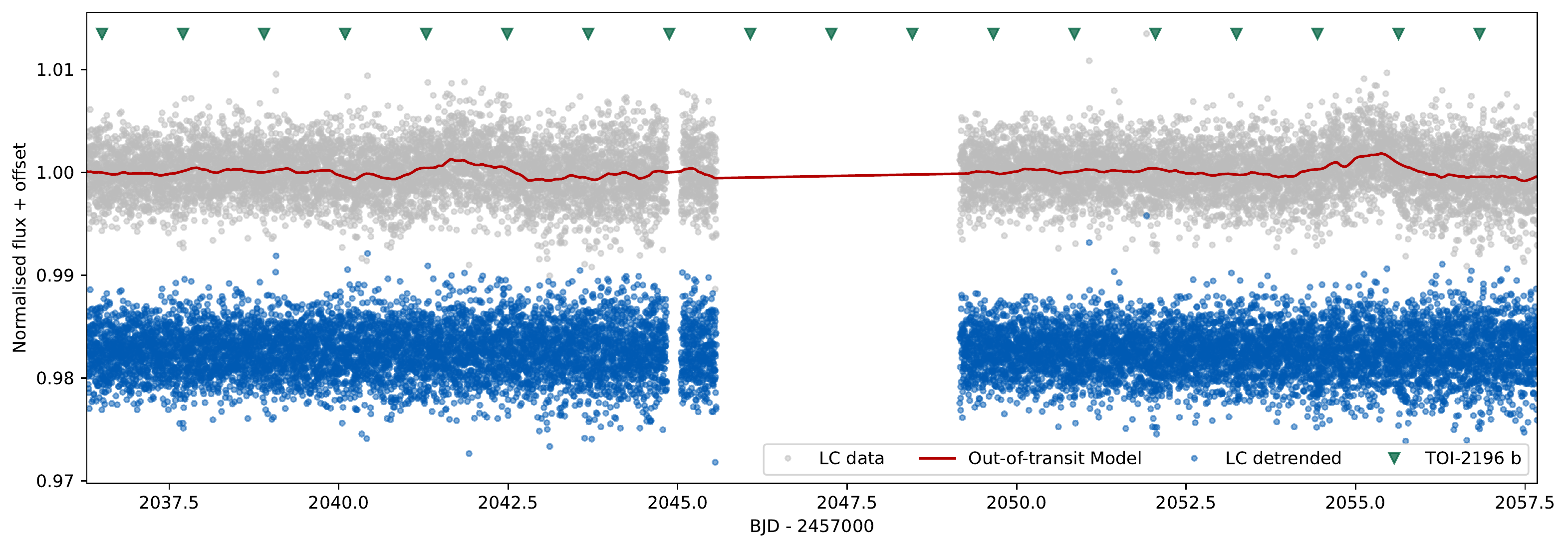}  
\caption{\textit{TESS} light curve for sector 27 (short cadence) is plotted in grey with the Gaussian Process model of the out-of-transit data overplotted in red. The detrended and normalised light curve is shown in blue, and, for the purposes of visualisation, a vertical offset has been applied. The triangles 
mark the locations of the individual transits of \targetb.}
\label{Fig: detrended sector27}
\end{figure*}

\section{Observations} \label{Section: Observations}
 
    \subsection{\emph{TESS} photometry} \label{Section: TESS photometry}
    Figure~\ref{Figure: DSS with TESS apertures} shows a \mbox{$3\arcmin \times 3\arcmin$} image from 
the Digitized Sky Survey 2 
(DSS2)  centered on \tic~(\targeta), marked with a cyan circle. 
\targeta~was  observed  by \textit{TESS}\footnote{\url{https://tess.mit.edu}} in sector~13, during the primary mission, 
and in sector~27  
in the first set of observations of the \textit{TESS} extended mission. 
The outline of the Science Processing Operations Center 
(SPOC) photometric aperture of sector~27 is overplotted in black in Fig.~\ref{Figure: DSS with TESS apertures}, while the red circles 
mark the positions of Gaia DR2 sources within 2\arcmin~from the target. 
The photometric dilution from other sources is negligible and   adjusted for by the pipeline. 

\targeta~was observed in sector~13  with camera 2 in full-frame images at a cadence of 30 min from 
19~June  to 17~July  2019. 
The observations of sector~27 spanned the interval 5 July   through 30 July 2020, with a gap in the middle 
of approximately one day, when the data were being downloaded. This produced 23.35~days of science data at  
two-minute cadence, including 14 transits of  \tic~monitored with camera~2, CCD~1.

Due to the combination of the short period of the planet (1.2~days) and the long cadence in sector~13, 
the planet candidate
\targetone~was not discovered orbiting its G-type host star until it was observed in sector~27 at a 2-min 
cadence and processed by  SPOC  at NASA Ames Research 
Center \citep{jenkins2016}. A search of the sector~27 data with an adaptive, wavelet-based matched filter 
\citep{2002ApJ...575..493J, 2010SPIE.7740E..0DJ,2020TPSkdph}
 identified the transit signature of \targetb~just 
above the detection threshold at 7.2~$\sigma$.  The data validation reports \citep[DVR;][]{Twicken:DVdiagnostics2018,Li:DVmodelFit2019}
process fit a limb-darkened transit model with a signal-to-noise ratio 
(S/N) of 9.2, a period of 1.1954~days, a duration of 1.6~hours, and a transit depth of 1127~ppm, corresponding 
to a preliminary planet radius of  $\sim 3.5$~\rearth. 
We independently detected the transit signal using the DST \citep[][]{Cabrera2012}  pipeline and found 
a planet candidate with an orbital period of $1.19387 \pm 0.00043$~days 
and a transit depth of $1359 \pm 147$~ppm.

We downloaded the light curves processed by the SPOC pipeline from the Mikulski Archive for Space Telescopes 
(MAST\footnote{\url{https://mast.stsci.edu/portal/Mashup/Clients/Mast/Portal.html}}) and
used the  Pre-search Data Conditioning Simple Aperture Photometry \mbox{(PDCSAP)} data. 
This was generated by the pipeline by identifying and correcting the SAP flux for instrumental signatures 
using cotrending basis vectors drawn from the light curves of an ensemble of quiet and highly temporally
 correlated stars long-term trends, thus resulting in a cleaner data set with fewer systematics \citep{Stumpe2014,Smith2020_PDCSAP}.

In order to remove any remaining low-frequency signals in preparation for the modelling
described in Sect.~\ref{Section: transit and RV modelling}, we further detrended the light curve by applying
a Gaussian Process (GP). This was achieved using the package 
{\tt citlalicue}\footnote{\url{https://github.com/oscaribv/citlalicue}} \citep[e.g.][]{2021MNRAS.505.4684G,2022MNRAS.509..866B},  
a {\tt PYTHON}  wrapper of {\tt george} \citep{Mackey2014} and {\tt pytransit}
\citep{2015MNRAS.450.3233P}. 
We masked out the transits of the planet and applied a squared exponential covariance function as well as a 
5~$\sigma$ clipping algorithm to remove outliers.
Figure~\ref{Fig: detrended sector27} displays both the PDCSAP data in sector 27 with the GP model overplotted, and the 
detrended normalised light curve. The latter is subsequently used in the joint   transit  and 
RV analysis   in Sect.~\ref{Section: transit and RV modelling}.

\subsection{Follow-up photometry from ground: LCOGT 1~m} \label{Follow-up photometry from ground: LCOGT}
The \textit{TESS} pixel scale is $\sim 21\arcsec$ pixel$^{-1}$  and photometric apertures typically extend out to roughly 1~arcmin. This generally results in multiple stars blending in the \textit{TESS} aperture. An eclipsing binary in one of the nearby blended stars could mimic a transit-like event in the large \textit{TESS} aperture. We therefore acquired ground-based transit follow-up photometry of TOI-2196 b as part of the \textit{TESS} Follow-up Observing Program Sub Group 1 \citep[TFOP SG1;][]{collins:2019}\footnote{https://tess.mit.edu/followup} to attempt to (1) rule out or identify nearby eclipsing binaries (NEBs) as potential sources of the detection in the \textit{TESS} data; (2) check for the transit-like event on-target using smaller photometric apertures than \textit{TESS} to confirm that the event is occurring on-target or, otherwise, in a star so close to TOI-2196 that it was not detectable by Gaia eDR3; (3) refine the \textit{TESS} ephemeris; and (4) place constraints on transit depth across optical filter bands.

  \begin{figure}[!ht]
 \centering
  \resizebox{\hsize}{!}
            {\includegraphics{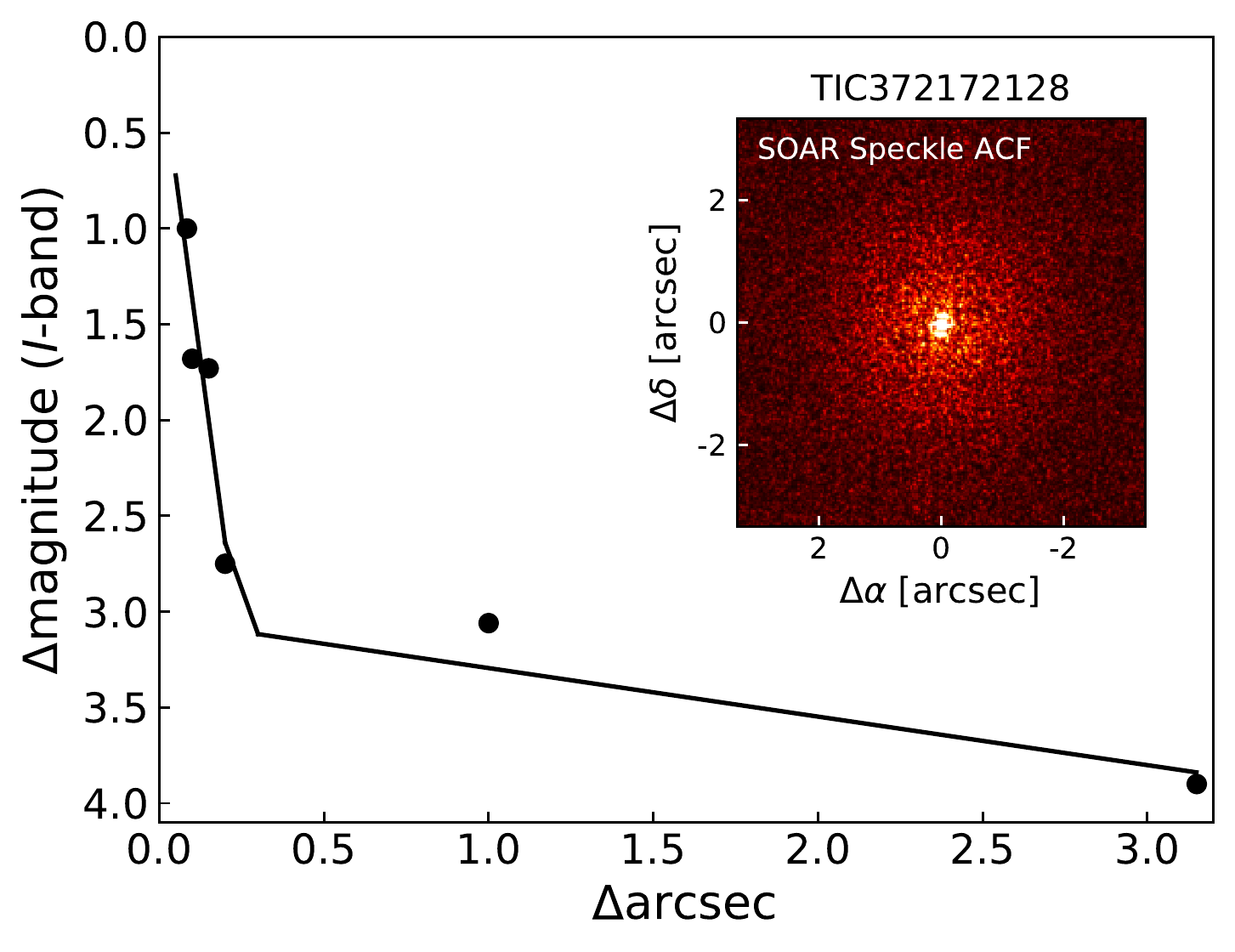}}
   \caption{Contrast curve    computed from observations in Cousins I-band  on the 4.1~m Southern 
Astrophysical Research   telescope. No bright companions are detected within   $3\arcsec$~of \targeta.
 }
      \label{Figure: speckle imaging}
 \end{figure}
 
We observed the transits of \targeta~from the Las Cumbres Observatory Global Telescope
\citep[LCOGT;][]{Brown:2013} 
1.0~m network on UTC 28~September 2020, 23~May 2021, and 6~June 2021 in Sloan $i'$ band 
and on UTC 20 June~2021 in 
Sloan $g'$ band. We used the {\tt TESS Transit Finder} to schedule our transit observations.
The 1.0~m telescopes are equipped 
with \mbox{$4096\times4096$} SINISTRO cameras having an image scale of $0\farcs389$ per pixel,
resulting in a 
\mbox{$26\arcmin\times26\arcmin$} field of view. The images were calibrated by the standard LCOGT 
{\tt BANZAI} pipeline \citep{McCully2018}. 
Photometric data were extracted using {\tt AstroImageJ} \citep{Collins:2017} and circular
photometric apertures with radii in the range 
$4\farcs7$ to $7\farcs0$. 
The TOI-2196 apertures exclude virtually all flux from the nearest Gaia eDR3 and \textit{TESS} Input Catalog neighbor (TIC 1988186200) $12\farcs5$ South. We find no evidence for an NEB within $2\farcm5$ of TOI-2196, and detect the transit event within the TOI-2196 photometric apertures.


\subsection{Follow-up speckle imaging from ground} \label{Follow-up speckle imaging from ground}
High-angular resolution imaging is needed to search for nearby sources that can contaminate
the \textit{TESS} photometry. This can 
result in an underestimated planetary radius or  may be the source of astrophysical false
positives such as background 
eclipsing binaries. We searched for stellar companions to \targeta~with speckle 
imaging on the 4.1~m Southern 
Astrophysical Research (SOAR) telescope \citep{2018PASP..130c5002T} on 31~October~2020,
observing in Cousins I-band, a visible bandpass similar to that of TESS. 
This observation was sensitive to a 3~magnitude
fainter star at an angular 
distance of 1\arcsec~from the target. More details of the observation are available in
\citet{2020AJ....159...19Z}. The 
5~$\sigma$ detection sensitivity and speckle auto-correlation functions from the 
observations are shown in 
Fig.~\ref{Figure: speckle imaging}. 
No nearby stars were detected within 3\arcsec~of TOI-2196 in the SOAR observations.
 
  \begin{figure}
\centering
\includegraphics[width=0.9\linewidth]{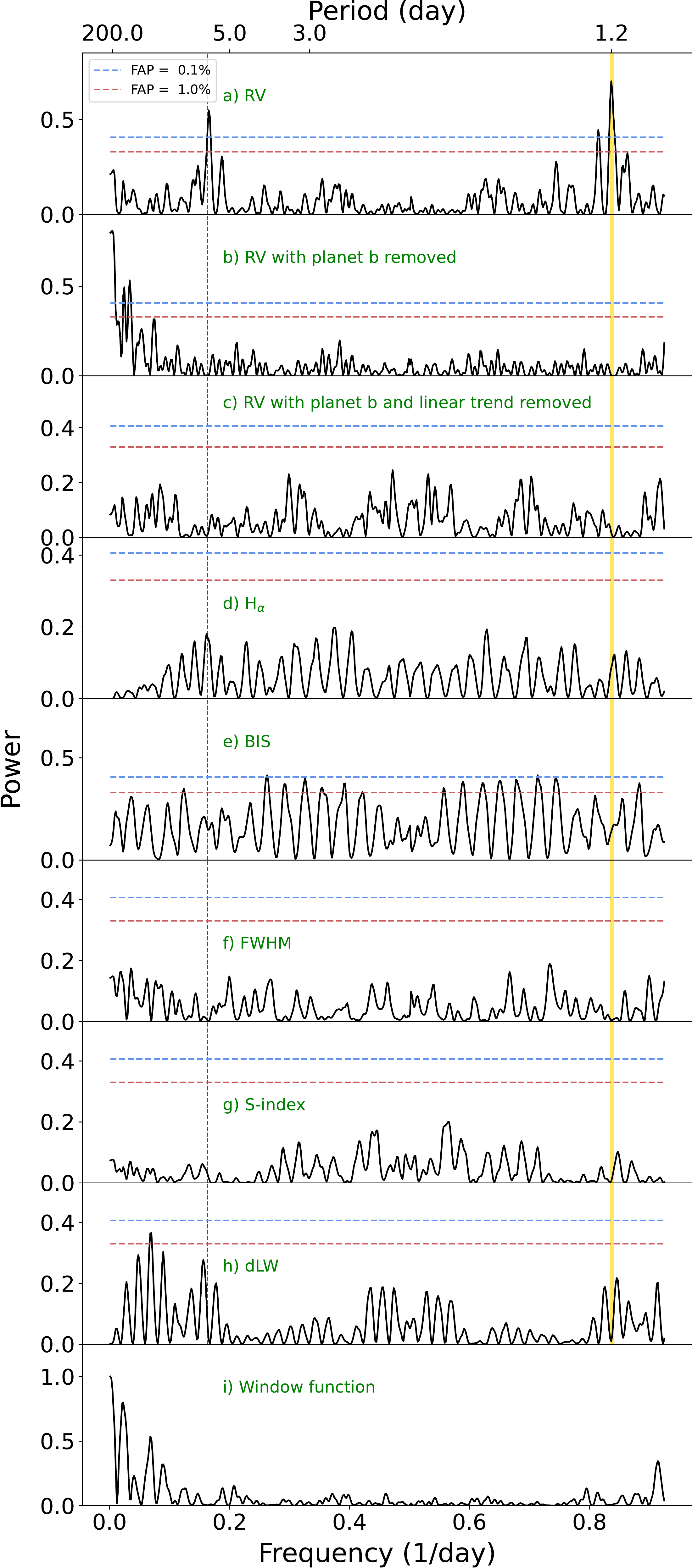}  
\caption{Generalised Lomb-Scargle periodogram of the HARPS RVs and stellar activity 
indicators. The horizontal lines mark the 
bootstrap false alarm probabilities at 0.1 and 1.0~\% as indicated in the legend. The orbital frequency 
of \targetb~($\nu_\mathrm{b}$\,=\,0.837~day$^{-1}$) is marked with a vertical thick yellow line, while the 
corresponding period is reported in the upper x-axis (\Porb\,$\approx$\, 1.2~day). The 1~day alias of planet~b 
is marked with a vertical dashed red line \mbox{($1-0.837=0.163$~day$^{-1}$)}.  \emph{a)} RV measurements. 
\emph{b)} RV residuals with the best-fitting Keplerian orbit of planet~b subtracted. \emph{c)} RV residuals 
following the subtraction of both the linear trend and the Doppler signal induced by planet~b. 
\emph{d-h)} Activity indicators and line profile variations. \emph{i)} The window function. }
    \label{Fig: TOI-2196 periodogram}
\end{figure}

 
\begin{table*}
\centering
 \caption{Spectroscopic  parameters for \targeta~modelled with {\tt {SME}} and {\tt {SpecMatch-Emp}}, 
posteriors from the {\tt {ARIADNE}},  
and the effective stellar temperature from Gaia DR2.}   
 \label{Table: stellar spectroscopic parameters}
\begin{tabular}{llccccccc }
 \hline
     \noalign{\smallskip}
Method  & $T_\mathrm{eff}$  & $\log(g)$ & [Fe/H]   & [Ca/H]  & [Mg/H]  & [Na/H]   & [Si/H]  & \vsini    \\  
& (K)  &(dex) &(cgs)  &(\kms)  \\
    \noalign{\smallskip}
     \hline
\noalign{\smallskip} 
{\tt {SME}}$\tablefootmark{a}$  & \steffsme    & \sloggCasme   &     \sfehsme &     \scahsme&     \smghsme&     \snasme&    \ssihsme & \svsini \\
 {\tt {SpecMatch-Emp}}   & \steffspechmatch & \sloggspechmatch   &  \sfehspechmatch  &\ldots&\ldots&\ldots &\ldots &\ldots  \\
{\tt {ARIADNE}}$\tablefootmark{b}$  &  \steffariadne& \sloggariadne       &     \sfehariadne  &\ldots&\ldots&\ldots &\ldots   &\ldots  \\
   Gaia DR2 &\steffgaia &\ldots &\ldots &\ldots&\ldots&\ldots &\ldots  &\ldots   \\
\noalign{\smallskip} 
\hline 
\end{tabular}
\tablefoot{
\tablefoottext{a}{Adopted as priors for the stellar mass and radius modelling in Sect.~\ref{Subsection: Stellar modelling}.} 
\tablefoottext{b}{Posteriors from  Bayesian Model Averaging  with ARIADNE.}
}
\end{table*}
  
  \begin{figure}[!ht]
 \centering
  \resizebox{\hsize}{!}
            {\includegraphics{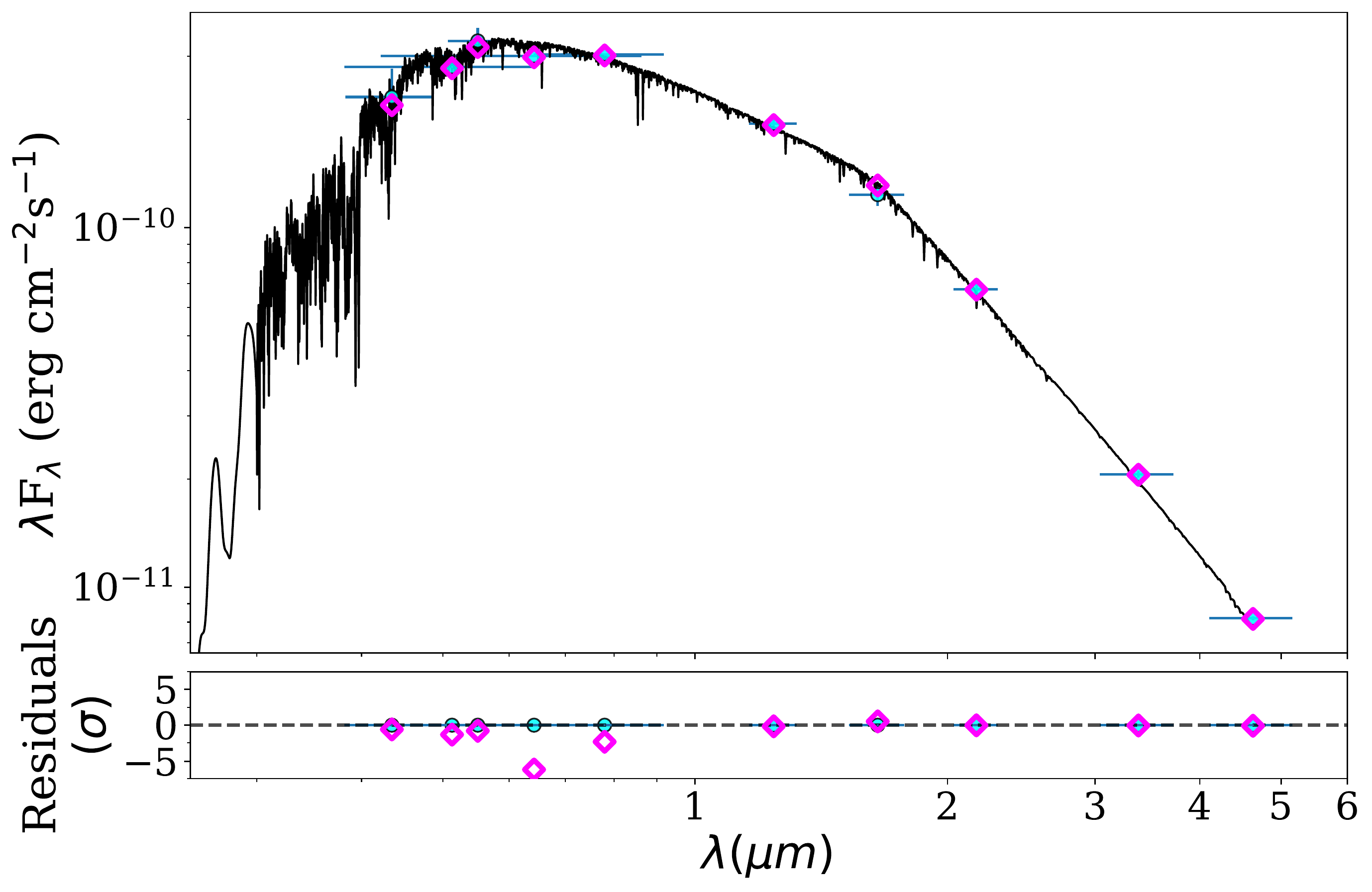}}
   \caption{Spectral energy distribution (SED) of \targeta~and the  model with highest probability 
    from  \citet[][{\tt {Phoenix~v2}}]{2013A&A...553A...6H}. We plot the 
synthetic      photometry   with magenta diamonds and the   observed 
photometry with blue points.  
The   1~$\sigma$ uncertainties are shown with vertical error bars, 
while the horizontal bars display the 
effective width of the passbands. 
In the lower panel we show the residuals  normalised to the errors of the photometry.}
      \label{Figure: SED}
 \end{figure}
 
\subsection{Radial velocity follow-up with HARPS} \label{Subsection: Radial velocity follow-up with HARPS}
We performed high-resolution ($R\approx115\,000$) spectroscopic observations of TOI-2196 using the High 
Accuracy Radial velocity Planet Searcher \citep[HARPS;][]{2003Msngr.114...20M} spectrograph mounted at the 
ESO 3.6~m telescope (La Silla observatory, Chile). We obtained a total of 41 spectra between 24~July~2021 and 
12~November~2021 UT as part of our HARPS large program (\mbox{ID: 106.21TJ.001, PI: Gandolfi}). 
All RVs and activity indicators are listed in Table~\ref{all_rv.tex} along with BJD$_\mathrm{TBD}$, exposure time, and S/N.
We reduced 
the data with the dedicated HARPS data reduction software (DRS) available at the observatory 
\citep{2007A&A...468.1115L} and extracted the radial velocity (RV) measurements using the code 
HARPS-TERRA \citep{2012ApJS..200...15A}, which employs a template-matching algorithm to derive precise 
relative velocities. We also extracted a variety of stellar activity and line profile variation indicators: the 
H$\alpha$ and S-index were extracted using TERRA; the FWHM  and the 
bisector inverse slope (BIS) were derived by cross-correlating the HARPS spectra with a G2 numerical 
mask \citep{Baranne1996,Pepe2002}; and the differential line width (dLW) was extracted 
using the code SERVAL \citep{Zechmeister2018}.


\section{Data analysis} \label{Section: Data analysis}

\subsection{Frequency analysis of HARPS data} \label{Subsubsection: Frequency analysis of HARPS data}
In order to search for the Doppler reflex motion induced by the transiting planet and unveil the presence of 
additional RV signals, we performed a frequency analysis of the HARPS RV measurements and activity indicators. 
To this aim, we computed the generalised Lomb-Scargle  \citep[GLS;][]{Zech09} periodograms of the HARPS 
time series (shown in Fig.~\ref{Fig: TOI-2196 periodogram}) and estimated the false alarm probabilities (FAPs) using 
the bootstrap technique described in \citet{1997A&A...320..831K}. We considered a peak to be significant if 
its FAP is less than 0.1\,\%. 

The GLS periodogram of the HARPS RVs displays a significant peak  at 
$\nu_\mathrm{b} = 0.837$~day$^{-1}$ which is  the transit frequency of TOI-2196\,b (Fig.~\ref{Fig: TOI-2196 periodogram}, 
upper panel, thick vertical yellow line). This peak is not significantly detected in any of the activity indicators 
confirming the planetary nature of the transit signal found in the \textit{TESS} light curve. We note the presence of a 
second significant peak at 0.163~day$^{-1}$ (vertical dashed  red line), which is an alias of the orbital frequency 
of the transiting planet due to the 1~day sampling of our observations.

The second panel of Fig.~\ref{Fig: TOI-2196 periodogram} displays the periodogram of the HARPS RV residuals 
following the subtraction of the best-fitting Doppler orbit of TOI-2196\,b (Sect.~\ref{Section: transit and RV modelling}). 
We found a significant excess of power at frequencies lower than the frequency resolution of our observations 
\mbox{($\approx$1/110~day = 0.009~day$^{-1}$}, where 110 days is the baseline of our observations). This power has no counterpart 
in any of the activity indicators, suggesting that it is likely caused by an outer orbiting companion. As described 
in Sect.~\ref{Section: transit and RV modelling}, we accounted for this long-period Doppler signal by adding a 
linear trend to the RV model. When both the Doppler signal of the transiting 
planet and the linear trend are subtracted from the HARPS RVs, no additional significant signals are found in 
the RV residuals (Fig.~\ref{Fig: TOI-2196 periodogram}, third panel).  


   \begin{figure*}[!ht]
 \centering
  \resizebox{\hsize}{!}
            {\includegraphics{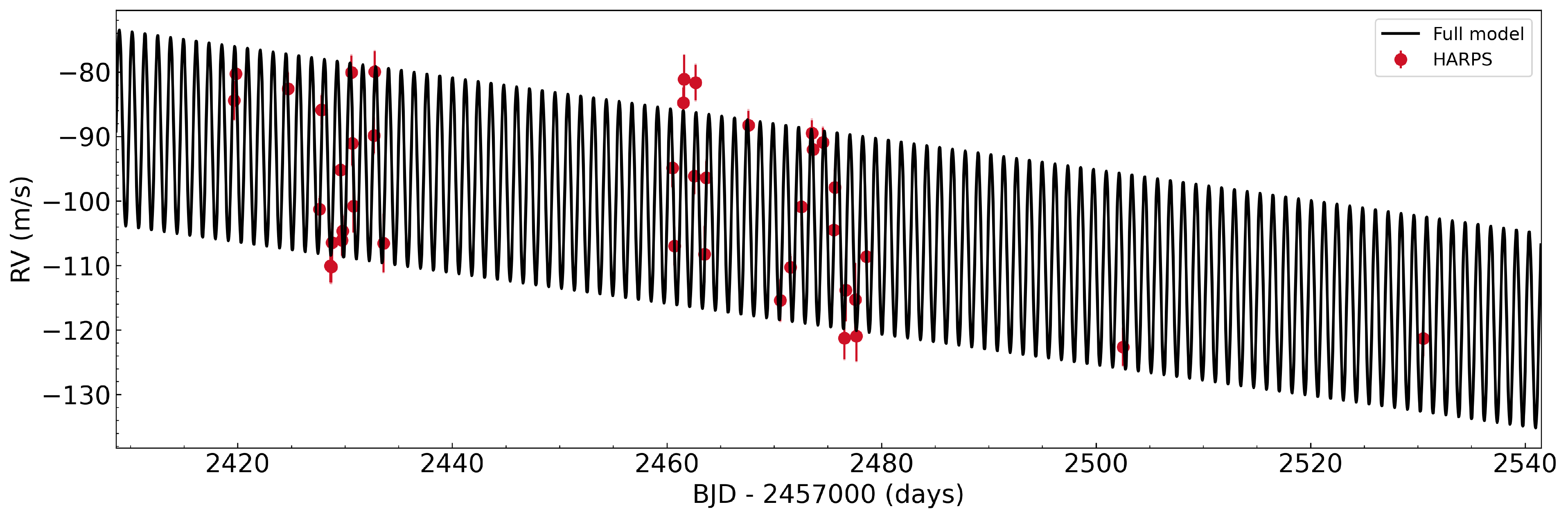}} 
   \caption{Radial velocity times series of \targeta~and the best-fitting RV model. The linear trend suggesting   the presence of an outer companion is clearly identifiable.}
      \label{Figure: RV time series}
 \end{figure*}
 


\subsection{Stellar properties} \label{Subsection: Stellar modelling}
 In order to derive the  fundamental parameters of the host star, we analysed  our co-added  
 high-resolution  HARPS spectra with two methods,  the    
 {\tt {SpecMatch-Emp}} code   \citep{2017ApJ...836...77Y} and 
 {\tt {SME}}\footnote{\url{http://www.stsci.edu/~valenti/sme.html}} 
 \citep[Spectroscopy Made Easy;][]{vp96, pv2017}. In particular, 
 {\tt {SpecMatch-Emp}} is an empirical code that 
compares observations of optical spectra to a dense   library of  
well-characterised FGKM stars,   while 
{\tt {SME}}  fits observed spectra to computed 
synthetic   spectra for a chosen set of parameters based on 
atomic and molecular line data from
VALD\footnote{\url{http://vald.astro.uu.se}}  \citep{Ryabchikova2015}  
and a  stellar atmosphere grid. We chose    Atlas12 \citep{Kurucz2013} and derived
 the stellar  effective temperature, \teff, the surface gravity, \logg, abundances, and
 the projected rotational velocity, \vsini. 
 Each parameter is modelled, one at a time, from specific spectral lines:
the broad line wings of H$_\alpha$ are  
particularly   sensitive to \teff, and the  line wings of the  \ion{Ca}{I} triplet   
6102, 6122, and 6162~\AA~are sensitive to the surface
gravity. Abundances and  the projected stellar rotational velocity, 
\vsini, were modelled from narrow and unblended lines between
6000~and 6600~\AA. We held the micro-  \citep{bruntt08} and macro-turbulent 
\citep{Doyle2014} velocities fixed to 1.0~\kms, and 2.8~\kms, respectively. Further details  on the modelling can be found
in \citet{2017A&A...604A..16F} and \citet{2018A&A...618A..33P}.
Results from both models  listed in Table~\ref{Table: stellar spectroscopic parameters}  are in good agreement within
the uncertainties and   with the effective temperature from
Gaia\footnote{\url{https://gea.esac.esa.int/archive/}} DR2.

  \begin{table}
 \centering
 \caption{Comparison of models of the stellar mass and radius.}
  \label{Table: stellar mass and radius}
 \resizebox{\columnwidth}{!}{%
\begin{tabular}{lccc}
 \hline\hline
     \noalign{\smallskip}
Method    & $M_\star$  &     $R_\star$   & $\rho_\star$     \\
  & ($M_{\odot}$)  & ($R_{\odot}$)  & (g~cm$^{-3}$)  \\
    \noalign{\smallskip}
     \hline
\noalign{\smallskip} 
{\tt {ARIADNE}}$\tablefootmark{a}$    & \smassariadne& \sradiusariadne & \srhoariadne     \\
Gravitational mass    & \smassariadnegrav&  \ldots & \ldots    \\
{\tt {BASTA}}    & \smassbasta    & \sradiusbasta & \srhobasta \\
{\tt {PARAM 1.3}}    & \smassparam    & \sradiusparam & \srhoparam \\
Gaia DR2  & \ldots    &  \srgaia & \ldots \\   
      \noalign{\smallskip} \noalign{\smallskip}
\hline 
\end{tabular}
}
\tablefoot{
\tablefoottext{a}{
Adopted for the modelling in Sect.~\ref{Section: transit and RV modelling}.} 
}
\end{table}

We used the spectral parameters from {\tt{SME}}  as priors to model the stellar radius and
mass with the python code 
{\tt {ARIADNE}}\footnote{\url{https://github.com/jvines/astroARIADNE}} \citep[][]{2022arXiv220403769V}.  
We fit the broadband photometry  bandpasses $G G_{\rm BP} G_{\rm RP}$ from  Gaia eDR3 and  
{\it WISE} W1-W2, along with     $JHK_S$ magnitudes from {\it 2MASS},
 the Johnson $B$ and $V$ magnitudes from APASS, and 
the Gaia eDR3 parallax,   to the 
{\tt {Phoenix~v2}} \citep{2013A&A...553A...6H}, {\tt {BtSettl}} \citep{2012RSPTA.370.2765A}, 
\citet{Castelli2004}, and \citet{1993yCat.6039....0K} atmospheric model grids.     
The dust maps of \citet{1998ApJ...500..525S} were used to obtain an upper limit on     $A_V$.
The relative probabilities of the models were used to compute a weighted average of each parameter  and  
the final  stellar radius is computed with Bayesian model averaging. 
Figure~\ref{Figure: SED} shows the SED model and the fitted bands. 
The {\tt {Phoenix~v2}}  model, which has the highest probability, was used to calculate the synthetic photometry.
We also obtained a luminosity of \Lumariadne~\Lsun,  an extinction that is consistent with zero \mbox{($A_\mathrm{V} =$ \Avariadne)}, as well as 
 the stellar mass   based on MIST \citep{2016ApJ...823..102C} isochrones. 
The model was checked with {\tt {BASTA}}\footnote{\url{https://github.com/timkahlke/BASTA}} 
\citep[the BAyesian STellar Algorithm;][]{2022MNRAS.509.4344A} using the stellar atmosphere grid from \citet{2018ApJ...856..125H}
and the same photometry passbands and priors as above, as well as 
{\tt {PARAM1.3}}\footnote{\url{http://stev.oapd.inaf.it/cgi-bin/param_1.3}} 
\citep{daSilva2006}. The latter model 
 uses Bayesian computation of stellar parameters based on PARSEC isochrones using $V$ magnitude, \teff, [Fe/H], 
and the Gaia eDR3 parallax as priors.

The results, listed in Table~\ref{Table: stellar mass and radius}, are 
in excellent agreement within the 1~$\sigma$ uncertainties. 
We  adopted the stellar parameters 
derived with {\tt {ARIADNE}} in our joint modelling of the radial velocities and light curves in Sect.~\ref{Section: transit and RV modelling}.

The stellar age was  estimated with  the observed NUV excess  
 utilising empirical activity-age relations. 
 We transformed the NUV and $B-V$ photometry to 
$\log R'_{\rm HK} = -4.92 \pm 0.08$ via the empirical relations of \citet{2011AJ....142...23F}, 
which implies an age of $\tau_\star = 5.2 \pm 1.3$~Gyr, according to the empirical relations of  \citet{2008ApJ...687.1264M}.
This is in agreement with \ageariadne~Gyr and \ageparam~Gyr derived from the MIST and PARSEC isochrones, respectively, in the  above modelling.
We note that there is no emission   in the  \ion{Ca}{I} H and K lines, suggesting that the star is not chromospherically active.


\subsection{Joint transit and radial velocity modelling} \label{Section: transit and RV modelling}

For the modelling of the \targeta~system, we turned to the open-source code
{\tt{pyaneti}}\footnote{\url{https://github.com/oscaribv/pyaneti}}
\citep{2019MNRAS.482.1017B, 2022MNRAS.509..866B} to sample the parameter
space using Markov chain Monte Carlo (MCMC) sampling combined with a
Bayesian approach. Following \citet{2022MNRAS.tmp..699B}, we used {\tt{pyaneti}}’s capability to perform multi-band fits and included  
the flattened \textit{TESS} light curve (Sect.~\ref{Section: TESS photometry}), 
the LCOGT light curves (Sect.~\ref{Follow-up photometry from ground: LCOGT}),  and  the  RVs 
listed in Table~\ref{all_rv.tex} in our joint model.

    \begin{figure}[!ht]
 \centering
  \resizebox{\hsize}{!}
            {\includegraphics{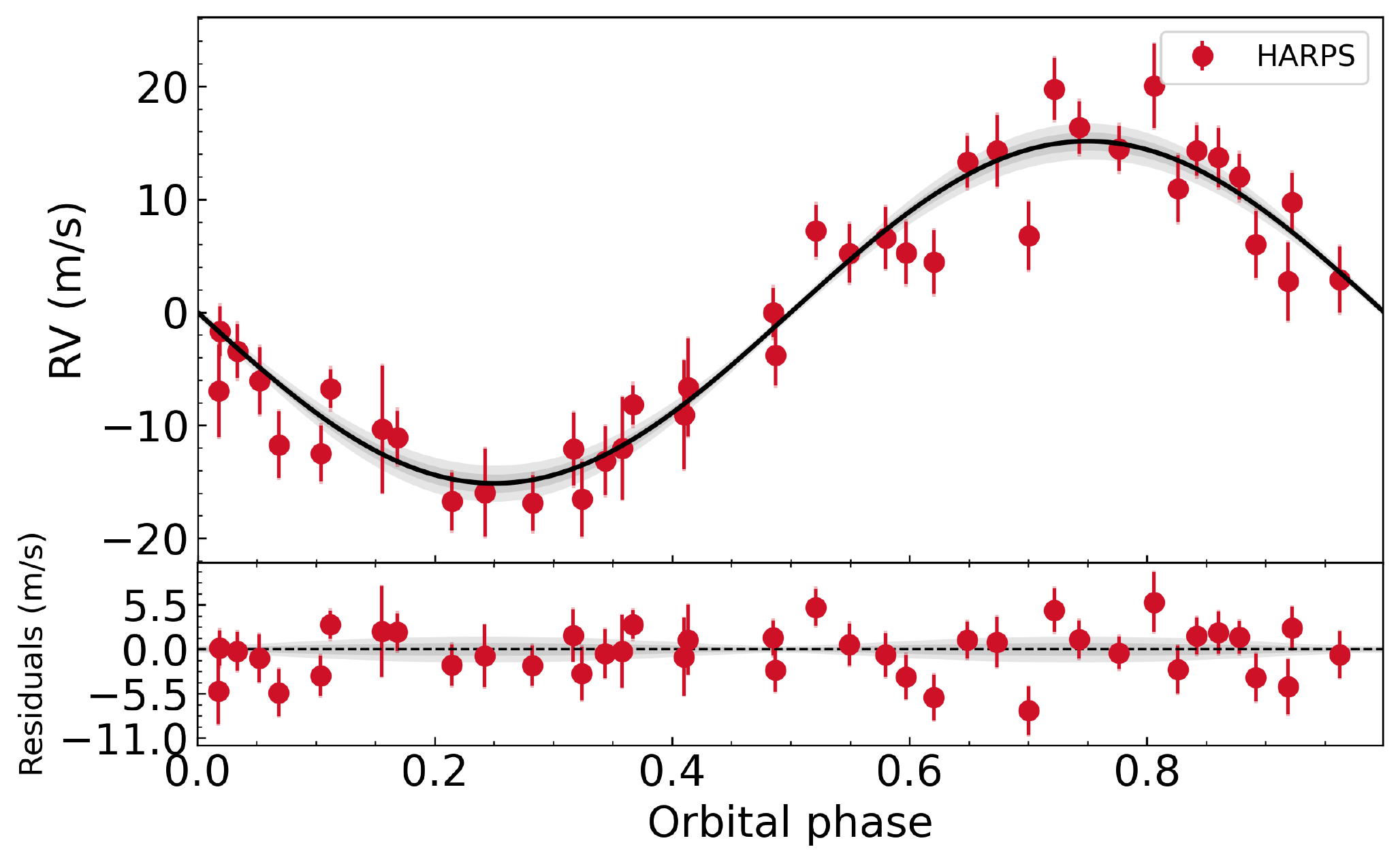}}  
   \caption{HARPS radial velocity  data phase-folded on the orbital period of \targetb~after subtraction of the stellar systemic velocity and
   the linear trend. The  solid black line is the RV model with 1~$\sigma$ 
   and 2~$\sigma$ credible intervals in shaded grey areas. 
   }
      \label{Figure: folded RVs}
 \end{figure}
 
\begin{figure}[!ht]
 \centering
  \resizebox{\hsize}{!}
            {\includegraphics{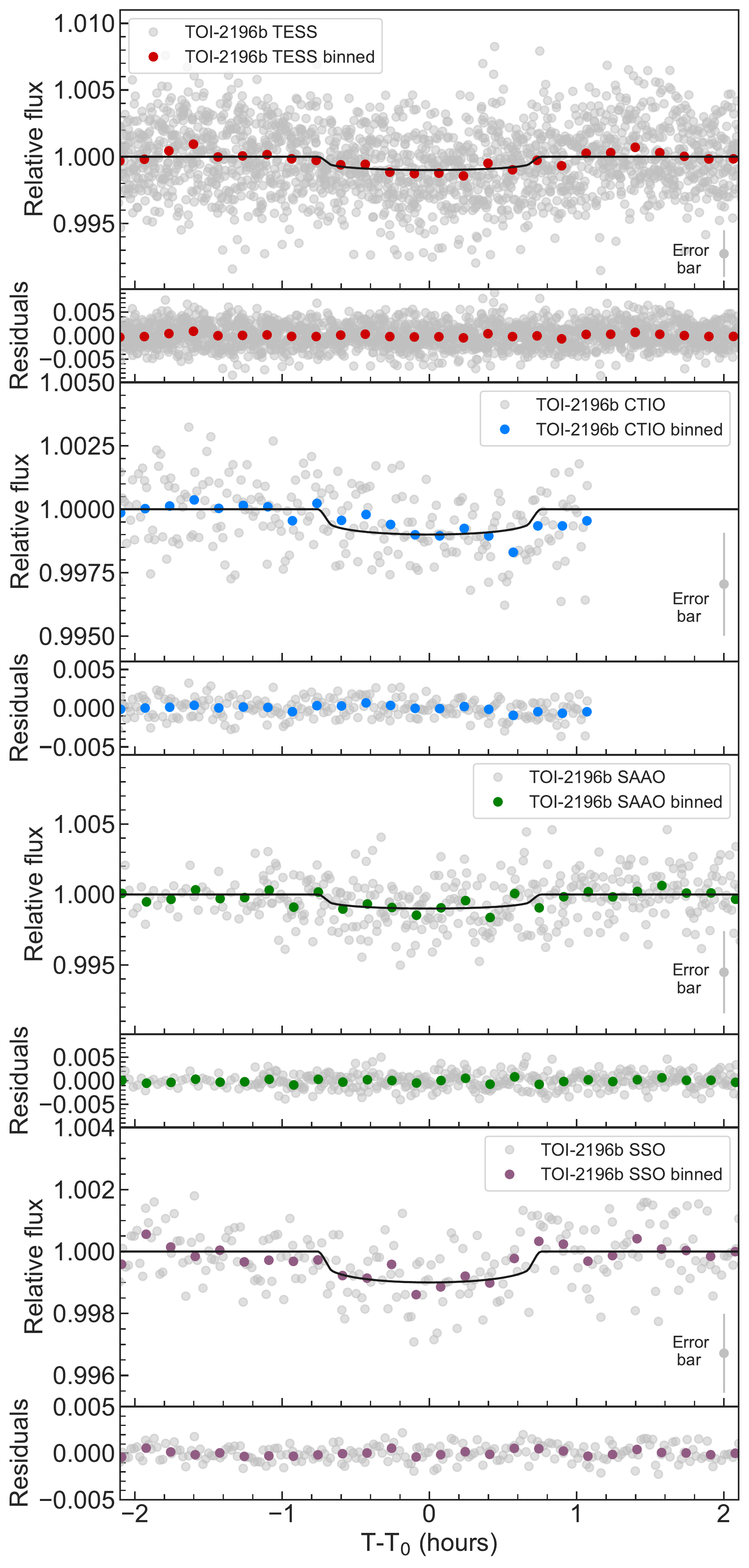}}
   \caption{Flattened and phase-folded \textit{TESS} light curve with the best-fitting transit model in black in the top panel, and follow-up photometry from ground performed with the LCOGT CTIO, SAAO, and SSO telescopes in the second to fourth panels as marked in the legends. The CTIO and SSO panels show the single transits those facilities observed, while the SAAO panel shows the two stacked transits detected with that telescope. The nominal short cadence data are plotted in grey in all panels and binned to 10 mins in colour.}
      \label{Figure: folded transit}
 \end{figure}
 
The parametrisation of the limb darkening coefficients $q_1$ and $q_2$ was
handled as per \citet{Kipping2013}, while the limb darkening model followed
the quadratic approach by \citet{Mandel2002}.
We placed a loose informative prior on the scaled semi-major axis, as well
as on $q_1$ and $q_2$, based on the tables by \citet{2017A&A...600A..30C} for the \textit{TESS} band, 
and \citet{2013A&A...552A..16C} for the ground-based photometry. 
For the
remaining parameters, we used uniform priors.
We tested a model in putting a beta prior on the eccentricity of the planet
and find an eccentricity consistent with zero. Given the short period of the
planet, we thus assumed a circular orbit  and sampled the parameter space
using 250 independent chains thinned with a factor of ten and created
posterior distributions with the last 5000 iterations. This translates to
125\,000 points for each sampled parameter per distribution.

We first carried out a sampling for the independent scaled planet
radii (\rplanet/\rstar) for the \textit{TESS} and each of the LCOGT bands. We obtain values of $R_{SAAO}$~= ~ \rpbSAAO, $R_{CTIO}$~=~\rpbCTIO, $R_{SSO}$~=~\rpbSSO,\ and $R_{TESS}$~=~\rpbSC\ individually, thus demonstrating independent detections from each facility and full consistency (within 1~$\sigma$) between the different estimates. 
We thereby assumed 
 the same transit depth for each band for our final model.  

The RV times series with the best-fitting RV model is shown in Fig.~\ref{Figure: RV time series}, which clearly 
 displays a  linear trend in the RVs  
pointing toward the presence of an outer companion (see Sect.~\ref{Section: rvtrend}). 
The model containing the linear trend is significantly favoured over the one with no trend ($\Delta$BIC = 54). 
 The  RVs folded to the orbital period of the planet are shown in Fig.~\ref{Figure: folded RVs}   with the RV model and  
1~$\sigma$ and 2~$\sigma$ credible intervals in shaded grey areas.  
The four light curves  of \targetb~are shown in Fig.~\ref{Figure: folded transit}, together with the best-fit transit model
with a single radius  plotted in solid black in each panel. 

All results and priors are listed in Table~\ref{Table: Orbital and planetary parameters}  along with 
the adopted stellar parameters derived in Sect.~ \ref{Subsection: Stellar modelling} used in the modelling. 
All adopted stellar parameters are also listed in Table~\ref{Table: Star basic parameters}.

\begin{table*}
\centering
\caption{Description of the {\tt{pyaneti}} model of \targetb~from Sect.~\ref{Section: transit and RV modelling} and the adopted stellar parameters used in the model from Sect.~\ref{Subsection: Stellar modelling} .}
\resizebox{1.75\columnwidth}{!}{%
\begin{tabular}{lcrr}
\hline
\hline
\noalign{\smallskip}
Parameter & Units & Priors$\tablefootmark{a}$ & Final value  \\
\noalign{\smallskip}
\hline

\noalign{\smallskip}
\multicolumn{3}{l}{\emph{Stellar parameters}}\\
\noalign{\smallskip}       
                ~~~$M_\star$   &Stellar mass (\Msun)  \dotfill &$\mathcal{F}$[1.032]     & \smass             \\
\noalign{\smallskip}                
                ~~~$R_\star$ &Stellar radius (\Rsun)  \dotfill &  $\mathcal{F}$[1.043]   &\sradius \\
\noalign{\smallskip}
                ~~~$T_{\rm eff}$  &Effective temperature (K) \dotfill & $\mathcal{F}$[5634]    & \stemp    \\
  
 \noalign{\smallskip}               
\multicolumn{3}{l}{\emph{Fitted parameters}}\\
\noalign{\smallskip}                
                ~~~$T_0$ &Transit epoch (\bjdtdb - 2\,457\,000)\dotfill &  $\mathcal{U}$[2036.4888, 2036.5288]   & \Tzerob             \\
\noalign{\smallskip}                
                ~~~$P_\mathrm{orb}$ &Orbital period (days)\dotfill &  $\mathcal{U}$[1.1943, 1.1963]  &\Pb \\
\noalign{\smallskip}
                ~~~$e$   &Eccentricity  \dotfill & $\mathcal{F}$[0]   & 0    \\
\noalign{\smallskip}
                ~~~$\omega$    &Argument of periastron (degrees)  \dotfill & $\mathcal{F}$[90]  & $90$   \\                
\noalign{\smallskip}                                
                ~~~$b$  &Impact parameter\dotfill &  $\mathcal{U}$[0, 1]   & \bb  \\
\noalign{\smallskip}                
                ~~~$a/R_\star$  & Scaled semi-major axis\dotfill &  $\mathcal{N}$[4.6, 0.1]   &\arb   \\
 \noalign{\smallskip}                                      
                ~~~$R_{\mathrm{p}}/R_\star$ & Scaled planet radius\dotfill & $\mathcal{U}$[0.01, 0.10]    &\rrb \\
\noalign{\smallskip}
                ~~~$K $ & Doppler semi-amplitude variation (\ms)\dotfill &   $\mathcal{U}$[0, 50]     &\kb     \\
 
\noalign{\smallskip}  

                ~~~$q_1$ &Limb-darkening coefficient, \textit{TESS} \dotfill &  $\mathcal{N}$[0.26, 0.10]   & \qoneSC    \\
                
\noalign{\smallskip}                              
                ~~~$q_2$  &Limb-darkening coefficient, \textit{TESS} \dotfill &  $\mathcal{N}$[0.47, 0.10]   & \qtwoSC        \\
                
\noalign{\smallskip}                
                ~~~$q_1$ &Limb-darkening coefficient, LCOGT CTIO \dotfill &  $\mathcal{N}$[0.52, 0.10]   & \qoneCTIO    \\
\noalign{\smallskip}                              
                ~~~$q_2$  &Limb-darkening coefficient, LCOGT CTIO \dotfill &  $\mathcal{N}$[0.17, 0.10]   & \qtwoCTIO        \\
                
\noalign{\smallskip}                
                ~~~$q_1$ &Limb-darkening coefficient, LCOGT SSO \dotfill &  $\mathcal{N}$[0.52, 0.10]   & \qoneSSO    \\
\noalign{\smallskip}                              
                ~~~$q_2$  &Limb-darkening coefficient, LCOGT SSO \dotfill &  $\mathcal{N}$[0.17, 0.10]   & \qtwoSSO        \\
                
\noalign{\smallskip}                
                ~~~$q_1$ &Limb-darkening coefficient, LCOGT SAAO \dotfill &  $\mathcal{N}$[0.52, 0.10]   & \qoneSAAO    \\
\noalign{\smallskip}                              
                ~~~$q_2$  &Limb-darkening coefficient, LCOGT SAAO \dotfill &  $\mathcal{N}$[0.17, 0.10]   & \qtwoSAAO        \\

\noalign{\smallskip}
\multicolumn{3}{l}{\emph{Derived Parameters}}\\               
\noalign{\smallskip}                                             
                  ~~~$M_\mathrm{b}$ & Planet mass (\mearth)\dotfill   & \dots     &     26.0\,$\pm$\,1.3  \\
\noalign{\smallskip}                
                   ~~~$R_\mathrm{b}$ & Planet radius (\rearth)\dotfill &  \dots    & \rpb \\
\noalign{\smallskip}
                ~~~$i\, \,\tablefootmark{b}$  &Inclination (degrees)\dotfill &   \dots   & \ib  \\                

 \noalign{\smallskip}                                              
                  ~~~$a$ &Semi-major axis (au)\dotfill &  \dots   & \ab   \\                
 \noalign{\smallskip}                
           ~~~$F$ &Instellation  ($F_\mathrm{\oplus}$)\dotfill & \dots     & 2000\,$\pm$\,100  \\
  
   \noalign{\smallskip}                
           ~~~$\rho_\mathrm{b}$ & Planet density  (g~cm$^{-3}$)\dotfill &  \dots   & \denpb \\

 \noalign{\smallskip}                
           ~~~$g_\mathrm{b}$ & Planet   surface gravity  (cm~s$^{-2}$) \dotfill & \dots    &    2100\,$\pm$\,200 \\
            
\noalign{\smallskip}                
           ~~~$T_\mathrm{eq}\,\tablefootmark{c}$ &Equilibrium temperature (K)\dotfill & \dots    &   1860\,$\pm$\,20 \\
           
 \noalign{\smallskip}                
           ~~~$\Lambda\,\tablefootmark{d} $ &  Jeans escape parameter \dotfill & \dots    &  \jspb \\ 
           
 \noalign{\smallskip}                
           ~~~TSM\tablefootmark{e} & Transmission spectroscopy metric \dotfill & \dots    &  \tsmb \\            
               
\noalign{\smallskip}                          
                ~~~$T_{14}$ &Total transit duration (hours) \dotfill &  \dots   & \ttotb     \\
 
 \noalign{\smallskip}                          
                ~~~$T_{23}$ &Full   transit  duration  (hours)  \dotfill & \dots    & \tfulb    \\               
  \noalign{\smallskip}                          
                ~~~$T_{12}$ &Ingress and egress   transit  duration  (hours)  \dotfill & \dots    &  \tegb    \\

\noalign{\smallskip}
\multicolumn{3}{l}{\emph{Additional Parameters}}\\      
\noalign{\smallskip}                              
                ~~~$\dot{\gamma_1}$  &Linear trend HARPS (\ms~days$^{-1}$)\dotfill &  $\mathcal{U}$[-100, 100]    &  \ltrend      \\

\noalign{\smallskip}                              
                ~~~$\dot{\gamma_1}$  &Systemic velocity HARPS (\kms)\dotfill &  $\mathcal{U}$[-1.0218, 1.0208]    &  \HARPSN     \\
               
\noalign{\smallskip}                              
                ~~~$\sigma_{F1}$ &RV jitter HARPS (\ms)\dotfill &$\mathcal{J}[10^{-3}, 10^{-1}]$ &\jHARPSN      \\        
  \noalign{\smallskip}                              
                ~~~$\sigma_{TESS}$ &\textit{TESS} light curve jitter\dotfill & $\mathcal{J}[10^{-2}, 10^{-3}]$ &   \jtrSC    \\    
  \noalign{\smallskip}                              
                ~~~$\sigma_{CTIO}$ & LCOGT CTIO light curve jitter\dotfill & $\mathcal{J}[10^{-2}, 10^{-3}]$ &   \jtrCTIO   \\   
                
  \noalign{\smallskip}                              
                ~~~$\sigma_{SSO}$ & LCOGT SSO light curve jitter\dotfill & $\mathcal{J}[10^{-2}, 10^{-3}]$ &   \jtrSSO   \\ 
                
  \noalign{\smallskip}                              
                ~~~$\sigma_{SAAO}$ & LCOGT SAAO light curve jitter\dotfill & $\mathcal{J}[10^{-2}, 10^{-3}]$ &   \jtrSAAO   \\ 
       
\noalign{\smallskip}                
\hline
\end{tabular}
}
\label{Table: Orbital and planetary parameters}
\tablefoot{
\tablefoottext{a}{$\mathcal{U}$[a,b] refers to uniform priors in the range  \emph{a} --  \emph{b}, $\mathcal{F}$[a] to a fixed value $a$,   $\mathcal{N}$[a,b] 
to Gaussian priors with mean \emph{a} and standard deviation  \emph{b}, and $\mathcal{J}$[a,b] to modified Jeffrey's priors \citep[Eq.~16 in][]{2005ApJ...631.1198G}.} 
\tablefoottext{b}{Orbit inclination   relative to the   plane of the sky.} 
\tablefoottext{c}{Dayside equilibrium temperature, assuming no heat redistribution and  zero albedo (Eq.~\ref{Teq}).}
\tablefoottext{d}{Jeans escape parameter defined as $\Lambda =  G M_\mathrm{p} m_\mathrm{H} / (k_\mathrm{B} T_\mathrm{eq} R_\mathrm{p})$ in \citet{2017A&A...598A..90F}.}
\tablefoottext{e}{\citet{2018PASP..130k4401K}.}
}
\end{table*}

     \begin{figure*}[!ht]
 \centering
\includegraphics[scale=0.45]{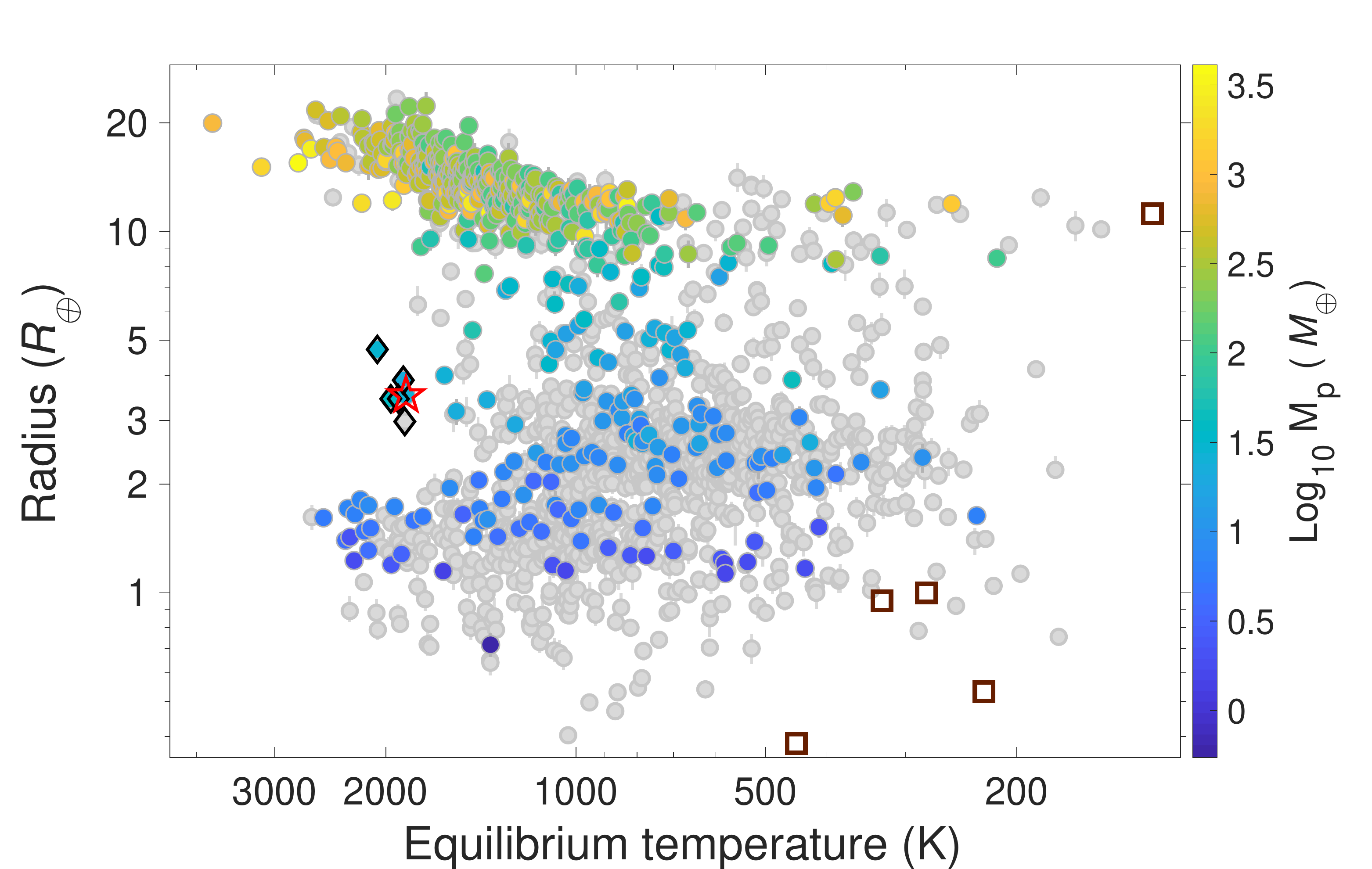} 
   \caption{Radius vs. equilibrium temperature diagram: known planets  
             with  a precision of 10~\% or better in  radius are plotted in grey. 
   About a third of these planets    have  also RV mass measurements with a precision of 30~\% or 
   better  and are colour-coded with planet mass. We set upper limits to planet masses at 13~\mjup.  
   \targetb~is marked with a red star symbol, the  five additional planets in the hot Neptune desert  
   with black diamonds,   and solar system planets   with brown squares. 
   The five additional planets located in the desert are: 
 K2-100~b \citep{2019MNRAS.490..698B},  TOI-849~b \citep{2020Natur.583...39A},   LTT 9779~b \citep{2020NatAs...4.1148J},  
 K2-278~b \citep{2018AJ....156..277L}, and Kepler-644~b \citep{2018ApJ...866...99B} of which the latter two have only  radius measurements.}
      \label{Figure: R-Teq}
 \end{figure*}


\subsection{Evidence of an outer companion}
\label{Section: rvtrend}

The assumption of a long-period  companion  is supported by the periodogram (panel~2 in Fig.~\ref{Fig: TOI-2196 periodogram}) and the linear trend   
in the RV data (Table~\ref{Table: Orbital and planetary parameters} and Fig.~\ref{Figure: RV time series}).
We computed the minimum mass of this outer companion, here denoted with $c$, using the measured linear trend 
of \mbox{\ltrend~m~s$^{-1}$~day$^{-1}$} from our RV analysis in Sect.~\ref{Section: transit and RV modelling}  and 
Eq.~2 of \citet{2016PASP..128j2001B}:
\begin{equation}
\frac{M_\mathrm{c}}{a^2_{\mathrm{c}}} > 0.0145 ~ \biggl |  \frac{\dot{\gamma}}{\mathrm{m}~\mathrm{s}^{-1}~\mathrm{yr}^{-1}} \biggr |  = 1.3~M_\mathrm{J}~\mathrm{au}^{-2}\ ,
\end{equation}
where $\dot{\gamma}$ is the slope of the linear trend  (acceleration), and $a_{\mathrm{c}}$ is the minimum semi-major 
axis compatible with the RV data. Following \citet{2017MNRAS.464.2708S}, we assumed zero eccentricity and a 
minimum orbital period of    twice the baseline of our RV measurements ($P_\mathrm{orb, c} > 222$~days), resulting in  
a minimum semi-major axis of $0.717~\pm~0.009$~au and a corresponding minimum mass of $0.65~\pm~0.05$~\mjup. 

We note that the Gaia renormalised unit weight error (RUWE) 
value\footnote{\url{https://gea.esac.esa.int/archive/documentation/GDR2/Gaia_archive/chap_datamodel/sec_dm_main_tables/ssec_dm_ruwe.html}.} 
is 0.99 for \targeta, corresponding to a low astrometric signal. This implies that a low-mass stellar companion scenario is unlikely. 
However, future long-term RV monitoring of the star is needed to firmly determine the nature of the signal.



 \begin{table*}
 \centering
 \caption{Known sub-Neptune   and Neptune planets in the hot Neptune desert ($T_\mathrm{eq} > 1800$~K)  
 with a precision in radius of 10~\% or better. In addition to \targetb, three of the planets have RV measurements with a precision of   30~\% or better.}    
\resizebox{2\columnwidth}{!}{%
 \label{Table: Hot Neptune desert planets}
\begin{tabular}{lcccccccccc}
\hline \hline \noalign{\smallskip}
Planet  &Radius &Mass   &  Bulk density  & $P_\mathrm{orb}$     &$T_\mathrm{eq}$ &  $T_\mathrm{eff}$ & [Fe/H] &   $\Lambda$\tablefootmark{a}   & Age & Ref   \\  
& (\rearth)  &(\mearth) & (\gc) &(d)    & (K) &    (K)&  &  & (Gyr) \\
    \noalign{\smallskip}
     \hline
\noalign{\smallskip}

\targetb                &\rpb                   & \mpb          &\denpb &   1.20  &1860&   5634 &  $0.14\pm0.05$  & 29.6&\ageariadne &\tablefootmark{b} \\ \noalign{\smallskip} 

K2-278~b&$2.98\pm0.23$  & \ldots& \ldots                &   3.33        &1867&   6747& $0.00\pm0.17$   & \ldots & \ldots       &\tablefootmark{c}\\ \noalign{\smallskip}  

K2-100~b                &$3.88\pm0.16$  & $21.8\pm6.2$& $2.05\pm0.64$            &   1.67        &1878&   5945   &$0.22\pm0.09$  &22.8&$0.75^{+0.004}_{-0.007}$  &\tablefootmark{d}\\ \noalign{\smallskip}  

TOI-849~b       &$3.44^{+0.16}_{-0.12}$ & $39.1 \pm 2.6 $&$5.26 \pm 0.71$         &   0.77        &1966&    5374&$0.19\pm0.03$   & 43.9&$6.7^{+2.8}_{-2.4}$               &\tablefootmark{e}\\ \noalign{\smallskip}  

Kepler-644~b            &$3.44^{+0.18}_{-0.35}$ & \ldots& \ldots                 &   3.17        &1912&     6540& $0.08\pm0.15$   & \ldots &$1.6^{+0.52}_{-0.32}$        &\tablefootmark{f}\\ \noalign{\smallskip}  
 
LTT~9779~b      &$4.72\pm0.23$  &$29.3 \pm 0.8$ &$1.53\pm0.23$          &   0.79          &2064&    5443 & $0.27\pm0.03$  &22.9& $1.9^{+1.7}_{-1.2}$  &\tablefootmark{g}\\  \noalign{\smallskip} 

\noalign{\smallskip} 

\noalign{\smallskip} 
\hline 
\end{tabular}
}
\tablefoot{
\tablefoottext{a}{\citet{2017A&A...598A..90F}.}    
\tablefoottext{b}{This work.} 
\tablefoottext{c}{ \citet{2018AJ....156..277L}.} 
\tablefoottext{d}{ \citet{2019MNRAS.490..698B}.}   
\tablefoottext{e}{\citet{2020Natur.583...39A}.}   
\tablefoottext{f}{ \citet{2018ApJ...866...99B}.}   
\tablefoottext{g}{\citet{2020NatAs...4.1148J}.}   
}
\end{table*}

\section{Discussion} \label{Section: Discussion}

\subsection{The hot Neptune desert} \label{Subsection: Hot Neptune desert}
 
\targetb~is one of very few planets found in the hot Neptune desert (shown in Fig.~\ref{Figure: R-Teq}). 
In this figure, we plot in grey the radius of all   known planets 
 with radius measurements from
 transit surveys   with a precision of 10~\% or better as a function of 
equilibrium temperature.  
About a third  of the planets in Fig.~\ref{Figure: R-Teq} also have   masses from  RV measurements  with  a precision of 30~\% or better and are colour-coded with mass. 
The data were downloaded  from the NASA Exoplanet archive\footnote{\url{https://exoplanetarchive.ipac.caltech.edu}} 
and we chose the latest results with the highest  precision for   planets with several entries or, if they share a similar precision, we chose the most recent results. 
Since the information about equilibrium temperature  is not always given by the references, we computed  $T_\mathrm{eq}$ in the same way
for all planets with \citep[e.g.][]{2005ApJ...626..523C}
\begin{equation} \label{Teq}
T_\mathrm{eq} =  \sqrt{\frac{R_\star}{2\,a}} \ T_\mathrm{eff}\  [f\,(1 - A_\mathrm{B})]^{1/4} 
,\end{equation}
where $a$ is the planet's semi-major axis (computed with Kepler III), $A_\mathrm{B}$ is the Bond albedo, and 
$f$ is the heat redistribution factor. The latter two parameters are here assumed to be zero and unity, respectively. We prefer to use $T_\mathrm{eq}$   over orbital period since the latter does not take into account differences of stellar types which may give  misleading results. 
With fixed Bond albedo and heat redistribution, an ultra-short period planet with $P_\mathrm{orb} < 1$~day around an M-dwarf  has a  lower 
equilibrium temperature than a planet with an orbital period of several days orbiting a sun-like star.
 
      \begin{figure}[!ht]
 \centering
  \resizebox{\hsize}{!}
            {\includegraphics[scale=0.5]{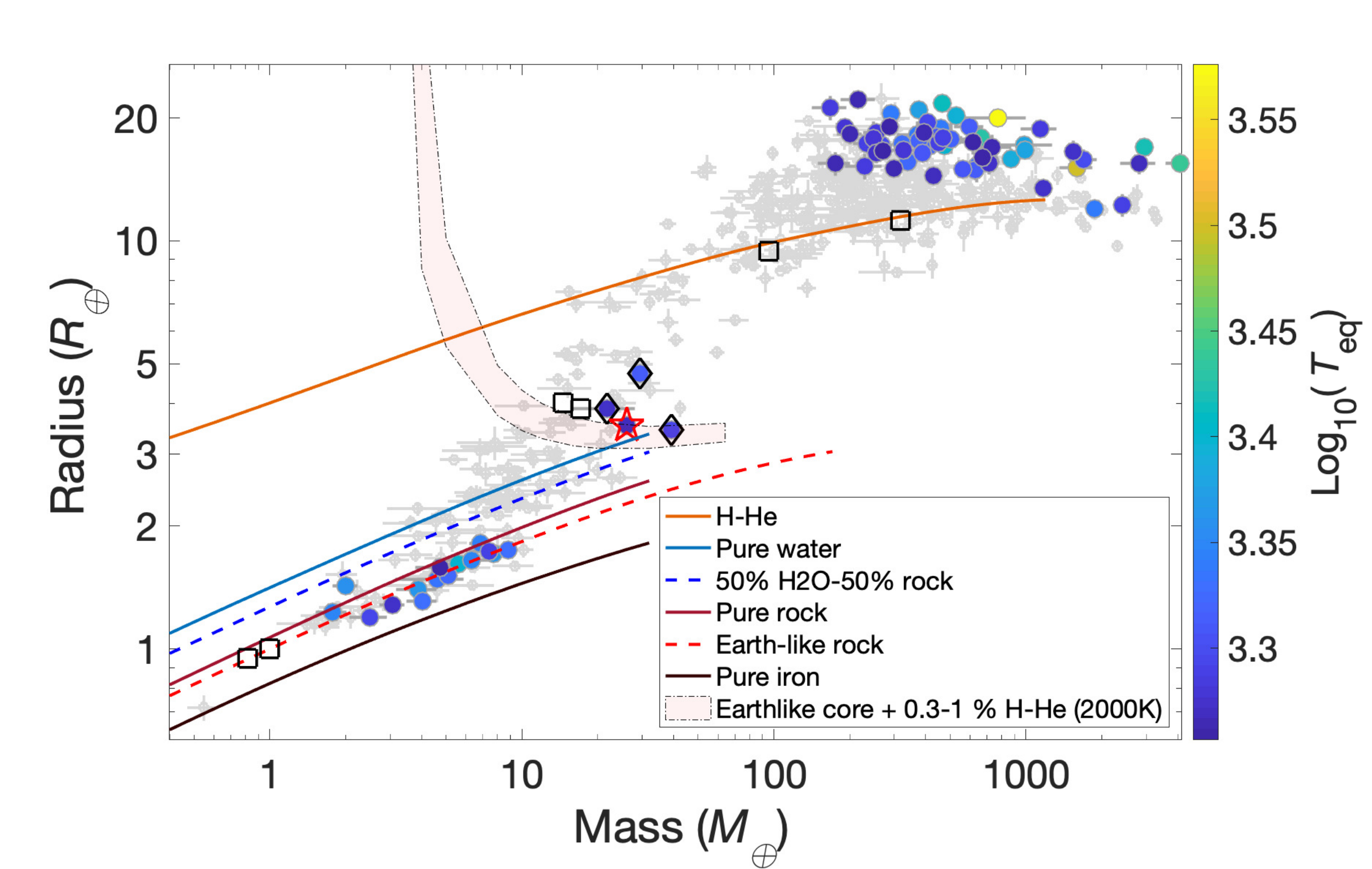}}. 
   \caption{Diagram of all  known planets with masses from radial velocity measurements up to 13~\mjup~and radii from transit photometry 
   with  30~\% and 10~\% uncertainties or lower in mass and radius, respectively. 
   Planets with   \mbox{$T_\mathrm{eq} \ge 1800$~K} 
   are colour-coded with   $T_\mathrm{eq}$, and the rest are plotted in light grey. In total, there are four  planets (including \targetb) with radii  between 
   \mbox{3 and 5 \rearth}~in this diagram that are identified as hot Neptune desert planets (cf. Fig.~\ref{Figure: R-Teq}).  
        \targetb~is marked with a red star symbol and the  three additional planets  
   with black diamonds. Solar system planets are marked  with brown squares.
 Interior models from \citet{2019PNAS..116.9723Z} are plotted as listed in the legend. 
The 100~\% water line is for a planet with condensed water phases. 
}
      \label{Figure: MR diagram Teq 1800}
 \end{figure}

     \begin{figure}[!ht]
 \centering
  \resizebox{\hsize}{!}
            {\includegraphics[scale=0.5]{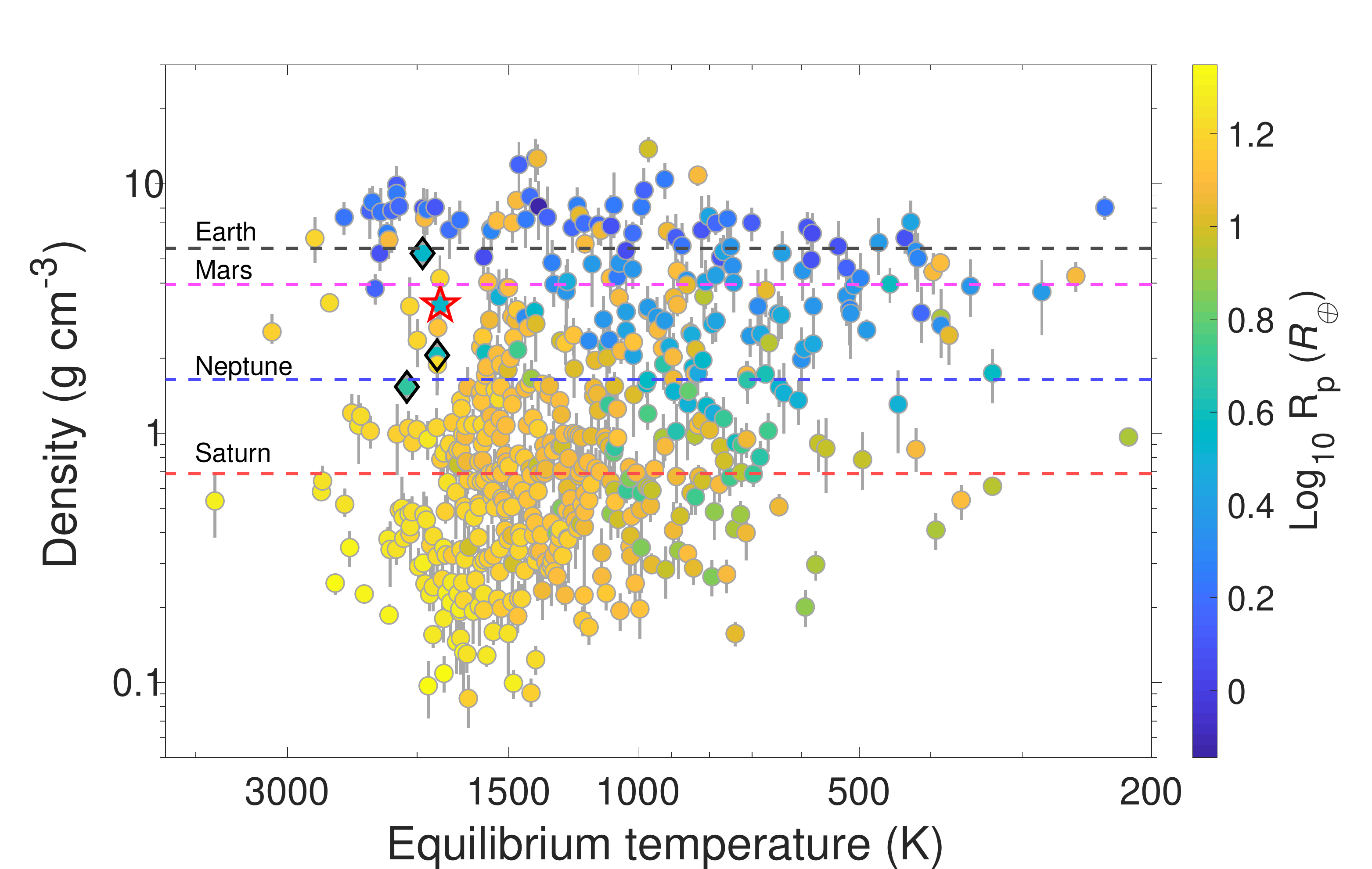}}
   \caption{Bulk density-equilibrium temperature diagram of the same planets as in Fig.~\ref{Figure: MR diagram Teq 1800}   colour-coded here with planet radius.
 The densities of Earth, Mars, Neptune, and Saturn are
   marked with dashed-coloured lines.
}
      \label{Figure: Density-Teq}
 \end{figure}

The  dearth of short-period  Neptunes is clearly seen in Fig.~\ref{Figure: R-Teq}.  
In this plot, the desert starts around 1600~K for medium-sized   planets, 
which corresponds to  orbital periods of  1.9~days around   sun-like stars.
For equilibrium temperatures higher than 1800~K, most planets have  $ R_{\mathrm{p}} \lesssim 1.8$~\rearth~or $R_{\mathrm{p}} \gtrsim 1$~\rjup.
In addition to \targetb, we identified five more planets  confirmed in the desert, which are  listed in Table~\ref{Table: Hot Neptune desert planets}: 
 K2-100~b \citep{2019MNRAS.490..698B}, TOI-849~b \citep{2020Natur.583...39A},   LTT 9779~b \citep{2020NatAs...4.1148J},  
 K2-278~b \citep{2018AJ....156..277L}, and Kepler-644~b \citep{2018ApJ...866...99B}   
of which the latter two  have no mass measurements.
This diagram suggests that this small group of planets  delimits two regimes: a hot sub-Neptune desert  
 for planets with radii between 1.8 and 3~\rearth, and a sub-Jovian desert for radii $5-12$~\rearth. 
 More planets in this parameter space are needed to
 establish whether this is  a selection effect or some kind of stability island in the desert.

In Fig.~\ref{Figure: MR diagram Teq 1800}, we show the position of \targetb~in 
a mass-radius diagram. We plot  the  same planets  with both   radii and RV masses as 
plotted in Fig.~\ref{Figure: R-Teq}. Here,  all planets are plotted in grey except for   
  planets with $T_\mathrm{eq} \ge1800$~K which are colour-coded with equilibrium temperature.
\targetb~has a radius smaller than Neptune but   an approximately 50~\% higher mass and, hence, twice Neptune's density.   
As already noted in Fig.~\ref{Figure: R-Teq}, it joins the small group of the three   planets found between small, rocky planets and  gaseous  giants.  
We also plot interior structure models from \citet{2019PNAS..116.9723Z}\footnote{\url{https://lweb.cfa.harvard.edu/~lzeng/planetmodels.html}}
 with and without the addition of atmospheres as
listed in the legend. 
According to these models, the composition of  \targetb~is consistent with an Earth-like core with an $0.3-1$~\% 
H-He atmosphere at an equilibrium temperature of 2000~K, and  
also lies slightly above a pure (condensed) water planet model. 
We investigated the   atmospheric loss and 
interior composition with a model that considers water or H/He 
phases for highly-irradiated planets such as \targetb~in Sects.~\ref{Section: atmospheric loss} and \ref{Section: Internal structure}, respectively.

  \begin{figure*}
     \centering
     \begin{subfigure}[b]{0.49\textwidth}
         \centering
         \includegraphics[width=\textwidth]{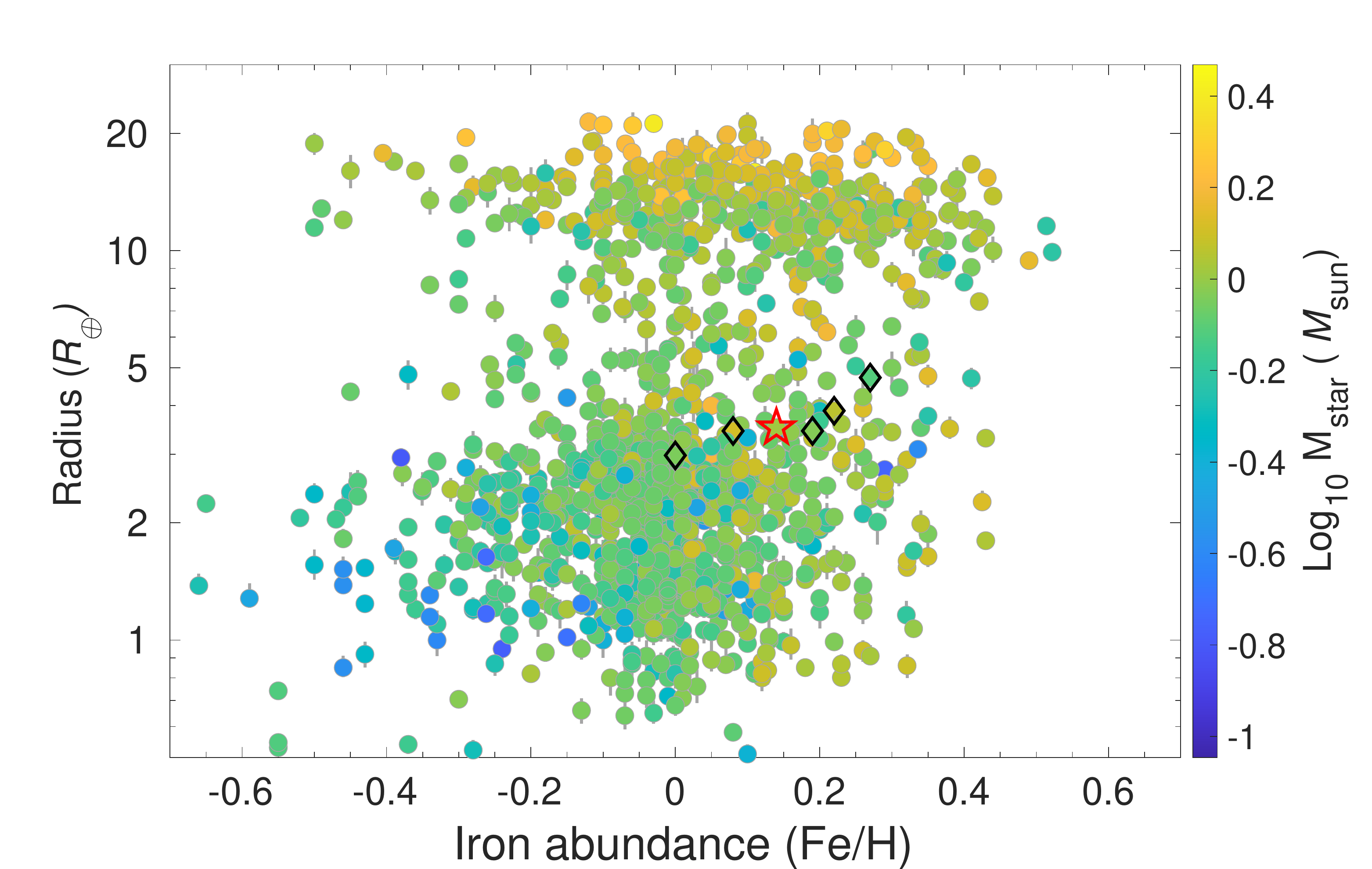}
            \caption{ }
         \label{Figure: Radius-feh-all}
     \end{subfigure}
     \begin{subfigure}[b]{0.49\textwidth}
         \centering
         \includegraphics[width=\textwidth]{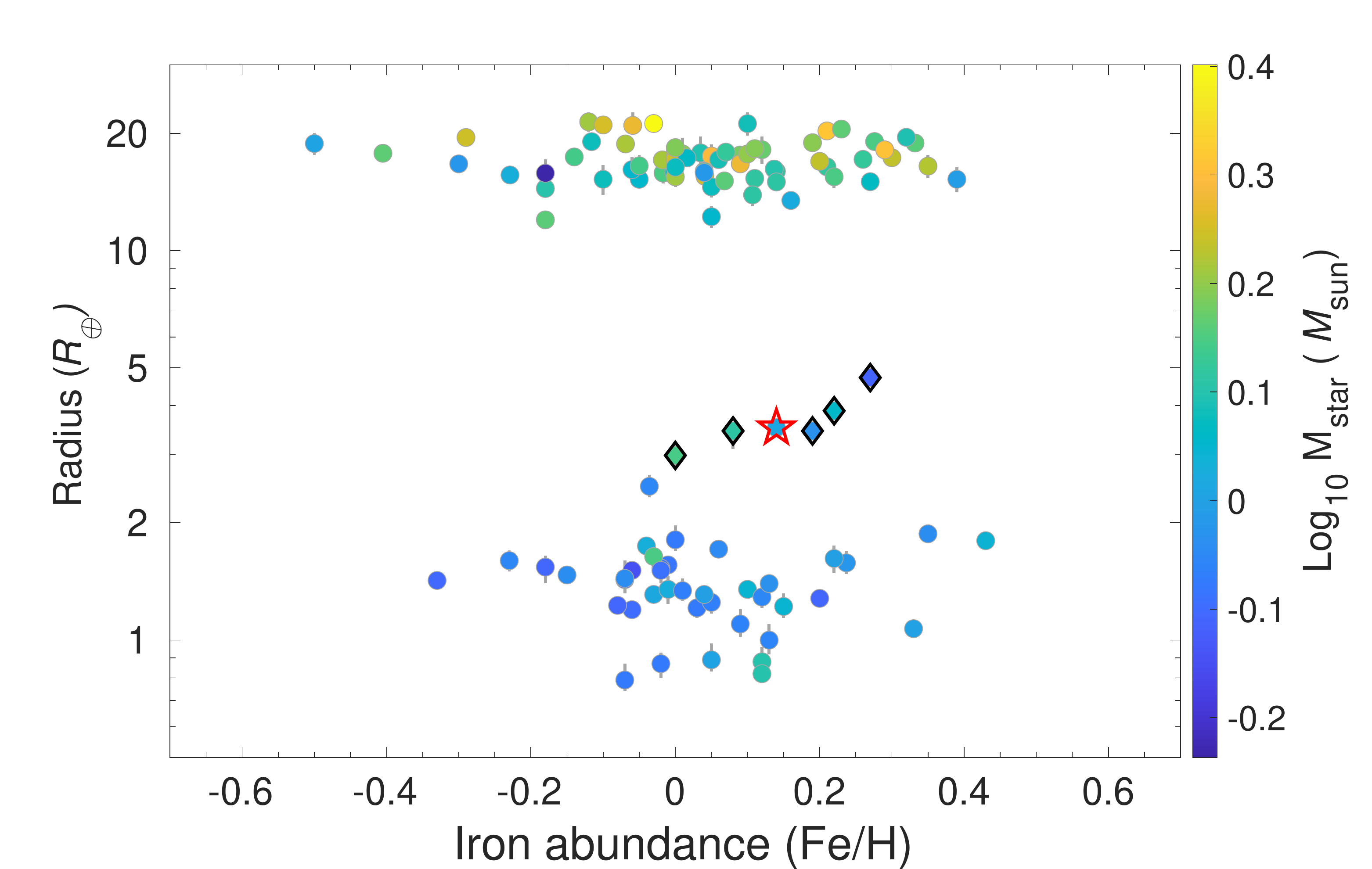}
            \caption{}
       \label{Figure: Radius-feh-Teq=1800}
     \end{subfigure}
                       \caption{Planet radius vs. iron abundance diagram of the same planets  
                       as in Fig.~\ref{Figure: R-Teq}: \emph{a)}  Colour-coded with stellar mass.    
             \emph{b)}~Same   but  only plotting planets with  \mbox{$T_\mathrm{eq} \ge 1800$~K,} 
            which corresponds to \mbox{$P_\mathrm{orb} \lesssim 1.3$~d} for a sun-like star assuming zero Bond
            albedo and a heat redistribution factor of unity, or 
            \mbox{$P_\mathrm{orb} \lesssim 0.43$~d} for a K5\,V star.}       
   \label{Figure: Radius-feh}
\end{figure*}

To allow for a comparison to both previous plots, we plotted planet bulk densities versus equilibrium temperatures 
in Fig.~\ref{Figure: Density-Teq}, colour-coded according to planetary radius. 
The hot Neptune desert is also visible  in this plot, albeit less prominent since planets with $\gtrsim 1$~\mjup~have increasingly
higher densities, thereby filling in the lower parts of the $\rho-T_\mathrm{eq}$ desert. Most planets in this plot  with high 
$T_\mathrm{eq}$  have either high  densities or large radii if the densities are low.

To investigate any potential correlations with stellar metallicity, we plotted the radius versus the iron abundance relative to
hydrogen in Fig.~\ref{Figure: Radius-feh}.  
In  panel a) we plot   all planets and in panel b) we plot planets  with  
 \mbox{$T_\mathrm{eq} \ge 1800$~K}. We note that all hot Neptune desert planets  
 have  greater than solar iron abundances, although the sample 
 is too small to draw any firm  conclusions.


 \subsection{Atmospheric loss} \label{Section: atmospheric loss}
Planets in close proximity to their host stars are exposed to  high levels of  X-ray and extreme UV radiation (XUV) that can 
erode planetary atmospheres.  If the escape is energy limited, the atmospheric mass loss rate of \targetb~can be written as \citep{Er07,2013ApJ...775..105O}: 
\begin{equation}
    \dot{M} = \epsilon \frac{\pi F_\mathrm{XUV} R_\mathrm{p}^3}{G M_\mathrm{p} }, \label{eq:atmospheric-loss}
\end{equation}
where $F_\mathrm{XUV}$ is the XUV flux received by the planet, $G$ the gravitational constant, and $\epsilon$ 
is an efficiency parameter. Planets with low densities in close orbits are most susceptible to mass 
loss, with $\epsilon$ that can also depend on the mass and radius of the planet \citep{Owen2012}. 
Since the mass of the atmosphere of   average sub-Neptunes is typically $1-10$~\%  of 
the planetary mass \citep{LopFor14}, atmospheric loss is possible and can transform  the planet 
 into a rocky super-Earth \citep{2013ApJ...776....2L,2013ApJ...775..105O}.

The XUV irradiation produced by solar-type stars is   the most prominent  at the earliest stage of their evolution (saturation regime). 
After this stage, the XUV luminosity decreases with time. For a  1~\Msun~star, we 
estimate that the saturation regime lasts 30 Myr and  emits $1.1\times 10^{31}$ erg~s$^{-1}$ of XUV 
\citep{Sanz2011},  resulting in a mass loss rate of \mbox{6.9~\mearth~Gyr$^{-1}$} for \targetb. 
While the XUV irradiation decreases with time after the saturation regime, it continues to contribute to atmospheric losses over   \mbox{$\sim 1$~Gyr}. To quantify this 
effect, we followed the approach of \cite{Agui2021} and integrated the average XUV radiation produced by a 
1~\Msun~star \citep{Sanz2011}, assuming that the planet properties remained roughly constant during its evolution. 
With an efficiency of 5~\% \citep[see Fig.~13 of][]{Owen2012}, we found that \targetb~has lost $\sim 0.8$~\mearth~of H/He. 
This implies that \targetb~could have formed with a volatile mass fraction of $4-6$~\%, and lost 50~\% to 90~\% of its 
atmosphere, a result that is consistent with the findings of \cite{Estrela2020}.

Rapidly rotating stars can have longer saturation regimes, lasting up to 300 Myr 
\citep{2015A&A...577L...3T,2021MNRAS.500.4560P}. When setting the duration of the saturation regime to 300 Myr, that is, ten times longer than in the computation above, the planet receives almost ten times more XUV energy, thus increasing the total lost mass by a factor of 10 (see Eq. \ref{eq:atmospheric-loss}). In this extreme case, we estimate a mass loss corresponding 
to $\sim 8~$\mearth~of H/He envelope, 
placing the initial volatile mass fraction of \targetb~at 35~\%. An envelope that 
massive contradicts the observed 1-10~\% range for sub-Neptunes, suggesting moderate or low activity levels for \targeta. 
However, these computations assume a constant planet mass and radius over time.
The mass loss estimates should consider the coupling of the mass loss and interior structure. 
The apparent correlation between the planet age and the   Jeans escape parameter $\Lambda$
in Table \ref{Table: Hot Neptune desert planets} is supported by the study of \cite{2017A&A...598A..90F}. 
They found  
that planets with $\Lambda \ge 15-35$ experience important mass loss until their mass or radius (or both) adapts to increase 
$\Lambda$. We conclude that \targetb~experienced  atmospheric mass loss, which may still be taking place with a present-day mass loss rate of 
0.01~\mearth~Gyr$^{-1}$. Its atmosphere has, however, not  yet  been removed  entirely, placing it in the sub-Neptune category 
of exoplanets. The decrease in the mass-loss rate that prevented \targetb~from losing its atmosphere 
could possibly be due to a change in the atmospheric composition. Atomic hydrogen, which can be 
produced by photo-dissociation of H\textsubscript{2} or H\textsubscript{2}O, tends to escape at high rates mostly because of its low mass, but also because of its long radiative cooling time. A longer radiative cooling time implies that the gas will lose its heat through mechanical work, that is, through evaporative flow, rather than through radiation \citep{Owen2012}. A selective loss of H could result in atmospheres that are dominated 
by He, O\textsubscript{2}, H\textsubscript{2}O, or other heavy species with much lower escape rates 
\citep[see][respectively]{2015ApJ...807....8H,2017MNRAS.464.3728B,Agui2021,2021MNRAS.502..750I}. 
Since this planet inhabits a sparsely populated part of the radius versus equilibrium temperature diagram, 
it is a good candidate to study exceptions to the evaporation valley and evaporating planets in general.


\subsection{Internal structure} \label{Section: Internal structure}
 
\begin{figure} 
\centering
{\includegraphics[scale=0.3]{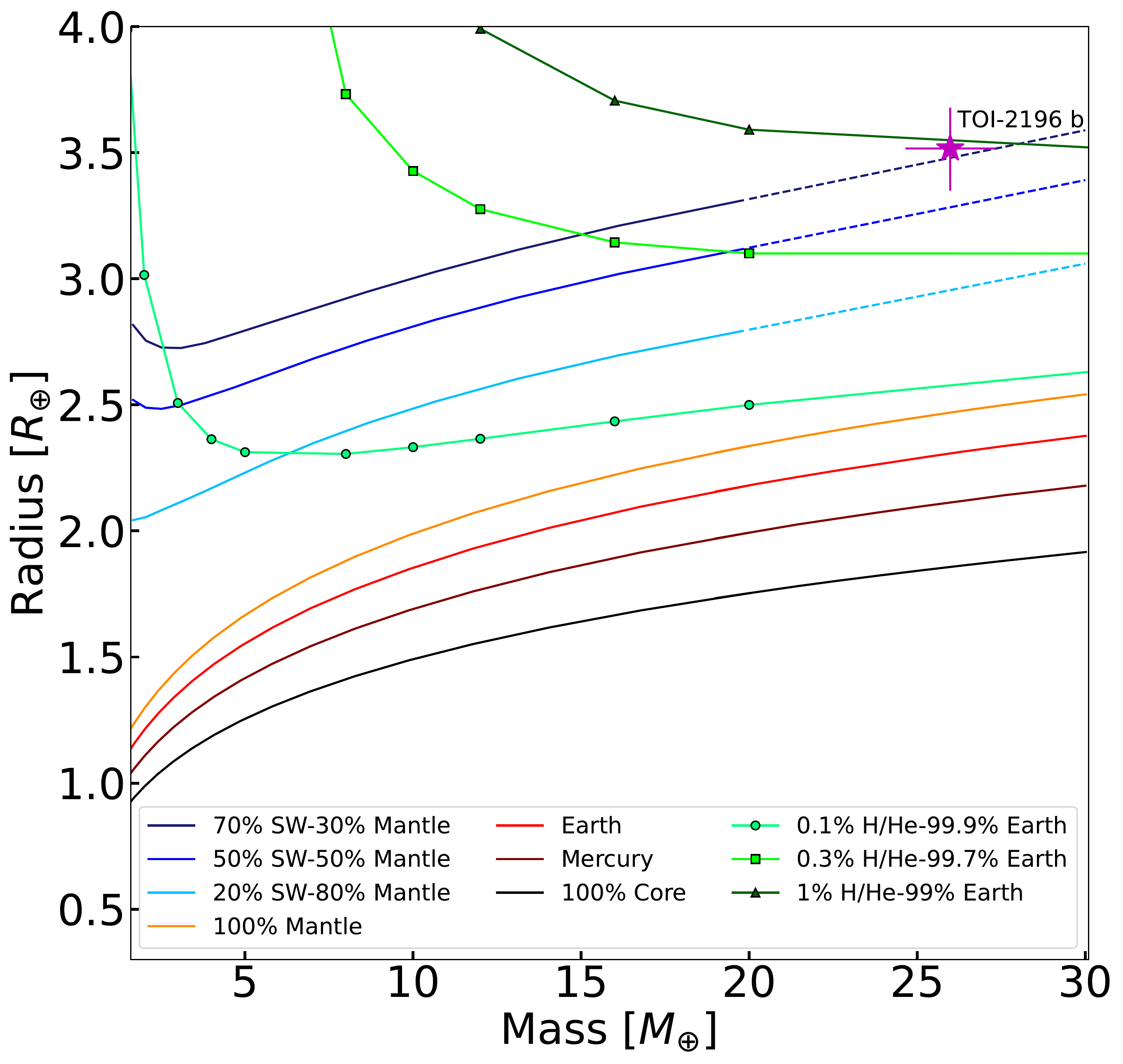}}
\caption{Position of \targetb~in the mass-radius diagram. We also show
mass-radius relationships for dry planets with different core mass fractions
\citep{Brugger16,Brugger17}, water planets with varying supercritical water
(SW) contents \citep{Mousis20,Acuna21} and Earth-like cores with different
H/He mass fractions \citep{2019PNAS..116.9723Z}. 
We assume equilibrium temperatures of 1200~K and 2000~K for   water
planets and rocky   planets with H/He atmospheres, respectively. 
Dashed lines indicate an
extrapolation of the SW relations beyond their  valid mass ranges. 
The lines for Earth, Mercury, and 100~\% core  are lines of constant composition of respective type of planet.}
\label{fig:mr_diag}
\end{figure}

If the composition of \targetb~would partly include  volatiles such as H/He or water, these elements 
would be in high-pressure and high-temperature phases due to the thickness of the volatile envelope.
In Fig.~\ref{fig:mr_diag}, we compare  \targetb~with mass-radius relationships of 
planets with several different amounts of water in a supercritical state calculated with the model 
presented in \citet{Mousis20} and \citet{Acuna21}, and with those of rocky planets with gaseous H/He atmospheres   \citep{2019PNAS..116.9723Z}. 
\targetb~lies
slightly above the 70\% supercritical water (SW) composition under the assumption
of a 100\% mantle bulk. If we were to assume a core mass fraction (CMF) in
agreement with its host stellar abundances (Table~\ref{tab:interior_table}), 
the water content would increase
since we would have to fit a similar total density with a more dense core.
A water mass fraction (WMF) of $\simeq$ 70\% is similar to the maximum
water content
found in Solar System bodies \citep{mckay19}. Thus, the atmosphere of  
\targetb~is likely to be dominated by H/He which are less dense than
water; otherwise, the
water content of \targetb~would be unrealistically high.

\begin{table}[H]
\centering
\caption{MCMC parameters of the interior structure analysis and their 1~$\sigma$ confidence intervals for 
scenario 1 (planet mass and radius as input) and scenario 2 (Fe/Si mole ratio in addition to mass and radius as input).}
\label{tab:interior_table}
\begin{tabular}{lcc}
\hline \hline \noalign{\smallskip}
Parameter  & Scenario 1    & Scenario 2     \\ 
\noalign{\smallskip}
\hline 
\noalign{\smallskip}
CMF\tablefootmark{a}        & $\mathcal{U}$(0,1)   & $0.244^{+0.057}_{-0.049}$ \\ \noalign{\smallskip}
CRF\tablefootmark{b}   & $0.45^{+0.05}_{-0.14}$ & $0.31^{+0.03}_{-0.02}$  \\ \noalign{\smallskip}
x$_{H/He}$\tablefootmark{c} & 0.0077$^{+0.0094}_{-0.0032}$ & 0.0066$^{+0.0031}_{-0.0020}$  \\ \noalign{\smallskip}
$M$ [$M_{\oplus}$]         & 26.031$\pm$1.389 & 25.998$^{+1.410}_{-1.360}$  \\ \noalign{\smallskip}
$R$ [$R_{\oplus}$]        & 3.517$\pm$0.170 & 3.470$^{+0.202}_{-0.127}$  \\ \noalign{\smallskip}
Fe/Si      & 6.261$^{+11.538}_{-6.261}$ & 0.747$\pm$0.175              \\ \noalign{\smallskip}
$z_{atm}$\tablefootmark{d} [km] & 8143$^{+1662}_{-1289}$ & 7185$^{+1298}_{-840}$ \\ 
\noalign{\smallskip}  
\hline 
\end{tabular}
\tablefoot{
\tablefoottext{a}{Core mass fraction.}    
\tablefoottext{b}{Core  radius fraction simultaneously computed with the CMF.} 
\tablefoottext{c}{Hydrogen and helium mass fraction.} 
\tablefoottext{d}{Planet atmosphere thickness.} 
}
\end{table}

We therefore performed a Markov chain Monte Carlo (MCMC)
Bayesian analysis \citep{Acuna21,Director17} of the internal composition of
\targetb~assuming a H/He atmosphere. Following  \citet{Brugger16} and  \citet{Brugger17}, 
we included two layers: an Fe-rich core and a silicate-rich mantle. 
To the calculated radius of the interior, we added the thickness of the
H/He atmosphere, $z_{atm}$, to estimate the total planetary radius. We obtained
the atmospheric thickness by subtracting the radius of a bare Earth-like core from
the mass-radius relationships of an Earth-like core with a H/He atmosphere,
presented by \cite{2019PNAS..116.9723Z}. 
The atmospheric thickness is then  a function of
the surface gravity, \mbox{$g_{0} = GM/R^{2}$}, where G is the gravitational constant, and the \mbox{H/He} mass fraction, 
\mbox{$z_{atm} = z_{atm}(g_{0},$x$_{H/He})$}. We set the surface conditions for the
interior model as 2000 K and 1 bar, since \cite{2019PNAS..116.9723Z} considered an 
isothermal temperature profile for their atmosphere. Changing our surface
pressure to other values would have a trivial effect on our interior bulk radius,
since this parameter does not have an influence on the total radius of solid
mantle-core planets \citep{Otegi20}.

We considered two scenarios to obtain the
interior structure of \targetb. Scenario 1   uses the  planet mass and radius
 as input to the MCMC analysis, while   scenario 2
also uses the stellar \mbox{Fe/Si} and \mbox{Mg/Si} mole
ratios  (Table~\ref{Table: stellar spectroscopic parameters}) as input. 
To compute the Fe/Si and Mg/Si mole ratios with the stellar
abundances, we follow the approach depicted in \citet{Brugger17} and
\citet{sotin07}  and obtain \mbox{Fe/Si = 0.775$\pm$0.170} and
\mbox{Mg/Si = 1.192$\pm$0.337}. 
The MCMC analysis yields posterior distribution functions (PDF) of the core mass fraction 
and the \mbox{H/He} mass fraction.

\begin{figure}
\centering
\resizebox{\hsize}{!}
{\includegraphics[scale=0.5]{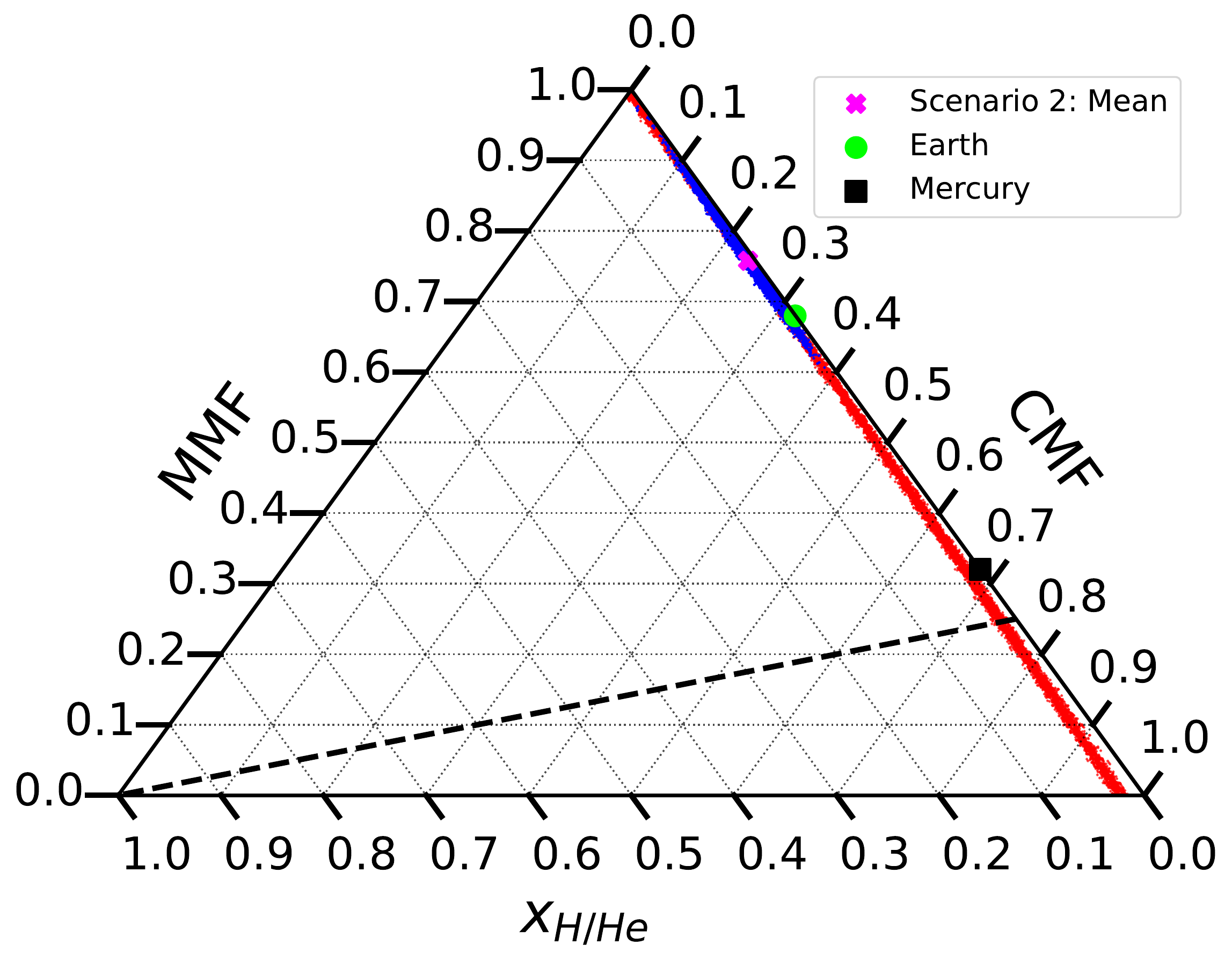}}
\caption{Ternary diagram of the MCMC models for TOI-2196 b in scenario 1 (red), with the planet mass and the radius  as input to the model, 
and in scenario 2 (blue), where also the stellar abundances are  included. 
The mantle mass fraction (MMF) is defined as MMF = 1 - CMF - x$_{H/He}$ where 
the latter parameter is the H/He mass fraction and CMF is the core mass fraction. The green 
circle and black square indicate the position of Earth and Mercury in the ternary diagram, respectively. 
The maximum CMF constrained by planet formation is limited by the black dashed line (see Sect.~\ref{Section: Internal structure}).}
\label{fig:ternary_diag}
\end{figure}

The CMF and the H/He mass fraction, $x_{H/He}$, represent our free parameters in the MCMC analysis. 
The CMF in scenario~1 is sampled as a uniform distribution between 0 and 1  which are the 
minimum and maximum values of the compositional parameters. In scenario~2, 
the CMF is constrained by the inclusion of the Fe/Si mole ratio as an input to the MCMC framework. 
This results in a mean value of the CMF  of $0.244^{+0.057}_{-0.049}$ (Table \ref{tab:interior_table}) 
which is slightly lower than the CMF of Earth (0.32). 
The Earth value  is, however, approximately at the limit of its 1~$\sigma$ confidence interval 
as seen in Fig.~\ref{fig:ternary_diag}. These results from scenario~2 are expected since the Fe/Si mole ratio of 
\targetb~is lower than the solar value (0.96). The  resulting mean value of the H/He mass fraction, 
x$_{H/He}$, is approximately 0.7~\% in both scenarios. In the case of scenario 2, the uncertainties 
are lower than in scenario~1. This is due to the use of the Fe/Si mole ratio, which breaks the degeneracy 
between the CMF and the H/He mass fraction in scenario 2. Both scenarios agree that the minimum 
volatile mass fraction is $\simeq$ 0.4~\%. In addition, we can consider scenario 1 as the most conservative 
one to derive the maximum volatile mass fraction for \targetb. The maximum CMF estimated 
from the protosolar nebula composition \citep{lodders09} is 0.75. We mark this limit as a black 
dashed line in the ternary diagram in Fig.~\ref{fig:ternary_diag}. We estimated the maximum volatile 
mass fraction as the x$_{H/He}$ at which this line crosses the red points that correspond to the 
MCMC simulations in scenario~1, which is approximately 3~\%.

In Sect.~\ref{Section: atmospheric loss}, we show that the past volatile mass  fraction of \targetb~was 
approximately $4-6$~\%, with an accompanying increase in mass of 0.8~\mearth. The radius for a planet of such  
mass and volatile mass fraction would be around 6~\rearth,~according to 
the mass-radius relation of 5~\% H/He from \citet{2019PNAS..116.9723Z}, which changes the density 
from 3.3~\gc~to 0.7~\gc. This shifts the position of \targetb~in 
the density diagram  in Fig.~\ref{Figure: Density-Teq} downwards to the Saturn density with the
difference of having a much lower mass than the gas giants in this parameter space. 
If we place the planet with the past radius and mass in the mass-radius 
diagram, it would be slightly above the small group of  planets that 
includes LTT 9779~b, K2-100~b, and TOI-849~b, but still located in the hot Neptune desert and  
would still be classified as a planet in the Neptune regime, not a  gaseous giant.

\section{Conclusions}\label{Section: Conclusions}
We present the detection and the analysis of the hot and volatile rich planet    \targetb. It is smaller   than  Neptune   
but  50~\% more massive resulting in a high bulk density for this type of planet. 
Another interesting result is the presence of a longer period body in this system detected in the radial velocity 
measurements as a linear trend with a minimum mass of $\sim$0.65~\mjup~, assuming zero eccentricity. 
The outer body may be a warm or cold gas-giant planet, although a brown dwarf, or a very low-mass stellar companion, cannot be fully excluded at the present stage. 
A future long-term RV monitoring of the star is needed to   determine the true nature of the signal. 
We estimate  the mass loss of  volatiles  for  planet~b at a young age  and find that while the 
mass loss could have been significant,  the planet has not changed in terms of its   character. 
It formed as a small volatile-rich planet and has remained one until today.  
The high equilibrium temperature of 1860~K, 
together with its radius and mass,  places \targetb~in the hot Neptune desert as a member of  a very small population found so far. 
This small population suggests that the   desert   is divided in two parts:  a hot sub-Neptune desert  
 ($R_\mathrm{p} \approx 1.8 -3$~\rearth) and a sub-Jovian desert  ($R_\mathrm{p} \approx 5-12$~\rearth). However, more planets are needed for further studies for this special region.


 \begin{acknowledgements}
 
 This paper includes data collected by the \textit{TESS} mission. Funding for the \textit{TESS} mission is provided by the NASA Explorer Program. We acknowledge the use of public TESS data from pipelines at the \textit{TESS} Science Office and at the \textit{TESS} Science Processing Operations Center.  Resources supporting this work were provided by the NASA High-End Computing (HEC) Program through the NASA Advanced Supercomputing (NAS) Division at Ames Research Center for the production of the SPOC data products. 
This work makes use of observations from the LCOGT network. Part of the LCOGT telescope time was granted by NOIRLab through the Mid-Scale Innovations Program (MSIP). 
This work uses observations made with ESO 3.6-m telescope at La Silla Observatory under programme ID 106.21TJ.001. 
We are grateful to the ESO staff members for their support during the observations, and to François Bouchy and Xavier Dumusque for coordinating the HARPS time sharing agreement. 
This work has made use of SME package, which benefits from the continuing development work by J. Valenti and N. Piskunov 
and we gratefully acknowledge their continued support. 
\citep{Kupka2000, Ryabchikova2015}.  
C.M.P., M.F., I.G., and J.K. gratefully acknowledge the support of the  Swedish National Space Agency (DNR 65/19, 174/18, 177/19, 2020-00104).  
L.M.S and D.G. gratefully acknowledge financial support from the CRT foundation under Grant No. 2018.2323 ``Gaseous or rocky? Unveiling the nature of small worlds''. 
P.K. acknowledges support from grant LTT-20015.
E.G. acknowledge the support of the  Th\"uringer Ministerium f\"ur Wirtschaft, Wissenschaft und Digitale Gesellschaft.  
J.S.J. greatfully acknowledges support by FONDECYT grant 1201371 and from the ANID BASAL projects ACE210002 and FB210003. H.J.D.  acknowledges support from the Spanish Research Agency of the Ministry of Science and Innovation (AEI-MICINN) under grant PID2019-107061GB-C66, DOI: 10.13039/501100011033. 
D.D. acknowledges support from the \textit{TESS} Guest Investigator Program grants 80NSSC21K0108 and 80NSSC22K0185.
M.E. acknowledges the support of the DFG priority program
SPP 1992 "Exploring the Diversity of Extrasolar Planets" (HA 3279/12-1). K.W.F.L. was supported by Deutsche Forschungsgemeinschaft grants RA714/14-1 within the DFG Schwerpunkt SPP 1992, Exploring the Diversity of Extrasolar Planets. 
N.N. acknowledges support from JSPS KAKENHI Grant Number JP18H05439, JST CREST Grant Number JPMJCR1761.
M.S.I.P. is funded by NSF.  
  \end{acknowledgements}

\bibliographystyle{aa}
\bibliography{references}

\begin{thebibliography}{108}
\expandafter\ifx\csname natexlab\endcsname\relax\def\natexlab#1{#1}\fi

\bibitem[{{Acu\~na} {et~al.}(2021){Acu\~na}, {Deleuil, Magali}, {Mousis,
  Olivier}, {Marcq, Emmanuel}, {Levesque, Ma\"eva}, \& {Aguichine,
  Artyom}}]{Acuna21}
{Acu\~na}, L., {Deleuil, Magali}, {Mousis, Olivier}, {et~al.} 2021, A\&A, 647,
  A53

\bibitem[{{Adams} \& {Laughlin}(2006)}]{2006ApJ...649.1004A}
{Adams}, F.~C. \& {Laughlin}, G. 2006, \apj, 649, 1004

\bibitem[{{Aguichine} {et~al.}(2021){Aguichine}, {Mousis}, {Deleuil}, \&
  {Marcq}}]{Agui2021}
{Aguichine}, A., {Mousis}, O., {Deleuil}, M., \& {Marcq}, E. 2021, \apj, 914,
  84

\bibitem[{{Aguirre B{\o}rsen-Koch} {et~al.}(2022){Aguirre B{\o}rsen-Koch},
  {R{\o}rsted}, {Justesen}, {Stokholm}, {Verma}, {Winther}, {Knudstrup},
  {Nielsen}, {Sahlholdt}, {Larsen}, {Cassisi}, {Serenelli}, {Casagrande},
  {Christensen-Dalsgaard}, {Davies}, {Ferguson}, {Lund}, {Weiss}, \&
  {White}}]{2022MNRAS.509.4344A}
{Aguirre B{\o}rsen-Koch}, V., {R{\o}rsted}, J.~L., {Justesen}, A.~B., {et~al.}
  2022, \mnras, 509, 4344

\bibitem[{{Allard} {et~al.}(2012){Allard}, {Homeier}, \&
  {Freytag}}]{2012RSPTA.370.2765A}
{Allard}, F., {Homeier}, D., \& {Freytag}, B. 2012, Philosophical Transactions
  of the Royal Society of London Series A, 370, 2765

\bibitem[{{Anglada-Escud{\'e}} \& {Butler}(2012)}]{2012ApJS..200...15A}
{Anglada-Escud{\'e}}, G. \& {Butler}, R.~P. 2012, \apjs, 200, 15

\bibitem[{{Armstrong} {et~al.}(2020){Armstrong}, {Lopez}, {Adibekyan}, {Booth},
  {Bryant}, {Collins}, {Deleuil}, {Emsenhuber}, {Huang}, {King}, {Lillo-Box},
  {Lissauer}, {Matthews}, {Mousis}, {Nielsen}, {Osborn}, {Otegi}, {Santos},
  {Sousa}, {Stassun}, {Veras}, {Ziegler}, {Acton}, {Almenara}, {Anderson},
  {Barrado}, {Barros}, {Bayliss}, {Belardi}, {Bouchy}, {Brice{\~n}o}, {Brogi},
  {Brown}, {Burleigh}, {Casewell}, {Chaushev}, {Ciardi}, {Collins},
  {Col{\'o}n}, {Cooke}, {Crossfield}, {D{\'\i}az}, {Delgado Mena}, {Demangeon},
  {Dorn}, {Dumusque}, {Eigm{\"u}ller}, {Fausnaugh}, {Figueira}, {Gan},
  {Gandhi}, {Gill}, {Gonzales}, {Goad}, {G{\"u}nther}, {Helled}, {Hojjatpanah},
  {Howell}, {Jackman}, {Jenkins}, {Jenkins}, {Jensen}, {Kennedy}, {Latham},
  {Law}, {Lendl}, {Lozovsky}, {Mann}, {Moyano}, {McCormac}, {Meru},
  {Mordasini}, {Osborn}, {Pollacco}, {Queloz}, {Raynard}, {Ricker}, {Rowden},
  {Santerne}, {Schlieder}, {Seager}, {Sha}, {Tan}, {Tilbrook}, {Ting}, {Udry},
  {Vanderspek}, {Watson}, {West}, {Wilson}, {Winn}, {Wheatley}, {Villasenor},
  {Vines}, \& {Zhan}}]{2020Natur.583...39A}
{Armstrong}, D.~J., {Lopez}, T.~A., {Adibekyan}, V., {et~al.} 2020, \nat, 583,
  39

\bibitem[{{Baranne} {et~al.}(1996){Baranne}, {Queloz}, {Mayor}, {Adrianzyk},
  {Knispel}, {Kohler}, {Lacroix}, {Meunier}, {Rimbaud}, \& {Vin}}]{Baranne1996}
{Baranne}, A., {Queloz}, D., {Mayor}, M., {et~al.} 1996, \aaps, 119, 373

\bibitem[{{Barrag{\'a}n} {et~al.}(2019{\natexlab{a}}){Barrag{\'a}n}, {Aigrain},
  {Kubyshkina}, {Gandolfi}, {Livingston}, {Fridlund}, {Fossati}, {Korth},
  {Parviainen}, {Malavolta}, {Palle}, {Deeg}, {Nowak}, {Rajpaul}, {Zicher},
  {Antoniciello}, {Narita}, {Albrecht}, {Bedin}, {Cabrera}, {Cochran}, {de
  Leon}, {Eigm{\"u}ller}, {Fukui}, {Granata}, {Grziwa}, {Guenther}, {Hatzes},
  {Kusakabe}, {Latham}, {Libralato}, {Luque},
  {Monta{\~n}{\'e}s-Rodr{\'\i}guez}, {Murgas}, {Nardiello}, {Pagano}, {Piotto},
  {Persson}, {Redfield}, \& {Tamura}}]{2019MNRAS.490..698B}
{Barrag{\'a}n}, O., {Aigrain}, S., {Kubyshkina}, D., {et~al.}
  2019{\natexlab{a}}, \mnras, 490, 698

\bibitem[{{Barrag{\'a}n} {et~al.}(2022{\natexlab{a}}){Barrag{\'a}n}, {Aigrain},
  {Rajpaul}, \& {Zicher}}]{2022MNRAS.509..866B}
{Barrag{\'a}n}, O., {Aigrain}, S., {Rajpaul}, V.~M., \& {Zicher}, N.
  2022{\natexlab{a}}, \mnras, 509, 866

\bibitem[{{Barrag{\'a}n} {et~al.}(2022{\natexlab{b}}){Barrag{\'a}n},
  {Armstrong}, {Gandolfi}, {Carleo}, {Vidotto}, {Villarreal D'Angelo},
  {Oklop{\v{c}}i{\'c}}, {Isaacson}, {Oddo}, {Collins}, {Fridlund}, {Sousa},
  {Persson}, {Hellier}, {Howell}, {Howard}, {Redfield}, {Eisner}, {Georgieva},
  {Dragomir}, {Bayliss}, {Nielsen}, {Klein}, {Aigrain}, {Zhang}, {Teske},
  {Twicken}, {Jenkins}, {Esposito}, {Van Eylen}, {Rodler}, {Adibekyan},
  {Alarcon}, {Anderson}, {Murphy}, {Barrado}, {Barros}, {Benneke}, {Bouchy},
  {Bryant}, {Butler}, {Burt}, {Cabrera}, {Casewell}, {Chaturvedi}, {Cloutier},
  {Cochran}, {Crane}, {Crossfield}, {Crouzet}, {Collins}, {Dai}, {Deeg},
  {Deline}, {Demangeon}, {Dumusque}, {Figueira}, {Furlan}, {Gnilka}, {Goad},
  {Goffo}, {Guti{\'e}rrez-Canales}, {Hadjigeorghiou}, {Hartman}, {Hatzes},
  {Harris}, {Henderson}, {Hirano}, {Hojjatpanah}, {Hoyer}, {Kab{\'a}th},
  {Korth}, {Lillo-Box}, {Luque}, {Marmier}, {Mo{\v{c}}nik}, {Muresan},
  {Murgas}, {Nagel}, {Osborne}, {Osborn}, {Osborn}, {Palle}, {Raimbault},
  {Ricker}, {Rubenzahl}, {Stockdale}, {Santos}, {Scott}, {Schwarz}, {Shectman},
  {Raimbault}, {Seager}, {S{\'e}gransan}, {Serrano}, {Skarka}, {Smith},
  {{\v{S}}ubjak}, {Tan}, {Udry}, {Watson}, {Wheatley}, {West}, {Winn}, {Wang},
  {Wolfgang}, \& {Ziegler}}]{2022MNRAS.tmp..699B}
{Barrag{\'a}n}, O., {Armstrong}, D.~J., {Gandolfi}, D., {et~al.}
  2022{\natexlab{b}}, \mnras [\eprint[arXiv]{2110.13069}]

\bibitem[{{Barrag{\'a}n} {et~al.}(2019{\natexlab{b}}){Barrag{\'a}n},
  {Gandolfi}, \& {Antoniciello}}]{2019MNRAS.482.1017B}
{Barrag{\'a}n}, O., {Gandolfi}, D., \& {Antoniciello}, G. 2019{\natexlab{b}},
  \mnras, 482, 1017

\bibitem[{{Ben{\'\i}tez-Llambay} {et~al.}(2011){Ben{\'\i}tez-Llambay},
  {Masset}, \& {Beaug{\'e}}}]{2011A&A...528A...2B}
{Ben{\'\i}tez-Llambay}, P., {Masset}, F., \& {Beaug{\'e}}, C. 2011, \aap, 528,
  A2

\bibitem[{{Berger} {et~al.}(2018){Berger}, {Huber}, {Gaidos}, \& {van
  Saders}}]{2018ApJ...866...99B}
{Berger}, T.~A., {Huber}, D., {Gaidos}, E., \& {van Saders}, J.~L. 2018, \apj,
  866, 99

\bibitem[{{Bolmont} {et~al.}(2017){Bolmont}, {Selsis}, {Owen}, {Ribas},
  {Raymond}, {Leconte}, \& {Gillon}}]{2017MNRAS.464.3728B}
{Bolmont}, E., {Selsis}, F., {Owen}, J.~E., {et~al.} 2017, \mnras, 464, 3728

\bibitem[{{Borucki} {et~al.}(2010){Borucki}, {Koch}, {Basri}, {Batalha},
  {Brown}, {Caldwell}, {Caldwell}, {Christensen-Dalsgaard}, {Cochran},
  {DeVore}, {Dunham}, {Dupree}, {Gautier}, {Geary}, {Gilliland}, {Gould},
  {Howell}, {Jenkins}, {Kondo}, {Latham}, {Marcy}, {Meibom}, {Kjeldsen},
  {Lissauer}, {Monet}, {Morrison}, {Sasselov}, {Tarter}, {Boss}, {Brownlee},
  {Owen}, {Buzasi}, {Charbonneau}, {Doyle}, {Fortney}, {Ford}, {Holman},
  {Seager}, {Steffen}, {Welsh}, {Rowe}, {Anderson}, {Buchhave}, {Ciardi},
  {Walkowicz}, {Sherry}, {Horch}, {Isaacson}, {Everett}, {Fischer}, {Torres},
  {Johnson}, {Endl}, {MacQueen}, {Bryson}, {Dotson}, {Haas}, {Kolodziejczak},
  {Van Cleve}, {Chandrasekaran}, {Twicken}, {Quintana}, {Clarke}, {Allen},
  {Li}, {Wu}, {Tenenbaum}, {Verner}, {Bruhweiler}, {Barnes}, \&
  {Prsa}}]{2010Sci...327..977B}
{Borucki}, W.~J., {Koch}, D., {Basri}, G., {et~al.} 2010, Science, 327, 977

\bibitem[{{Bowler}(2016)}]{2016PASP..128j2001B}
{Bowler}, B.~P. 2016, \pasp, 128, 102001

\bibitem[{{Brown} {et~al.}(2013){Brown}, {Baliber}, {Bianco}, {Bowman},
  {Burleson}, {Conway}, {Crellin}, {Depagne}, {De Vera}, {Dilday}, {Dragomir},
  {Dubberley}, {Eastman}, {Elphick}, {Falarski}, {Foale}, {Ford}, {Fulton},
  {Garza}, {Gomez}, {Graham}, {Greene}, {Haldeman}, {Hawkins}, {Haworth},
  {Haynes}, {Hidas}, {Hjelstrom}, {Howell}, {Hygelund}, {Lister}, {Lobdill},
  {Martinez}, {Mullins}, {Norbury}, {Parrent}, {Paulson}, {Petry}, {Pickles},
  {Posner}, {Rosing}, {Ross}, {Sand}, {Saunders}, {Shobbrook}, {Shporer},
  {Street}, {Thomas}, {Tsapras}, {Tufts}, {Valenti}, {Vander Horst}, {Walker},
  {White}, \& {Willis}}]{Brown:2013}
{Brown}, T.~M., {Baliber}, N., {Bianco}, F.~B., {et~al.} 2013, Publications of
  the Astronomical Society of the Pacific, 125, 1031

\bibitem[{{Brugger} {et~al.}(2017){Brugger}, {Mousis}, {Deleuil}, \&
  {Deschamps}}]{Brugger17}
{Brugger}, B., {Mousis}, O., {Deleuil}, M., \& {Deschamps}, F. 2017, The
  Astrophysical Journal, 850, 93

\bibitem[{Brugger {et~al.}(2016)Brugger, Mousis, Deleuil, \&
  Lunine}]{Brugger16}
Brugger, B., Mousis, O., Deleuil, M., \& Lunine, J.~I. 2016, The Astrophysical
  Journal, 831, L16

\bibitem[{{Bruntt} {et~al.}(2008){Bruntt}, {De Cat}, \& {Aerts}}]{bruntt08}
{Bruntt}, H., {De Cat}, P., \& {Aerts}, C. 2008, \aap, 478, 487

\bibitem[{{Cabrera} {et~al.}(2012){Cabrera}, {Csizmadia}, {Erikson}, {Rauer},
  \& {Kirste}}]{Cabrera2012}
{Cabrera}, J., {Csizmadia}, S., {Erikson}, A., {Rauer}, H., \& {Kirste}, S.
  2012, \aap, 548, A44

\bibitem[{{Castelli} \& {Kurucz}(2004)}]{Castelli2004}
{Castelli}, F. \& {Kurucz}, R.~L. 2004, astro-ph/0405087
  [\eprint{astro-ph/0405087}]

\bibitem[{{Charbonneau} {et~al.}(2005){Charbonneau}, {Allen}, {Megeath},
  {Torres}, {Alonso}, {Brown}, {Gilliland}, {Latham}, {Mandushev}, {O'Donovan},
  \& {Sozzetti}}]{2005ApJ...626..523C}
{Charbonneau}, D., {Allen}, L.~E., {Megeath}, S.~T., {et~al.} 2005, \apj, 626,
  523

\bibitem[{{Chen} \& {Rogers}(2016)}]{2016ApJ...831..180C}
{Chen}, H. \& {Rogers}, L.~A. 2016, \apj, 831, 180

\bibitem[{{Choi} {et~al.}(2016){Choi}, {Dotter}, {Conroy}, {Cantiello},
  {Paxton}, \& {Johnson}}]{2016ApJ...823..102C}
{Choi}, J., {Dotter}, A., {Conroy}, C., {et~al.} 2016, \apj, 823, 102

\bibitem[{{Claret}(2017)}]{2017A&A...600A..30C}
{Claret}, A. 2017, \aap, 600, A30

\bibitem[{{Claret} {et~al.}(2013){Claret}, {Hauschildt}, \&
  {Witte}}]{2013A&A...552A..16C}
{Claret}, A., {Hauschildt}, P.~H., \& {Witte}, S. 2013, \aap, 552, A16

\bibitem[{{Collins}(2019)}]{collins:2019}
{Collins}, K. 2019, in American Astronomical Society Meeting Abstracts, Vol.
  233, American Astronomical Society Meeting Abstracts \#233, 140.05

\bibitem[{{Collins} {et~al.}(2017){Collins}, {Kielkopf}, {Stassun}, \&
  {Hessman}}]{Collins:2017}
{Collins}, K.~A., {Kielkopf}, J.~F., {Stassun}, K.~G., \& {Hessman}, F.~V.
  2017, \aj, 153, 77

\bibitem[{{da Silva} {et~al.}(2006){da Silva}, {Girardi}, {Pasquini},
  {Setiawan}, {von der L{\"u}he}, {de Medeiros}, {Hatzes}, {D{\"o}llinger}, \&
  {Weiss}}]{daSilva2006}
{da Silva}, L., {Girardi}, L., {Pasquini}, L., {et~al.} 2006, \aap, 458, 609

\bibitem[{Director {et~al.}(2017)Director, Gattiker, Lawrence, \&
  Wiel}]{Director17}
Director, H.~M., Gattiker, J., Lawrence, E., \& Wiel, S.~V. 2017, Journal of
  Statistical Computation and Simulation, 87, 3521

\bibitem[{{Doyle} {et~al.}(2014){Doyle}, {Davies}, {Smalley}, {Chaplin}, \&
  {Elsworth}}]{Doyle2014}
{Doyle}, A.~P., {Davies}, G.~R., {Smalley}, B., {Chaplin}, W.~J., \&
  {Elsworth}, Y. 2014, \mnras, 444, 3592

\bibitem[{{Erkaev} {et~al.}(2007){Erkaev}, {Kulikov}, {Lammer}, {Selsis},
  {Langmayr}, {Jaritz}, \& {Biernat}}]{Er07}
{Erkaev}, N.~V., {Kulikov}, Y.~N., {Lammer}, H., {et~al.} 2007, \aap, 472, 329

\bibitem[{{Estrela} {et~al.}(2020){Estrela}, {Swain}, {Gupta}, {Sotin}, \&
  {Valio}}]{Estrela2020}
{Estrela}, R., {Swain}, M.~R., {Gupta}, A., {Sotin}, C., \& {Valio}, A. 2020,
  \apj, 898, 104

\bibitem[{{Findeisen} {et~al.}(2011){Findeisen}, {Hillenbrand}, \&
  {Soderblom}}]{2011AJ....142...23F}
{Findeisen}, K., {Hillenbrand}, L., \& {Soderblom}, D. 2011, \aj, 142, 23

\bibitem[{Foreman-Mackey {et~al.}(2014)Foreman-Mackey, Hoyer, Bernhard, \&
  Angus}]{Mackey2014}
Foreman-Mackey, D., Hoyer, S., Bernhard, J., \& Angus, R. 2014, george: George
  (v0.2.0)

\bibitem[{{Fossati} {et~al.}(2017){Fossati}, {Erkaev}, {Lammer}, {Cubillos},
  {Odert}, {Juvan}, {Kislyakova}, {Lendl}, {Kubyshkina}, \&
  {Bauer}}]{2017A&A...598A..90F}
{Fossati}, L., {Erkaev}, N.~V., {Lammer}, H., {et~al.} 2017, \aap, 598, A90

\bibitem[{{Fridlund} {et~al.}(2017){Fridlund}, {Gaidos}, {Barrag{\'a}n},
  {Persson}, {Gandolfi}, {Cabrera}, {Hirano}, {Kuzuhara}, {Csizmadia}, {Nowak},
  {Endl}, {Grziwa}, {Korth}, {Pfaff}, {Bitsch}, {Johansen}, {Mustill},
  {Davies}, {Deeg}, {Palle}, {Cochran}, {Eigm{\"u}ller}, {Erikson}, {Guenther},
  {Hatzes}, {Kiilerich}, {Kudo}, {MacQueen}, {Narita}, {Nespral},
  {P{\"a}tzold}, {Prieto-Arranz}, {Rauer}, \& {Van
  Eylen}}]{2017A&A...604A..16F}
{Fridlund}, M., {Gaidos}, E., {Barrag{\'a}n}, O., {et~al.} 2017, \aap, 604, A16

\bibitem[{{Fulton} {et~al.}(2017){Fulton}, {Petigura}, {Howard}, {Isaacson},
  {Marcy}, {Cargile}, {Hebb}, {Weiss}, {Johnson}, {Morton}, {Sinukoff},
  {Crossfield}, \& {Hirsch}}]{2017AJ....154..109F}
{Fulton}, B.~J., {Petigura}, E.~A., {Howard}, A.~W., {et~al.} 2017, \aj, 154,
  109

\bibitem[{{Georgieva} {et~al.}(2021){Georgieva}, {Persson}, {Barrag{\'a}n},
  {Nowak}, {Fridlund}, {Locci}, {Palle}, {Luque}, {Carleo}, {Gandolfi}, {Kane},
  {Korth}, {Stassun}, {Livingston}, {Matthews}, {Collins}, {Howell}, {Serrano},
  {Albrecht}, {Bieryla}, {Brasseur}, {Ciardi}, {Cochran}, {Colon},
  {Crossfield}, {Csizmadia}, {Deeg}, {Esposito}, {Furlan}, {Gan}, {Goffo},
  {Gonzales}, {Grziwa}, {Guenther}, {Guerra}, {Hirano}, {Jenkins}, {Jensen},
  {Kab{\'a}th}, {Knudstrup}, {Lam}, {Latham}, {Levine}, {Matson}, {McDermott},
  {Osborne}, {Paegert}, {Quinn}, {Redfield}, {Ricker}, {Schlieder}, {Scott},
  {Seager}, {Smith}, {Tenenbaum}, {Twicken}, {Vanderspek}, {Van Eylen}, \&
  {Winn}}]{2021MNRAS.505.4684G}
{Georgieva}, I.~Y., {Persson}, C.~M., {Barrag{\'a}n}, O., {et~al.} 2021,
  \mnras, 505, 4684

\bibitem[{{Ginzburg} {et~al.}(2018){Ginzburg}, {Schlichting}, \&
  {Sari}}]{2018MNRAS.476..759G}
{Ginzburg}, S., {Schlichting}, H.~E., \& {Sari}, R. 2018, \mnras, 476, 759

\bibitem[{{Gregory}(2005)}]{2005ApJ...631.1198G}
{Gregory}, P.~C. 2005, \apj, 631, 1198

\bibitem[{{Gupta} \& {Schlichting}(2019)}]{2019MNRAS.487...24G}
{Gupta}, A. \& {Schlichting}, H.~E. 2019, \mnras, 487, 24

\bibitem[{{Hidalgo} {et~al.}(2018){Hidalgo}, {Pietrinferni}, {Cassisi},
  {Salaris}, {Mucciarelli}, {Savino}, {Aparicio}, {Silva Aguirre}, \&
  {Verma}}]{2018ApJ...856..125H}
{Hidalgo}, S.~L., {Pietrinferni}, A., {Cassisi}, S., {et~al.} 2018, \apj, 856,
  125

\bibitem[{{Hu} {et~al.}(2015){Hu}, {Seager}, \& {Yung}}]{2015ApJ...807....8H}
{Hu}, R., {Seager}, S., \& {Yung}, Y.~L. 2015, \apj, 807, 8

\bibitem[{{Husser} {et~al.}(2013){Husser}, {Wende-von Berg}, {Dreizler},
  {Homeier}, {Reiners}, {Barman}, \& {Hauschildt}}]{2013A&A...553A...6H}
{Husser}, T.~O., {Wende-von Berg}, S., {Dreizler}, S., {et~al.} 2013, \aap,
  553, A6

\bibitem[{{Ito} \& {Ikoma}(2021)}]{2021MNRAS.502..750I}
{Ito}, Y. \& {Ikoma}, M. 2021, \mnras, 502, 750

\bibitem[{{Jenkins}(2002)}]{2002ApJ...575..493J}
{Jenkins}, J.~M. 2002, \apj, 575, 493

\bibitem[{{Jenkins} {et~al.}(2010){Jenkins}, {Chandrasekaran}, {McCauliff},
  {Caldwell}, {Tenenbaum}, {Li}, {Klaus}, {Cote}, \&
  {Middour}}]{2010SPIE.7740E..0DJ}
{Jenkins}, J.~M., {Chandrasekaran}, H., {McCauliff}, S.~D., {et~al.} 2010, in
  Society of Photo-Optical Instrumentation Engineers (SPIE) Conference Series,
  Vol. 7740, Software and Cyberinfrastructure for Astronomy, ed. N.~M.
  {Radziwill} \& A.~{Bridger}, 77400D

\bibitem[{{Jenkins} {et~al.}(2020{\natexlab{a}}){Jenkins}, {Tenenbaum},
  {Seader}, {Burke}, {McCauliff}, {Smith}, {Twicken}, \&
  {Chandrasekaran}}]{2020TPSkdph}
{Jenkins}, J.~M., {Tenenbaum}, P., {Seader}, S., {et~al.} 2020{\natexlab{a}},
  {Kepler Data Processing Handbook: Transiting Planet Search}, Kepler Science
  Document KSCI-19081-003

\bibitem[{{Jenkins} {et~al.}(2016){Jenkins}, {Twicken}, {McCauliff},
  {Campbell}, {Sanderfer}, {Lung}, {Mansouri-Samani}, {Girouard}, {Tenenbaum},
  {Klaus}, {Smith}, {Caldwell}, {Chacon}, {Henze}, {Heiges}, {Latham},
  {Morgan}, {Swade}, {Rinehart}, \& {Vanderspek}}]{jenkins2016}
{Jenkins}, J.~M., {Twicken}, J.~D., {McCauliff}, S., {et~al.} 2016, in Society
  of Photo-Optical Instrumentation Engineers (SPIE) Conference Series, Vol.
  9913, \procspie, 99133E

\bibitem[{{Jenkins} {et~al.}(2020{\natexlab{b}}){Jenkins}, {D{\'\i}az},
  {Kurtovic}, {Espinoza}, {Vines}, {Rojas}, {Brahm}, {Torres},
  {Cort{\'e}s-Zuleta}, {Soto}, {Lopez}, {King}, {Wheatley}, {Winn}, {Ciardi},
  {Ricker}, {Vanderspek}, {Latham}, {Seager}, {Jenkins}, {Beichman}, {Bieryla},
  {Burke}, {Christiansen}, {Henze}, {Klaus}, {McCauliff}, {Mori}, {Narita},
  {Nishiumi}, {Tamura}, {de Leon}, {Quinn}, {Villase{\~n}or}, {Vezie},
  {Lissauer}, {Collins}, {Collins}, {Isopi}, {Mallia}, {Ercolino}, {Petrovich},
  {Jord{\'a}n}, {Acton}, {Armstrong}, {Bayliss}, {Bouchy}, {Belardi}, {Bryant},
  {Burleigh}, {Cabrera}, {Casewell}, {Chaushev}, {Cooke}, {Eigm{\"u}ller},
  {Erikson}, {Foxell}, {G{\"a}nsicke}, {Gill}, {Gillen}, {G{\"u}nther}, {Goad},
  {Hooton}, {Jackman}, {Louden}, {McCormac}, {Moyano}, {Nielsen}, {Pollacco},
  {Queloz}, {Rauer}, {Raynard}, {Smith}, {Tilbrook}, {Titz-Weider}, {Turner},
  {Udry}, {Walker}, {Watson}, {West}, {Palle}, {Ziegler}, {Law}, \&
  {Mann}}]{2020NatAs...4.1148J}
{Jenkins}, J.~S., {D{\'\i}az}, M.~R., {Kurtovic}, N.~T., {et~al.}
  2020{\natexlab{b}}, Nature Astronomy, 4, 1148

\bibitem[{{Jin} {et~al.}(2014){Jin}, {Mordasini}, {Parmentier}, {van Boekel},
  {Henning}, \& {Ji}}]{2014ApJ...795...65J}
{Jin}, S., {Mordasini}, C., {Parmentier}, V., {et~al.} 2014, \apj, 795, 65

\bibitem[{{Kempton} {et~al.}(2018){Kempton}, {Bean}, {Louie}, {Deming}, {Koll},
  {Mansfield}, {Christiansen}, {L{\'o}pez-Morales}, {Swain}, {Zellem},
  {Ballard}, {Barclay}, {Barstow}, {Batalha}, {Beatty}, {Berta-Thompson},
  {Birkby}, {Buchhave}, {Charbonneau}, {Cowan}, {Crossfield}, {de Val-Borro},
  {Doyon}, {Dragomir}, {Gaidos}, {Heng}, {Hu}, {Kane}, {Kreidberg}, {Mallonn},
  {Morley}, {Narita}, {Nascimbeni}, {Pall{\'e}}, {Quintana}, {Rauscher},
  {Seager}, {Shkolnik}, {Sing}, {Sozzetti}, {Stassun}, {Valenti}, \& {von
  Essen}}]{2018PASP..130k4401K}
{Kempton}, E. M.~R., {Bean}, J.~L., {Louie}, D.~R., {et~al.} 2018, \pasp, 130,
  114401

\bibitem[{{Kipping}(2013)}]{Kipping2013}
{Kipping}, D.~M. 2013, \mnras, 435, 2152

\bibitem[{{Kubyshkina} {et~al.}(2018){Kubyshkina}, {Lendl}, {Fossati},
  {Cubillos}, {Lammer}, {Erkaev}, \& {Johnstone}}]{2018A&A...612A..25K}
{Kubyshkina}, D., {Lendl}, M., {Fossati}, L., {et~al.} 2018, \aap, 612, A25

\bibitem[{{Kuerster} {et~al.}(1997){Kuerster}, {Schmitt}, {Cutispoto}, \&
  {Dennerl}}]{1997A&A...320..831K}
{Kuerster}, M., {Schmitt}, J.~H.~M.~M., {Cutispoto}, G., \& {Dennerl}, K. 1997,
  \aap, 320, 831

\bibitem[{{Kupka} {et~al.}(2000){Kupka}, {Ryabchikova}, {Piskunov}, {Stempels},
  \& {Weiss}}]{Kupka2000}
{Kupka}, F.~G., {Ryabchikova}, T.~A., {Piskunov}, N.~E., {Stempels}, H.~C., \&
  {Weiss}, W.~W. 2000, Baltic Astronomy, 9, 590

\bibitem[{{Kurucz}(1993)}]{1993yCat.6039....0K}
{Kurucz}, R.~L. 1993, VizieR Online Data Catalog, VI/39

\bibitem[{{Kurucz}(2013)}]{Kurucz2013}
{Kurucz}, R.~L. 2013, {ATLAS12: Opacity sampling model atmosphere program},
  Astrophysics Source Code Library

\bibitem[{{Li} {et~al.}(2019){Li}, {Tenenbaum}, {Twicken}, {Burke}, {Jenkins},
  {Quintana}, {Rowe}, \& {Seader}}]{Li:DVmodelFit2019}
{Li}, J., {Tenenbaum}, P., {Twicken}, J.~D., {et~al.} 2019, \pasp, 131, 024506

\bibitem[{{Livingston} {et~al.}(2018){Livingston}, {Crossfield}, {Petigura},
  {Gonzales}, {Ciardi}, {Beichman}, {Christiansen}, {Dressing}, {Henning},
  {Howard}, {Isaacson}, {Fulton}, {Kosiarek}, {Schlieder}, {Sinukoff}, \&
  {Tamura}}]{2018AJ....156..277L}
{Livingston}, J.~H., {Crossfield}, I. J.~M., {Petigura}, E.~A., {et~al.} 2018,
  \aj, 156, 277

\bibitem[{{Lodders} {et~al.}(2009){Lodders}, {Palme}, \& {Gail}}]{lodders09}
{Lodders}, K., {Palme}, H., \& {Gail}, H.~P. 2009, Landolt B\&ouml;rnstein, 4B,
  712

\bibitem[{{Lopez} \& {Fortney}(2013)}]{2013ApJ...776....2L}
{Lopez}, E.~D. \& {Fortney}, J.~J. 2013, \apj, 776, 2

\bibitem[{{Lopez} \& {Fortney}(2014)}]{LopFor14}
{Lopez}, E.~D. \& {Fortney}, J.~J. 2014, \apj, 792, 1

\bibitem[{{Lovis} \& {Pepe}(2007)}]{2007A&A...468.1115L}
{Lovis}, C. \& {Pepe}, F. 2007, \aap, 468, 1115

\bibitem[{{Lundkvist} {et~al.}(2016){Lundkvist}, {Kjeldsen}, {Albrecht},
  {Davies}, {Basu}, {Huber}, {Justesen}, {Karoff}, {Silva Aguirre}, {van
  Eylen}, {Vang}, {Arentoft}, {Barclay}, {Bedding}, {Campante}, {Chaplin},
  {Christensen-Dalsgaard}, {Elsworth}, {Gilliland}, {Handberg}, {Hekker},
  {Kawaler}, {Lund}, {Metcalfe}, {Miglio}, {Rowe}, {Stello}, {Tingley}, \&
  {White}}]{2016NatCo...711201L}
{Lundkvist}, M.~S., {Kjeldsen}, H., {Albrecht}, S., {et~al.} 2016, Nature
  Communications, 7, 11201

\bibitem[{{Mamajek} \& {Hillenbrand}(2008)}]{2008ApJ...687.1264M}
{Mamajek}, E.~E. \& {Hillenbrand}, L.~A. 2008, \apj, 687, 1264

\bibitem[{{Mandel} \& {Agol}(2002)}]{Mandel2002}
{Mandel}, K. \& {Agol}, E. 2002, \apjl, 580, L171

\bibitem[{{Mayor} {et~al.}(2003){Mayor}, {Pepe}, {Queloz}, {Bouchy},
  {Rupprecht}, {Lo Curto}, {Avila}, {Benz}, {Bertaux}, {Bonfils}, {Dall},
  {Dekker}, {Delabre}, {Eckert}, {Fleury}, {Gilliotte}, {Gojak}, {Guzman},
  {Kohler}, {Lizon}, {Longinotti}, {Lovis}, {Megevand}, {Pasquini}, {Reyes},
  {Sivan}, {Sosnowska}, {Soto}, {Udry}, {van Kesteren}, {Weber}, \&
  {Weilenmann}}]{2003Msngr.114...20M}
{Mayor}, M., {Pepe}, F., {Queloz}, D., {et~al.} 2003, The Messenger, 114, 20

\bibitem[{{Mazeh} {et~al.}(2016){Mazeh}, {Holczer}, \&
  {Faigler}}]{2016A&A...589A..75M}
{Mazeh}, T., {Holczer}, T., \& {Faigler}, S. 2016, \aap, 589, A75

\bibitem[{{Mazeh} {et~al.}(2013){Mazeh}, {Nachmani}, {Holczer}, {Fabrycky},
  {Ford}, {Sanchis-Ojeda}, {Sokol}, {Rowe}, {Zucker}, {Agol}, {Carter},
  {Lissauer}, {Quintana}, {Ragozzine}, {Steffen}, \& {Welsh}}]{Mazeh13}
{Mazeh}, T., {Nachmani}, G., {Holczer}, T., {et~al.} 2013, \apjs, 208, 16

\bibitem[{{McCully} {et~al.}(2018){McCully}, {Volgenau}, {Harbeck}, {Lister},
  {Saunders}, {Turner}, {Siiverd}, \& {Bowman}}]{McCully2018}
{McCully}, C., {Volgenau}, N.~H., {Harbeck}, D.-R., {et~al.} 2018, in Society
  of Photo-Optical Instrumentation Engineers (SPIE) Conference Series, Vol.
  10707, \procspie, 107070K

\bibitem[{McKay {et~al.}(2019)McKay, DiSanti, Kelley, Knight, Womack,
  Wierzchos, Pinto, Bonev, Villanueva, Russo, Cochran, Biver, Bauer, Ronald
  J.~Vervack, Gibb, Roth, \& Kawakita}]{mckay19}
McKay, A.~J., DiSanti, M.~A., Kelley, M. S.~P., {et~al.} 2019, The Astronomical
  Journal, 158, 128

\bibitem[{Mousis {et~al.}(2020)Mousis, Deleuil, Aguichine, Marcq, Naar,
  Aguirre, Brugger, \& Gon{\c{c}}alves}]{Mousis20}
Mousis, O., Deleuil, M., Aguichine, A., {et~al.} 2020, The Astrophysical
  Journal, 896, L22

\bibitem[{{Otegi} {et~al.}(2020){Otegi}, {Dorn, C.}, {Helled, R.}, {Bouchy,
  F.}, {Haldemann, J.}, \& {Alibert, Y.}}]{Otegi20}
{Otegi}, J.~F., {Dorn, C.}, {Helled, R.}, {et~al.} 2020, A\&A, 640, A135

\bibitem[{{Owen} \& {Jackson}(2012)}]{Owen2012}
{Owen}, J.~E. \& {Jackson}, A.~P. 2012, \mnras, 425, 2931

\bibitem[{{Owen} \& {Lai}(2018)}]{2018MNRAS.479.5012O}
{Owen}, J.~E. \& {Lai}, D. 2018, \mnras, 479, 5012

\bibitem[{{Owen} \& {Wu}(2013)}]{2013ApJ...775..105O}
{Owen}, J.~E. \& {Wu}, Y. 2013, \apj, 775, 105

\bibitem[{{Parviainen}(2015)}]{2015MNRAS.450.3233P}
{Parviainen}, H. 2015, \mnras, 450, 3233

\bibitem[{{Pepe} {et~al.}(2002){Pepe}, {Mayor}, {Galland}, {Naef}, {Queloz},
  {Santos}, {Udry}, \& {Burnet}}]{Pepe2002}
{Pepe}, F., {Mayor}, M., {Galland}, F., {et~al.} 2002, \aap, 388, 632

\bibitem[{{Persson} {et~al.}(2018){Persson}, {Fridlund}, {Barrag{\'a}n}, {Dai},
  {Gandolfi}, {Hatzes}, {Hirano}, {Grziwa}, {Korth}, {Prieto-Arranz},
  {Fossati}, {Van Eylen}, {Justesen}, {Livingston}, {Kubyshkina}, {Deeg},
  {Guenther}, {Nowak}, {Cabrera}, {Eigm{\"u}ller}, {Csizmadia}, {Smith},
  {Erikson}, {Albrecht}, {Sobrino}, {Cochran}, {Endl}, {Esposito}, {Fukui},
  {Heeren}, {Hidalgo}, {Hjorth}, {Kuzuhara}, {Narita}, {Nespral}, {Palle},
  {P{\"a}tzold}, {Rauer}, {Rodler}, \& {Winn}}]{2018A&A...618A..33P}
{Persson}, C.~M., {Fridlund}, M., {Barrag{\'a}n}, O., {et~al.} 2018, \aap, 618,
  A33

\bibitem[{{Petigura} {et~al.}(2022){Petigura}, {Rogers}, {Isaacson}, {Owen},
  {Kraus}, {Winn}, {MacDougall}, {Howard}, {Fulton}, {Kosiarek}, {Weiss},
  {Behmard}, \& {Blunt}}]{2022arXiv220110020P}
{Petigura}, E.~A., {Rogers}, J.~G., {Isaacson}, H., {et~al.} 2022, arXiv
  e-prints, arXiv:2201.10020

\bibitem[{{Piskunov} \& {Valenti}(2017)}]{pv2017}
{Piskunov}, N. \& {Valenti}, J.~A. 2017, \aap, 597, A16

\bibitem[{{Poppenhaeger} {et~al.}(2021){Poppenhaeger}, {Ketzer}, \&
  {Mallonn}}]{2021MNRAS.500.4560P}
{Poppenhaeger}, K., {Ketzer}, L., \& {Mallonn}, M. 2021, \mnras, 500, 4560

\bibitem[{{Ricker} {et~al.}(2015){Ricker}, {Winn}, {Vanderspek}, {Latham},
  {Bakos}, {Bean}, {Berta-Thompson}, {Brown}, {Buchhave}, {Butler}, {Butler},
  {Chaplin}, {Charbonneau}, {Christensen-Dalsgaard}, {Clampin}, {Deming},
  {Doty}, {De Lee}, {Dressing}, {Dunham}, {Endl}, {Fressin}, {Ge}, {Henning},
  {Holman}, {Howard}, {Ida}, {Jenkins}, {Jernigan}, {Johnson}, {Kaltenegger},
  {Kawai}, {Kjeldsen}, {Laughlin}, {Levine}, {Lin}, {Lissauer}, {MacQueen},
  {Marcy}, {McCullough}, {Morton}, {Narita}, {Paegert}, {Palle}, {Pepe},
  {Pepper}, {Quirrenbach}, {Rinehart}, {Sasselov}, {Sato}, {Seager},
  {Sozzetti}, {Stassun}, {Sullivan}, {Szentgyorgyi}, {Torres}, {Udry}, \&
  {Villasenor}}]{2015JATIS...1a4003R}
{Ricker}, G.~R., {Winn}, J.~N., {Vanderspek}, R., {et~al.} 2015, Journal of
  Astronomical Telescopes, Instruments, and Systems, 1, 014003

\bibitem[{{Ryabchikova} {et~al.}(2015){Ryabchikova}, {Piskunov}, {Kurucz},
  {Stempels}, {Heiter}, {Pakhomov}, \& {Barklem}}]{Ryabchikova2015}
{Ryabchikova}, T., {Piskunov}, N., {Kurucz}, R.~L., {et~al.} 2015, \physscr,
  90, 054005

\bibitem[{{Sanchis-Ojeda} {et~al.}(2014){Sanchis-Ojeda}, {Rappaport}, {Winn},
  {Kotson}, {Levine}, \& {El Mellah}}]{2014ApJ...787...47S}
{Sanchis-Ojeda}, R., {Rappaport}, S., {Winn}, J.~N., {et~al.} 2014, \apj, 787,
  47

\bibitem[{{Sanz-Forcada} {et~al.}(2011){Sanz-Forcada}, {Micela}, {Ribas},
  {Pollock}, {Eiroa}, {Velasco}, {Solano}, \&
  {Garc{\'\i}a-{\'A}lvarez}}]{Sanz2011}
{Sanz-Forcada}, J., {Micela}, G., {Ribas}, I., {et~al.} 2011, \aap, 532, A6

\bibitem[{{Schlegel} {et~al.}(1998){Schlegel}, {Finkbeiner}, \&
  {Davis}}]{1998ApJ...500..525S}
{Schlegel}, D.~J., {Finkbeiner}, D.~P., \& {Davis}, M. 1998, \apj, 500, 525

\bibitem[{{Smith} {et~al.}(2017){Smith}, {Gandolfi}, {Barrag{\'a}n}, {Bowler},
  {Csizmadia}, {Endl}, {Fridlund}, {Grziwa}, {Guenther}, {Hatzes}, {Nowak},
  {Albrecht}, {Alonso}, {Cabrera}, {Cochran}, {Deeg}, {Cusano},
  {Eigm{\"u}ller}, {Erikson}, {Hidalgo}, {Hirano}, {Johnson}, {Korth}, {Mann},
  {Narita}, {Nespral}, {Palle}, {P{\"a}tzold}, {Prieto-Arranz}, {Rauer},
  {Ribas}, {Tingley}, \& {Wolthoff}}]{2017MNRAS.464.2708S}
{Smith}, A.~M.~S., {Gandolfi}, D., {Barrag{\'a}n}, O., {et~al.} 2017, \mnras,
  464, 2708

\bibitem[{Smith {et~al.}(2020)}]{Smith2020_PDCSAP}
Smith, J.~C. {et~al.} 2020, in Kepler Data Processing Handbook: KSCI-19081-003,
  ed. J.~M. Jenkins, 131--197

\bibitem[{{Sotin} {et~al.}(2007){Sotin}, {Grasset}, \& {Mocquet}}]{sotin07}
{Sotin}, C., {Grasset}, O., \& {Mocquet}, A. 2007, \icarus, 191, 337

\bibitem[{{Stumpe} {et~al.}(2014){Stumpe}, {Smith}, {Catanzarite}, {Van Cleve},
  {Jenkins}, {Twicken}, \& {Girouard}}]{Stumpe2014}
{Stumpe}, M.~C., {Smith}, J.~C., {Catanzarite}, J.~H., {et~al.} 2014, \pasp,
  126, 100

\bibitem[{{Szab{\'o}} \& {Kiss}(2011)}]{2011ApJ...727L..44S}
{Szab{\'o}}, G.~M. \& {Kiss}, L.~L. 2011, \apjl, 727, L44

\bibitem[{{Tokovinin}(2018)}]{2018PASP..130c5002T}
{Tokovinin}, A. 2018, \pasp, 130, 035002

\bibitem[{{Tu} {et~al.}(2015){Tu}, {Johnstone}, {G{\"u}del}, \&
  {Lammer}}]{2015A&A...577L...3T}
{Tu}, L., {Johnstone}, C.~P., {G{\"u}del}, M., \& {Lammer}, H. 2015, \aap, 577,
  L3

\bibitem[{{Twicken} {et~al.}(2018){Twicken}, {Catanzarite}, {Clarke},
  {Girouard}, {Jenkins}, {Klaus}, {Li}, {McCauliff}, {Seader}, {Tenenbaum},
  {Wohler}, {Bryson}, {Burke}, {Caldwell}, {Haas}, {Henze}, \&
  {Sanderfer}}]{Twicken:DVdiagnostics2018}
{Twicken}, J.~D., {Catanzarite}, J.~H., {Clarke}, B.~D., {et~al.} 2018, \pasp,
  130, 064502

\bibitem[{{Valenti} \& {Piskunov}(1996)}]{vp96}
{Valenti}, J.~A. \& {Piskunov}, N. 1996, \aaps, 118, 595

\bibitem[{{Van Eylen} {et~al.}(2018){Van Eylen}, {Agentoft}, {Lundkvist},
  {Kjeldsen}, {Owen}, {Fulton}, {Petigura}, \& {Snellen}}]{VanEylen2018}
{Van Eylen}, V., {Agentoft}, C., {Lundkvist}, M.~S., {et~al.} 2018, \mnras,
  479, 4786

\bibitem[{{Van Eylen} {et~al.}(2021){Van Eylen}, {Astudillo-Defru}, {Bonfils},
  {Livingston}, {Hirano}, {Luque}, {Lam}, {Justesen}, {Winn}, {Gandolfi},
  {Nowak}, {Palle}, {Albrecht}, {Dai}, {Campos Estrada}, {Owen},
  {Foreman-Mackey}, {Fridlund}, {Korth}, {Mathur}, {Forveille}, {Mikal-Evans},
  {Osborne}, {Ho}, {Almenara}, {Artigau}, {Barrag{\'a}n}, {Barros}, {Bouchy},
  {Cabrera}, {Caldwell}, {Charbonneau}, {Chaturvedi}, {Cochran}, {Csizmadia},
  {Damasso}, {Delfosse}, {De Medeiros}, {D{\'\i}az}, {Doyon}, {Esposito},
  {F{\H{u}}r{\'e}sz}, {Figueira}, {Georgieva}, {Goffo}, {Grziwa}, {Guenther},
  {Hatzes}, {Jenkins}, {Kabath}, {Knudstrup}, {Latham}, {Lavie}, {Lovis},
  {Mennickent}, {Mullally}, {Murgas}, {Narita}, {Pepe}, {Persson}, {Redfield},
  {Ricker}, {Santos}, {Seager}, {Serrano}, {Smith}, {Su{\'a}rez Mascare{\~n}o},
  {Subjak}, {Twicken}, {Udry}, {Vanderspek}, \& {Zapatero
  Osorio}}]{2021MNRAS.507.2154V}
{Van Eylen}, V., {Astudillo-Defru}, N., {Bonfils}, X., {et~al.} 2021, \mnras,
  507, 2154

\bibitem[{{Vines} \& {Jenkins}(2022)}]{2022arXiv220403769V}
{Vines}, J.~I. \& {Jenkins}, J.~S. 2022, arXiv e-prints, arXiv:2204.03769

\bibitem[{{Yee} {et~al.}(2017){Yee}, {Petigura}, \& {von
  Braun}}]{2017ApJ...836...77Y}
{Yee}, S.~W., {Petigura}, E.~A., \& {von Braun}, K. 2017, \apj, 836, 77

\bibitem[{{Zechmeister} \& {K{\"u}rster}(2009)}]{Zech09}
{Zechmeister}, M. \& {K{\"u}rster}, M. 2009, \aap, 496, 577

\bibitem[{{Zechmeister} {et~al.}(2018){Zechmeister}, {Reiners}, {Amado},
  {Azzaro}, {Bauer}, {B{\'e}jar}, {Caballero}, {Guenther}, {Hagen}, {Jeffers},
  {Kaminski}, {K{\"u}rster}, {Launhardt}, {Montes}, {Morales}, {Quirrenbach},
  {Reffert}, {Ribas}, {Seifert}, {Tal-Or}, \& {Wolthoff}}]{Zechmeister2018}
{Zechmeister}, M., {Reiners}, A., {Amado}, P.~J., {et~al.} 2018, \aap, 609, A12

\bibitem[{{Zeng} {et~al.}(2019){Zeng}, {Jacobsen}, {Sasselov}, {Petaev},
  {Vanderburg}, {Lopez-Morales}, {Perez-Mercader}, {Mattsson}, {Li}, {Heising},
  {Bonomo}, {Damasso}, {Berger}, {Cao}, {Levi}, \&
  {Wordsworth}}]{2019PNAS..116.9723Z}
{Zeng}, L., {Jacobsen}, S.~B., {Sasselov}, D.~D., {et~al.} 2019, Proceedings of
  the National Academy of Science, 116, 9723

\bibitem[{{Ziegler} {et~al.}(2020){Ziegler}, {Tokovinin}, {Brice{\~n}o},
  {Mang}, {Law}, \& {Mann}}]{2020AJ....159...19Z}
{Ziegler}, C., {Tokovinin}, A., {Brice{\~n}o}, C., {et~al.} 2020, \aj, 159, 19

\end{thebibliography}
 
 \appendix
 
\section{HARPS data}
\clearpage
\begin{sidewaystable*}
\begin{small}
\begin{center}
  \caption{Radial velocities and spectral activity indicators extracted from  HARPS spectra.
    \label{all_rv.tex}}
    \resizebox{0.95\columnwidth}{!}{%
  \begin{tabular}{rrrrrrrrrrrcc}
    \hline
    \hline
     \noalign{\smallskip}
    \multicolumn{1}{c}{BJD$_\mathrm{TBD}$ (d)} &
    \multicolumn{1}{c}{RV} &
    \multicolumn{1}{c}{$\sigma_\mathrm{RV}$} &
    \multicolumn{1}{c}{BIS\tablefootmark{a}}&
    \multicolumn{1}{c}{FHWM} &
    \multicolumn{1}{c}{$\sigma_\mathrm{FWHM}$} &
    \multicolumn{1}{c}{dlW} &
    \multicolumn{1}{c}{$\sigma_\mathrm{dlW}$} &
    \multicolumn{1}{c}{S-index} &
    \multicolumn{1}{c}{$\sigma_\mathrm{S-index}$} &   
    \multicolumn{1}{c}{$\mathrm{H_{\alpha}}$\tablefootmark{a}} &
    \multicolumn{1}{c}{$\mathrm{T_{exp}}$} &
    \multicolumn{1}{c}{SNR}    \\ 
    \noalign{\smallskip}
    \multicolumn{1}{c}{-2457000} &
    \multicolumn{1}{c}{(\ms)} &
   \multicolumn{1}{c}{(\ms)} &
   \multicolumn{1}{c}{(\ms)} &
   \multicolumn{1}{c}{(\kms)} &
   \multicolumn{1}{c}{(\kms)} &
   \multicolumn{1}{c}{(\kms)} &
   \multicolumn{1}{c}{(\kms)} &
    \multicolumn{1}{c}{---} &
    \multicolumn{1}{c}{---} &
    \multicolumn{1}{c}{---} &
    \multicolumn{1}{c}{(s)} &
    \multicolumn{1}{c}{$@550$~nm} \\ 
    \hline
2419.661798 &  0.016377394& 	0.002998808& 	-0.029284737	& 7.140376417		& 0.357018821	& 	11.22278765	& 	1.846614321& 	0.172559779& 	0.003497890& 	0.998197146& 	2100	& 	28.4\\
2419.812144& 	0.020510551& 		0.002956139& 	-0.014619464	& 7.138481028		& 0.356924051	& 	11.05843113	& 	2.062412649& 	0.181116594& 	0.003877305& 	0.999901855& 	2100	& 	29.1\\
2424.705788& 	0.018156264& 		0.002574676& 	-0.006295520	& 7.137545136		& 0.356877257	& 	-1.018335758	& 	1.512092568& 	0.160211164& 	0.002998846& 	0.995506908& 	2400	& 	33.5\\
2427.626567& 	-0.000481393& 	0.001746737& 	-0.014103206	& 7.137688356		& 0.356884418	& 	-8.910686514	& 	1.508117865& 	0.162656713& 	0.002592005& 	1.005319679& 	2400	& 	38.5\\
2427.810834& 	0.014869853& 		0.002302020& 	-0.014228541	& 7.151375312		& 0.357568766	& 	-7.928578176	& 	1.652097812& 	0.172320637& 	0.003198341& 	1.003490323& 	2400	& 	37.9\\
2428.639266& 	-0.009267622& 	0.002545470& 	-0.011727208	& 7.143755477		& 0.357187774	& 	2.109340183	& 	1.791528912& 	0.161061807& 	0.003508646& 	0.989901071& 	2400	& 	28.1\\
2428.720190& 	-0.009409333& 	0.002468478& 	-0.014294956	& 7.143700306		& 0.357185015	& 	-5.089153308	& 	1.945891805& 	0.164078903& 	0.003161212& 	0.985814569& 	2100	& 	25.7\\
2428.793337& 	-0.005702919& 	0.003047684& 	-0.017037973	& 7.147611628		& 0.357380581	& 	-0.909110563	& 	1.543522493& 	0.171465389& 	0.004068059& 	0.988770164& 	2100	& 	28.1\\
2429.600426& 	0.005566032& 		0.002201269& 	-0.022909545	& 7.142051926		& 0.357102596	& 	-3.451494525	& 	1.456484985& 	0.177564159& 	0.003168727& 	0.990316656& 	2400	& 	32.4\\
2429.701720& 	-0.005281992& 	0.002460195& 	-0.018750634	& 7.154825057		& 0.357741253	& 	-2.086622561	& 	1.290738998& 	0.172611870& 	0.003022147& 	0.986268731& 	2400	& 	31.1\\
2429.779264& 	-0.003881037& 	0.002363704& 	-0.003667828	& 7.149787004		& 0.357489350	& 	-2.847121560	& 	1.483978769& 	0.182034670& 	0.003499536& 	0.999352266& 	2400	& 	31.7\\
2430.605449& 	0.020690196& 		0.002615154& 	-0.026428800	& 7.140959158		& 0.357047958	& 	-0.876773808	& 	1.898119925& 	0.170006651& 	0.003561435& 	0.988708161& 	2400	& 	28.5\\
2430.675216& 	0.009714846& 		0.003448313& 	-0.029337161	& 7.142366435		& 0.357118322	& 	3.997268509	& 	2.841797354& 	0.156868483& 	0.004172708& 	1.009883115& 	2400	& 	22.7\\
2430.793963& 	0&             		0.004101949& 	-0.014566066	& 7.154824695		& 0.357741235	& 	17.76124298	& 	2.793467679& 	0.152186686& 	0.005026666& 	0.990507323& 	2400	& 	19.5\\
2432.708547& 	0.010934787& 		0.002813860& 	-0.017938111	& 7.142515648		& 0.357125782	& 	3.522072813	& 	2.007823948& 	0.173702481& 	0.004060536& 	0.996595505& 	2100	& 	26.7\\
2432.771624& 	0.020807163& 		0.003165590& 	0.005396245	& 7.146888465		& 0.357344423	& 	1.482052001	& 	2.180986562& 	0.197696011& 	0.004567345& 	1.000737242& 	2400	& 	25.7\\
2433.589404& 	-0.005759122& 	0.004543362& 	-0.012638667	& 7.141966902		& 0.357098345	& 	13.73210968	& 	3.238231507& 	0.156301540& 	0.004919294& 	1.005853124& 	2400	& 	18.4\\
2460.511224& 	0.005895298& 		0.002957171& 	-0.023272945	& 7.138558338		& 0.356927917	& 	0.173120003	& 	2.294055305& 	0.160858365& 	0.003977862& 	0.990624216& 	2400	& 	24.2\\
2460.703136& 	-0.006202675& 	0.002962588& 	-0.018801690	& 7.132139304		& 0.356606965	& 	1.994551964	& 	2.507608654& 	0.167471110& 	0.004007204& 	0.996410287& 	2400	& 	26.6\\
2461.528107& 	0.015996184& 		0.002325161& 	-0.025352893	& 7.120917671		& 0.356045884	& 	-4.356761078	& 	1.674535666& 	0.167688344& 	0.003392494& 	0.993608739& 	2400	& 	29.1\\
2461.603775& 	0.019667017& 		0.003719634& 	-0.019049084	& 7.139453642		& 0.356972682	& 	6.047329254	& 	2.393000701& 	0.234380548& 	0.005077442& 	0.985384534& 	2700	& 	23.5\\
2462.548288& 	0.004650384& 		0.002753746& 	-0.013112217	& 7.148418697		& 0.357420935	& 	-2.298627181	& 	1.835342272& 	0.163604136& 	0.003269482& 	1.003634973& 	2400	& 	28.8\\
2462.697690& 	0.019120366& 		0.002733685& 	-0.023693674	& 7.140938077		& 0.357046904	& 	1.663468315	& 	2.176539325& 	0.163773867& 	0.003736499& 	1.005965119& 	2400	& 	28.4\\
2463.523956& 	-0.007480005& 	0.004354283& 	0.003485374	& 7.122732003		& 0.356136600	& 	12.53839580	& 	2.886964792& 	0.246356354& 	0.005634937& 	0.984976420& 	2400	& 	21.3\\
2463.686332& 	0.004361950& 		0.002607838& 	-0.022706813	& 7.151924899		& 0.357596245	& 	4.120020246	& 	2.327407159& 	0.186998648& 	0.004292668& 	0.996282723& 	2400	& 	25.9\\
2467.619853& 	0.012521115& 		0.002237602& 	-0.018435484	& 7.134447548		& 0.356722377	& 	-4.351583321	& 	1.134570545& 	0.170492054& 	0.003035039& 	0.998282998& 	2100	& 	36.2\\
2470.576694& 	-0.014622498& 	0.003224917& 	-0.006783539	& 7.135561560		& 0.356778078	& 	0.830073320	& 	2.446677634& 	0.222651785& 	0.004940337& 	0.994695320& 	2400	& 	24.7\\
2471.526954& 	-0.009509969& 	0.001711479& 	-0.016340895	& 7.135242022		& 0.356762101	& 	-9.899564167	& 	1.551739631& 	0.169208331& 	0.002841745& 	0.994933456& 	2400	& 	38.4\\
2472.542615& 	-0.000106014& 	0.002917387& 	-0.016347214	& 7.121805322		& 0.356090266	& 	5.079370467	& 	2.249824118& 	0.170382224& 	0.003877015& 	0.994651399& 	2400	& 	26.9\\
2473.515372& 	0.011278635& 		0.001998208& 	-0.014749743	& 7.145702914		& 0.357285146	& 	-4.351123275	& 	1.618986012& 	0.162113008& 	0.002889922& 	0.991353982& 	2400	& 	35.4\\
2473.636500& 	0.008755525& 		0.002003567& 	-0.021534768	& 7.138369046		& 0.356918452	& 	-9.531662223	& 	1.359621503& 	0.187348591& 	0.003337473& 	0.994049878& 	2400	& 	37.2\\
2474.557662& 	 0.009871217& 	0.002272828& 	-0.020330444	& 7.141533353		& 0.357076668	& 	-0.237590000	& 	1.540770964& 	0.171545292& 	0.003304167& 	1.000336629& 	2400	& 	32.2\\
2475.556563& 	-0.003711372& 	0.002152913& 	-0.015830215	& 7.147198245		& 0.357359912	& 	 0.211648000	& 	1.715704638& 	0.174905076& 	0.003290315& 	1.011519561& 	2400	& 	32.1\\
2475.670168& 	 0.002865511& 		0.002723374& 	-0.010147531	& 7.136697027		& 0.356834851	& 	8.052287871	& 	2.305851198& 	0.175846935& 	0.004053048& 	0.992356017& 	2400	& 	27.7\\
2476.559248& 	-0.020460820& 	0.003324717& 	-0.008768004	& 7.143915814		& 0.357195791	& 	8.236826883	& 	2.036921846& 	0.160440241& 	0.003685363& 	0.998942580& 	2400	& 	26.4\\
2476.661638& 	-0.013027107& 	0.004830712& 	-0.014629884	& 7.111509280		& 0.355575464	& 	45.07848938	& 	3.356598711& 	0.183341839& 	0.005507660& 	0.999140509& 	2400	& 	19.6\\
2477.552408& 	-0.014509429& 	0.005655293& 	-0.023830532	& 7.148861028		& 0.357443051	& 	31.61015391	& 	3.608882476& 	0.211182628& 	0.006363658& 	1.000245782& 	2400	& 	16.9\\
2477.656210& 	-0.020161994& 	0.003897063& 	-0.038365216	& 7.144018458		& 0.357200923	& 	16.59729732	& 	2.786390384& 	0.201276239& 	0.005205999& 	0.988922086& 	2400	& 	22.3\\
2478.601501& 	-0.007845141& 	0.002402883& 	-0.017684734	& 7.153895769		& 0.357694788	& 	5.732488697	& 	2.000829919& 	0.173492875& 	0.003579341& 	0.993562249& 	2400	& 	30\\
2502.537972& 	-0.021841648& 	0.002927285& 	-0.014627725	& 7.121132243		& 0.356056612	& 	1.435156202	& 	1.865134741& 	0.194001634& 	0.004109944& 	0.998458563& 	2400	& 	28.6\\
2530.516620& 	-0.020554079& 	0.002689375& 	-0.015004302	& 7.132953081		& 0.356647654	& 	-2.396505538	& 	1.873603833& 	0.183751932& 	0.003612815& 	0.993572861& 	2400	& 	31.3\\
    \hline
  \end{tabular}
  }
  \tablefoot{
\tablefoottext{a}{We used 5~\% of the activity indicator as  uncertainties.}
}
\end{center}
\end{small}
\end{sidewaystable*}

\end{document}